\documentclass[a4paper,12pt,authoryear,twoside,sort&compress,prc]{revtex4-1}

\usepackage{fancyhdr}
\bibpunct{[}{]}{,}{n}{}{;}

\fancypagestyle{plain}{
      \fancyhead{}
      
}
\usepackage{color}
\usepackage{latexsym}
\usepackage{amsbsy}
\usepackage{amsmath}
\usepackage{amssymb}
\usepackage[varg]{txfonts}
\usepackage{mathrsfs}

\usepackage{multirow}
\usepackage{booktabs}
\usepackage{tabularx}
\usepackage{upgreek}
\usepackage[small]{caption}
\usepackage[]{subfig}
\usepackage{verbatim}
\usepackage{array}
\usepackage{color}
\usepackage{graphicx}
\usepackage{chemarr}
\usepackage{caption}

\DeclareMathAlphabet{\mathpzc}{OT1}{pzc}{m}{it}

\renewcommand\arraystretch{1.2}

\begin{document}


\title{Quantitative study of coherent pairing modes with two neutron transfer: Sn-isotopes}
\author{G. Potel}
\email{gregory.potel@gmail.com}
\affiliation{Dipartimento di Fisica, Universit\`{a} di Milano, Via Celoria 16, 20133 Milano, Italy.}
\affiliation{INFN, Sezione di Milano Via Celoria 16, 20133 Milano, Italy.}
\affiliation{Departamento de Fisica Atomica, Molecular y Nuclear, Universidad de Sevilla, Facultad de Fisica, Avenida Reina Mercedes s/n, Sevilla, Spain}

\author{A. Idini}
\email{andrea.idini@mi.infn.it}
\affiliation{Dipartimento di Fisica, Universit\`{a} di Milano,Via Celoria 16, 20133 Milano, Italy.}
\affiliation{INFN, Sezione di Milano Via Celoria 16, 20133 Milano, Italy.}

\author{F. Barranco}
\email{barranco@us.es}
\affiliation{Departamento de Fisica Aplicada III, Universidad de Sevilla, Escuela Superior de Ingenieros,
Sevilla, 41092 Camino de los Descubrimientos s/n,
Spain.}

\author{E. Vigezzi}
\email{enrico.vigezzi@mi.infn.it}
\affiliation{INFN, Sezione di Milano Via Celoria 16, 20133 Milano, Italy.}

\author{\mbox{R. A. Broglia}}
\email{broglia@mi.infn.it}
\thanks{Corresponding author. Tel:(+39)3388959875}
\affiliation{Dipartimento di Fisica, Universit\`{a} di Milano,
Via Celoria 16, 20133 Milano, Italy.}
\affiliation{INFN, Sezione di Milano Via Celoria 16, 20133 Milano, Italy.}
\affiliation{The Niels Bohr Institute, University of Copenhagen, Blegdamsvej 17,
2100 Copenhagen {\O}, Denmark.}


\begin{abstract}
Pairing rotations and pairing vibrations are collective modes associated with a field, the pair field, which changes  the number of particles
by two. Consequently, they can be studied at profit with the help of two-particle transfer reactions on superfluid and in normal nuclei, respectively. 
The advent of exotic beams has opened, for the first time, the possibility to carry out such studies in medium heavy nuclei, within 
the same isotopic chain. In the case studied in the present paper that of the Sn-isotopes (essentially from closed ($Z=N=50$) to closed ($Z=50$,$N=82$) shells).
The static and dynamic off-diagonal, long range order phase coherence in gauge space displayed by pairing rotations and vibrations respectively, leads to coherent states which behave almost classically. Consequently, these modes are amenable to an accurate nuclear structure description in terms of simple models containing the right physics, in particular BCS plus QRPA and HF 
mean field plus RPA respectively. The associated two- nucleon transfer spectroscopic amplitudes predicted by such model calculations can thus 
be viewed as essentially ``exact''.
This fact, together with the availability of optical potentials for the different real and virtual channels involved in the reactions considered, namely $^{A+2}$Sn$+ p$, $^{A+1}$Sn$+ d$ and $^{A}$Sn$+ t$, allows for the calculation of the associated absolute cross sections without, arguably, 
free parameters.
The numerical predictions of the absolute differential cross sections, obtained making use of the above mentioned nuclear structure and optical potential inputs, within the framework of second order DWBA, taking into account simultaneous, successive and non-orthogonality contributions provide, within experimental errors in general, and below $10\%$ uncertainty in particular, an overall account of the experimental findings for all of the measured $^{A+2}$Sn$(p,t){}^{A}$Sn$(gs)$ reactions, for which absolute cross sections have been reported to date.
\end{abstract}



\maketitle

\section{Introduction}
Customarily, the fingerprint of shell closure in nuclei is associated with a sharp, step function--like distinction between occupied and empty single--particle states, in correspondence with magic numbers \cite{Mayer:55} (for a recent example concerning $^{132}$Sn, see \cite{Cottle:10} and \cite{Jones:10}).
At variance with the case of infinite, fermionic systems, in which there essentially exists a continuum of states at the Fermi energy, in finite--many-body (FMB) systems, like e.g. the atomic nucleus, $\varepsilon_F$ is not, in principle, well defined, at least not in closed shell nuclei. This is because a sizable energy gap 
is observed between the last occupied ($j_<$) and the first unoccupied ($j_>$) orbitals.
This is also the case for other FMB fermionic systems, e.g. $C_{60}$ fullerene, which displays a relatively large HOMO--LUMO gap, of the order of 1.6 eV as compared to $\varepsilon_F \approx 15$ eV \cite{Haufler:91}. In such cases, one can use  as a working definition of $\varepsilon_F$ \cite{Mahaux:85}, $(\varepsilon_{j_>}-\varepsilon_{j_<})/2$ ($(\varepsilon_{HOMO}-\varepsilon_{LUMO})/2$ in the case of $C_{60}$). 
Away from closed shells, medium-heavy nuclei become, as a rule, superfluid, the distinction between occupied and empty states being blurred around $\varepsilon_F$.
The Fermi energy is, in such situation, well defined and equal to the energy for which the occupancy probability attains the value of one half. In keeping with 
this result, in the case of closed shell nuclei, $\varepsilon_F$ can be properly  defined as the minimum of the dispersion relation associated 
with  pair addition and pair removal modes.

\section{Pair--spin formalism}
The mixing taking place in superfluid nuclei between particle and holes is economically embodied in the Bogoliubov--Valatin \cite{Bogoliubov:58,Valatin:58} quasiparticle transformation
\begin{subequations}
\begin{equation}
 \alpha_{\nu} = U_{\nu}a_{\nu} - V_{\nu}a^{\dagger}_{\bar{\nu}} = U'_{\nu}a'_{\nu} - V'_{\nu}a'^{\dagger}_{\bar{\nu}},
\label{eq.alpha(a)}
\end{equation} 
\begin{equation}
 \alpha^{\dagger}_{\nu} = U^{*}_{\nu}a^{\dagger}_{\nu} - V^{*}_{\nu}a_{\bar{\nu}} = U'_{\nu}a'^{\dagger}_{\nu} - V'_{\nu}a'_{\bar{\nu}},
\label{eq.alpha}
\end{equation} 
\end{subequations}
where 
\begin{equation}
U_{\nu} = U'_{\nu} e^{i \phi} \quad , \quad V_{\nu} = V'_{\nu} e^{-i \phi},
\label{eq.UV}
\end{equation}
are the BCS occupation amplitudes, $U'_{\nu}$ and $V'_{\nu}$ being real quantities, while $a'^{\dagger}_{\bar \nu}= \mathcal{G}(\phi)a^{\dagger}_{\bar \nu} \mathcal{G}^{-1}(\phi) = e^{-i \phi}a^{\dagger}_{\bar \nu}$ and $a'_{\bar \nu}= \mathcal{G}(\phi)a_{\bar \nu} \mathcal{G}^{-1}(\phi) = e^{i \phi}a_{\bar \nu}$ are creation and annihilation operators referred to the intrinsic system of reference in gauge space (i.e. body fixed BCS deformed state), $\mathcal{G} = \textrm{exp}(-iN\phi)$ inducing a rotation of the angle $\phi$ in this (2-D) space (gauge transformation), $N$ being the number of particle operator (see Apps. A, B and C). The states $\vert \nu \rangle$ and $\vert \bar{\nu} \rangle$, connected by the time reversal operator, have the same energy (Kramers degeneracy).
In keeping with the fact that (1) is a unitary transformation, 
\begin{equation}
U_{\nu}U^*_{\nu} + V_{\nu}V^*_{\nu} = U^{'2}_{\nu} + V^{'2}_{\nu} =1 
\label{eq.Norm}
\end{equation}
The transformation (1) provides the rotation in Hilbert space of the creation  and annihilation fermion operators, which diagonalizes the mean field, BCS pairing Hamiltonian 
(angle $\theta_{\nu}$, see Figs. A.1(b) and (c)),
\begin{equation}
(H_p - \lambda N)_{MF} = \sum_{\nu>0} \epsilon_{\nu} N_{\nu} - G \alpha'_0 (P^{' \dagger}+P') + G \alpha_0^{'2} = 
\sum_{\nu>0} E_{\nu} \tilde N_{\nu} + E_{gs},
\label{eq.4}
\end{equation}
where, $\epsilon_{\nu}= \varepsilon_{\nu} - \lambda$ and
\begin{equation}
E_{gs} = \sum_{\nu>0} (\epsilon_{\nu} -E_{\nu}) + \frac{\Delta^{'2}}{G},
\end{equation}
is the ground state energy (see Appendix \ref{Appendix:Pair}), while $\lambda = \varepsilon_F$. In the above expression we find 
the particle number operators
\begin{equation}
 N_{\nu} = a^{\dagger}_{\nu}a^{}_{\nu} + a^{\dagger}_{\bar{\nu}}a_{\bar{\nu}},
\end{equation}
the pair creation and annihilation operators,
\begin{equation}
 P^{\dagger}_{\nu} = a^{\dagger}_{\nu}a^{\dagger}_{\bar{\nu}}, \qquad P^{}_{\nu} = a^{}_{\bar \nu}a^{}_{\nu},
\end{equation}
the complex ($\phi$ indicating the gauge angle) condensed (superfluid) Cooper field (see Appendix \ref{Appendix:B})
\begin{equation}
 \alpha_0=\langle BCS \vert P \vert BCS \rangle =  \langle BCS \vert P^{\dagger} \vert BCS \rangle ^{*} = e^{- 2 i \phi}\alpha'_0 =
e^{- 2 i \phi} \sum_{\nu>0} U'_{\nu} V'_{\nu}=\sum_{\nu>0} U^{*}_{\nu} V_{\nu},
 \label{eq.5}
\end{equation}
the quasiparticle energy
\begin{equation}
E_{\nu} = (\epsilon_{\nu}^2 + \Delta^{'2})^{1/2},
\end{equation}
the absolute value (modulus) of the pairing gap
\begin{equation}
 \Delta'= G \alpha'_0 = G \sum_{\nu>0} U'_{\nu} V'_{\nu}= G e^{2 i \phi} \sum_{\nu>0} U^*_{\nu}V_{\nu} = \Delta e^{2 i \phi},
\end{equation}
and the quasiparticle number operators,
\begin{equation}
 \widetilde{N}_{\nu} = \alpha^{\dagger}_{\nu}\alpha^{}_{\nu} +  \alpha^{\dagger}_{\bar{\nu}}\alpha_{\bar{\nu}}.
\end{equation}

For the case of $^{120}$Sn, the quantities $\epsilon_\nu$, $E_\nu$, $U'_\nu$, $V'_\nu$ and $U'_\nu V'_\nu$ introduced above are given in Tables 1 and 2.

The ground state of the system referred to the laboratory system of reference $\mathcal{K}$ as well as to the intrinsic (body-fixed) frame $\mathcal{K'}$ (see Appendices A and B), can be written as
\begin{align}
   \vert BCS (\phi) \rangle_{\mathcal{K}} & = \frac{1}{\textrm{Norm}} \prod_{\nu>0} \alpha_{\nu} \alpha_{\bar{\nu}} \vert 0 \rangle 
                                            = \prod_{\nu>0} (U_{\nu} + V_{\nu}a^{\dagger}_{\nu}a^{\dagger}_{\bar{\nu}})\vert 0 \rangle 	\notag \\
& = \prod_{\nu>0} e^{i \phi}         (U'_{\nu}+V'_{\nu} e^{-2i\phi}  a^{\dagger}_{\nu} a^{\dagger}_{\bar{\nu}}) \vert 0 \rangle 
  = e^{i \Omega \phi} \prod_{\nu>0}  (U'_{\nu}+V'_{\nu}             a'^{\dagger}_{\nu}a'^{\dagger}_{\bar{\nu}}) \vert 0 \rangle         \notag \\
& =  \vert BCS (\phi = 0) \rangle_{\mathcal{K'}},
\label{eq.BCS}
\end{align}
where $\Omega$ is the pair degeneracy of the single-particle subspace, leading to an overall (trivial) phase.

The central feature of the $\vert BCS \rangle$ wavefunction, that is condensation of largely overlapping Cooper pairs, is captured by the pairspin (quasispin) formulation of superconductivity (see \cite{Anderson:58,Bohr:89} and refs. therein), which ascribes values (see Fig. \ref{fig:1}(a) and Appendix \ref{Appendix:Pair}) 
    \begin{equation}
   s_x(\nu) = s_y(\nu)= 0, \qquad s_z(\nu)=-1/2, \\
\label{eq:14a}
    \end{equation}
to empty states, and
    \begin{equation}
   s_x(\nu) = s_y(\nu)= 0, \qquad s_z(\nu)=+1/2,
\label{eq:14b}
    \end{equation}
to occupied states of the non-interacting (normal) system. The symmetry axis in pairspin  space (the $z$-axis) is referred to as the gauge axis (see Appendix \ref{Appendix:Pair}, Fig. \ref{fig:A1}(a)).
A superfluid system (see Fig. \ref{fig:1}(b), see also Fig. \ref{fig:A3} Appendix \ref{Appendix:Pair}) is characterized by a  collective pairspin $\overrightarrow{S_{\bot}} \equiv \{S_x,S_y\}$, 
which points in a direction perpendicular to $z$, associated with the azimuthal angle 2 $\phi$. 
This direction defines an intrinsic reference frame $\mathcal{K'}$ (see Fig. B1, Appendix B), in which 
the components of the average  total pairspin $ \langle \vec{S} \rangle = \sum_{\nu>0} \langle \vec s(\nu) \rangle $ in the mean field ground state,
take the values
(see Fig. \ref{fig:A1}(b) and Appendix \ref{Appendix:Pair}, see also (\ref{eq.B28a}) and \ref{eq.B29a})
\begin{subequations}
   \begin{eqnarray}
& &      \langle S_{y'} \rangle = 0, \\
& &      \langle S_{x'} \rangle = \alpha'_{0} = \sum_{\nu>0}U'_{\nu}V'_{\nu}, \label{eq.12b} \\
& &   \langle S_{z} \rangle \equiv   \langle S_{z'} \rangle  = \frac{1}{2} \sum_{\nu>0} (V^{'2}_{\nu} - U^{'2}_{\nu}).
   \end{eqnarray}
\end{subequations}

Thus, $S_{\bot}$ gives a measure of the mixing of empty and occupied states of the BCS solution (see Eqs. (\ref{eq:14a},\ref{eq:14b}). 
This is tantamount to saying that $S_{\bot}$ defines a privileged orientation in gauge space (see Appendix \ref{Appendix:Pair}), in keeping with the fact that (\ref{eq.BCS}) is a wavepacket in the number of pairs of particles. An emergent property of the associated symmetry breaking phase transition is generalized rigidity in gauge space (see Appendix \ref{Appendix:D}). That is, the system can be set into rotation (or change its rotational frequency) in gauge space in terms of two-particle transfer reactions.
The pairspin polarization may rotate collectively around the gauge axis, and the azimuthal angle $2 \phi$ of $S_{\bot}$  (see Appendix \ref{Appendix:Pair} and Fig. \ref{fig:A1}, see also (\ref{eq.B28a}) and (\ref{eq.B29a})) is therefore a dynamical variable associated with pairing rotational bands.

\subsection{Order Parameter}

The modulus of the order parameter (\ref{eq.5}), that is, of the quantity  $\alpha'_0 \approx \sum_{\nu>0} U'_{\nu}V'_{\nu}$ $= \Delta'/G$, is, for medium heavy nuclei 
($A \approx 120$) of the order of 1.4 MeV/$G \approx$ 7, in keeping with the fact that $G \approx 25 \textrm{ MeV}/A$ (major j-shell approximation, see \cite{Brink:05} Chapters 2 and 3 and refs. therein). 
In other words, roughly of the order of ten  $(\nu, \bar \nu)$ Cooper pairs contribute to the nuclear condensate in superfluid nuclei. 
Consequently, large fluctuations are expected for $\alpha'_0$.

These fluctuations are generated by the residual interaction acting between the quasiparticles (cf. \cite{Anderson:58}, cf. also \cite{Brink:05} in particular Ch. 4 and Appendices I and J of this reference).
In the harmonic (QRPA) approximation the two associated pair fields are: 1) $(U^{2}_{\nu}-V^{2}_{\nu})(\Gamma^{\dagger}_{\nu}+\Gamma^{ }_{\nu})$ leading essentially to a bound two-quasiparticle like state (pairing vibration mixed to $\beta$-vibrations, in deformed nuclei, cf. \cite{Bes:66} and refs. therein) lying on top of the pairing gap, $\Gamma^{\dagger}_{\nu}= \alpha^{\dagger}_{\nu}\alpha^{\dagger}_{\bar{\nu}}$ being the two-quasiparticle creation quasiboson operator; 2) $(U^{2}_{\nu}+V^{2}_{\nu})(\Gamma^{\dagger}_{\nu}-\Gamma^{ }_{\nu})$ which sets the $\vert BCS \rangle_{\mathcal{K}'}$ intrinsic state into rotation in gauge space, and whose fluctuations diverge in the long-wavelength limit, in just such a way that the resulting ground state
\begin{equation}
 \vert N_0 \rangle \sim \int \textrm{d}\phi e^{{iN_0}\phi} \vert BCS (\phi)\rangle_{\mathcal{K}} \sim (\sum_{\nu>0} c(\nu) a^{\dagger}_{\nu}a^{\dagger}_{\bar{\nu}})^{N_0/2} \vert 0 \rangle,
\label{eq.13}
\end{equation}
transforms irreducibly under gauge transformation
\begin{equation}
 \mathcal{G}(\phi) = e^{-{i {N} \phi}}.
\end{equation}
The states (\ref{eq.13}) are the members of a pairing rotational band build out of a condensation of $N/2$ Cooper pairs each described by $\vert \tilde{0} \rangle = \sum_{\nu > 0} c(\nu) a^{\dagger}_{\nu}a^{\dagger}_{\bar{\nu}}\vert 0 \rangle$ (see Appendix \ref{Appendix:B}, in particular Eq. (\ref{pairband}),(\ref{eq.B17}) and (\ref{eq.B23})).
It is of notice that the presence of rotational bands in the spectra of many--body systems is the fingerprint of deformation both in real (3D) and abstract (e.g. gauge) spaces 
(see Fig. \ref{fig:analogy}, see also Table XI in ref. \cite{Broglia:73}).

Fluctuations of the pair field are, of course, already present in the normal (correlated) ground state $|0 \rangle = |gs(A_0) \rangle$ 
of closed shell nuclei (mass number $A_0$). That is in systems in which, while $\alpha'_0=0$, the (dynamical) value of the order parameter, i.e. the zero-point
fluctuation of $\alpha_0'$ around its zero value, 
\begin{eqnarray}
 \alpha_{dyn}
      & = &\langle (\alpha-\alpha'_0)^2 \rangle = 1/2 \left( \langle 0 \vert P^{\dagger} P\vert 0 \rangle + \langle 0 \vert P^{} P^{\dagger} \vert 0 \rangle\right)= \notag \\
      & = & \frac{1}{2} \left( \sum_{int} \vert \langle int(A_0-2)\vert P \vert 0 \rangle \vert^{2} + \vert \langle int(A_0+2)\vert P^{\dagger} \vert 0 \rangle \vert^{2} \right) \notag \\
      & \approx & \frac{1}{2} \left( \langle gs(A_0 -2)|P|0 \rangle + \langle gs(A+2)|P^{\dagger}| 0 \rangle \right),
\label{sigma}
\end{eqnarray}
displays finite values (see \cite{Bes:66,Nikam:87a,Nikam:87b,Shimizu:89} and refs. therein), directly related to the 
quantity $E_{corr}/G$,
where $E_{corr}$ is the average correlation energy of the pair addition ($|gs(A_{0}+2)>$ state) and of the pair removal  ($|gs(A_{0}-2)>$ state) modes.
In other words, the binding energy of the two-nucleon Cooper pair moving on top of the $A_0$ closed shell system, and of the two-hole Cooper pair moving in the $A_0$ Fermi sea. 

In the case of $^{132}$Sn ($A_0 =132$), we obtain (see Table \ref{Table:Sn132_PV} and Fig. \ref{fig:dispers})

\begin{subequations}
\begin{equation}
E_{corr}(A+2) = 2 | \varepsilon_{j>} - \lambda| - W(A+2) = 1.17 {\rm MeV},
\end{equation}
\begin{equation}
E_{corr}(A-2) = 2 | \varepsilon_{j<} - \lambda| - W(A-2) = 2.14 {\rm MeV},
\end{equation}
\end{subequations}
and thus $\bar E_{corr}  =1.66 $ MeV. It is of notice that in the case in which the energies $W \to 0$, the system becomes superfluid, the BCS $\lambda$ parameter coinciding with the minimum
value of the dispersion relation shown in Fig. \ref{fig:dispers}(a). Consequently $E_{corr}/G \approx \frac{1}{2}$ $\left [ \frac{E_{corr}(A_0+2)}{G(A_0+2)} + 
\frac{E_{corr}(A_0-2)}{G(A_0-2)} \right] $ $   \approx \frac{1}{2} (8.9 +13.6) \approx 11$,  a value which is not very different from that of $\Delta'/G$ associated with the superfluid
nucleus $^{120}$Sn. 

The quantity (\ref{sigma}) can be also estimated from the ratio $(\Lambda/G)$ between the (two-particle)-(two-hole) (pairing)  vibration coupling strength 
(see e.g. \cite{Bes:66}) and the pairing coupling constant. 
In the two-level model \cite{Brink:05}, this quantity is given by $(\Lambda/G) = 2 \sqrt{\Omega}$, $\Omega$ being  the pair degeneracy of the single-particle space in which 
the two nucleons (two nucleon holes) participating in the pair addition and removal modes are allowed to correlate. 
As a rule, these are the valence shells of the closed shell system $A_0$.

In the case of the newly discovered $^{132}$Sn, doubly magic nucleus, $\Omega \approx 30$ and $\alpha_{dyn} \approx (\Lambda/G) \approx 2 \sqrt{30} \approx 11$. Making use of the actual values of $\Lambda$ (see Table \ref{Table:Sn132_PV}) one obtains $\frac{1}{2} \left[ \frac{1.08}{0.131} + \frac{1.60}{0.157} \right] \approx 9$. Summing up, $\alpha_{dyn}(^{132}$Sn) $ \geq \alpha (^{120}$Sn).

This result embodies the very difference between gauge spontaneous symmetry breaking in atomic nuclei and in condensed matter. 
In a chunk of e.g. Pb, of which more than 50\% is the atom built on the isotope
$^{208}$Pb, (or of Sn, of which none is the atom built on the isotope $^{132}$Sn, the corresponding nucleus  being highly unstable), 
at a temperature  below the critical temperature $T_c = 7.2 K (3.72 K)$ at which
the metal becomes superconducting but in the presence of a magnetic  field stronger than the critical value {$H_c = 0.08 \textrm{ T}$} (0.03 T), 
of the order of $10^3$ times the earth magnetic field,  Cooper pairs break as soon  as they are formed, leading to a hardly 
observable  effect, in particular concerning the structure and stability of the crystal.

On the other hand, in the case of their ground state  and thus at absolute zero temperature -- as it is the case for all natural occurring 
nuclear species on earth -- the Cooper
pairs  associated with the normal (non superfluid) system  displaying two nucleons above or two holes below  closed shell, like e.g. $|gs(_{82}^{210}{\rm Pb}_{128})>$ and $|gs (_{82}^{206}{\rm Pb}_{124})>$ respectively, these fermions are  strongly correlated, as evidenced by the large two-nucleon transfer cross sections with which they are excited 
(see e.g. \cite{Broglia:73} and refs. therein). In keeping with the fact that a consistent fraction of this cross section arises, e.g. in the case of $^{208}$Pb$(t,p)^{210}$Pb$(gs)$
by the transfer of two particles to levels below the Fermi energy of $^{208}$Pb (ground state correlations), 
Cooper pair correlations  blur dynamically the difference between occupied and empty single-particle states thought to be a trademark of closed shell systems.

The same arguments presented above, can be used for $^{132}_{50}$Sn$_{82}$, in which case  the summed backwardsgoing amplitudes amount to $\sum_i Y_i \approx 0.5$, 
as reported in Table \ref{Table:Sn132_PV}. Arguably, one  can posit,  that in the nuclear case it is not, or at least not only, the condensed (superfluid) state  which is peculiar, but  the normal
state\footnote{Within this context one can also mention a different (although not directly pertinent to the Sn-isotopes studied in the present paper) analogy between nuclear  and metallic normal state properties which has important consequences  on Cooper pair stability. Bad conductors, that is bad single-particle electronic metals (like e.g. Pb, Sn and Hg), display stable Cooper pair condensation at low temperatures, becoming superconductors, while good conductors , independent electron motion metals, (e.g. Au, Ar, Cu) do not. In the nuclear case,  arguably, one of the best nuclear embodiments of Cooper's model is the {$ |gs({}^{11} $Li$ )> $} state, pair addition mode of $^9$Li. The associated "single particle" system ($^{10}$Li) is not only unbound, but more revealing, single-particle mean field levels are so strongly dressed by the coupling to the bosonic (vibrational) modes of the medium, that a new magic number becomes operative in this case, namely $N=6$ instead of $N=8$ (parity inversion phenomenon) (see \cite{Brink:05}, Ch.11, \cite{Potel:10} and refs. therein). }, in  which pair addition and pair removal Cooper are virtually poised in the ground state of the closed shell system $A_0$, ready to condense (i.e. $W(A_0 +2), W(A_0 -2) \to 0$, see Fig. \ref{fig:dispers}).
inducing a (fluctuating) alignment of  pairspins perpendicular to the gauge ($z-$axis), and thus to a domain wall and associated generalized rigidity (see Fig. \ref{fig:A3}). This can also be seen from   the pairspin states in normal systems (see Figs. 4 and 5), defined as,
\begin{subequations}
\begin{align}
\vert 0 \rangle_{\nu} & = \frac{1}{Norm.} \left( Y_{add} (\nu) \vert s_z(\nu) = -1/2 \rangle + X_{rem}(\nu) |s_z (\nu) = 1/2\rangle \right) \nonumber \\ 
                      & = \frac{1}{Norm.} \left( Y_{add}(\nu) |2\rangle_{\nu} + X_{rem}(\nu) |1\rangle_{\nu} \right),  \quad \quad \textrm{for } \; \epsilon_{\nu} < \epsilon_F,  
\label{pairspin_a}
\end{align}
and 
\begin{align}
|0\rangle_{\nu} & = \frac{1}{Norm.} \left( Y_{rem} (\nu) |s_z(\nu) = +1/2 \rangle + X_{add}(\nu) |s_z (\nu) = -1/2 \rangle \right) \nonumber \\
                & = \frac{1}{Norm.} \left( Y_{rem} |1 \rangle_{\nu} + X_{add}(\nu) |2 \rangle_{\nu} \right), \quad \quad \textrm{for } \; \epsilon_{\nu} > \epsilon_F,
\label{pairspin_b}
\end{align}
\end{subequations}
where $X,Y$ are the forwardsgoing and backwardsgoing RPA pair vibration amplitudes (see insets of Fig.\ref{fig:dispers})
\begin{equation} \left.
\begin{array}{l} 
  X_{n}(\nu;\beta) \\ Y_n(\nu;\beta) \end{array}  \right\}
= \frac{(\sqrt{\Omega_{\nu}}/2) \Lambda_n(\beta)}{2\epsilon_{\nu} \mp W_n(\beta)}.
\label{eq.XY}
\end{equation}
The quantity $\Omega_{\nu} = (2j_{\nu}+1)/2$ is the pair degeneracy of orbital $\nu$, $\beta= - 2$ and $\beta=+2$ label the pair addition and pair removal modes respectively, while $n$ (= 1,2,... ) 
numbers the solutions of the RPA dispersion relation in subsequent order of excitation energy (we deal here only with the $n=1$, lowest energy pairing vibrations).
%
%
%
%
Consequently, as soon as the single-particle field $H_{sp}$ -- field  which acts on the nuclear pairspin along the gauge  ($z$)-axis in a similar way  
in which a magnetic field  acts in the case of a metallic superconductor (see App. A, discussion after Eq. (A27); see also
discussion after Eq, (A35)) -- is decreased, a fact that takes  place  moving away from magic numbers (in the case of e.g. $^{132}$Sn this implies reducing the single-particle 
gap  $| \varepsilon_{j<} - \varepsilon_{j>}| $ from about $\approx 5$ MeV to few hundreds of keV, se Tables \ref{Table:2} and \ref{Table:Sn132_PV}), nuclear 
Cooper pairs condense, the ground state of the corresponding nucleus becoming
amenable to a BCS-type description, a fact which already takes place with the presence of two removal modes in $A_0$. 

It is of notice the presence of strong fluctuations in pairspin observed  in Fig. 5, as compared to the smooth variations shown in Fig. 1, idealization 
of  the situation representative of a high purity, metallic crystal (within this context see Fig. \ref{fig:A3} of Appendix \ref{Appendix:Pair}).  In fact, finite nuclei display orbitals which contribute
very differently to pairing corrleations, in particular hot orbitals (see e.g. \cite{Broglia:73,Broglia:71,Broglia:72b}), related  to the different pair 
degeneracy $\Omega_{\nu}$ and with the relative amount of $s-$component of the different $j_{\nu}^2(0)$ pure two-particle configurations.
One can view such an imperfect pairspin alignment as a limitation of its applicability to finite many-body systems. Conversely, one can interpret it as a reflection 
of the richness with which these systems in general, and finite nuclei in particular, embody symmetry breaking phase transitions. Namely, among other things, in terms
of very non-conventional normal phases. Normal phases which display (virtually) traces of e.g. domain walls with varied degree of stability, dynamically violating 
the symmetry in question (gauge symmetry in the present case). These properties are precisely those which render the study of pairing correlations 
in nuclei central, in the quest for the mechanism which are at the basis of the stability of nuclear species, in particular along the drip lines \footnote{Within this context  one can mention the fact that if instead of pairspins 
with two projections, one studies the properties of finite systems which depend on the alignment of a pairspin with twenty components , like protein evolution and structure in which each projection corresponds to one of the twenty natually occurring aminoacids, the fluctuations of pair spin and thus of emergent properties, become even richer
and subtler than in the case shown in Fig. 5 (see e.g. ref. \cite{Broglia:12ArXiv} and refs. therein). From a technical point of view, but not only, the situation may be analogous  to  that resulting moving from tinkering with the Ising model, to confront oneself  with Potts model.}. 

In keeping with the fact that the $P^{\dagger}$ and $P$ are the basic operators entering both the pairing interaction ($H_p= -GP'^{\dagger}P'$) and the pair mean field ($-G\alpha'_0(P^{\dagger}+P)+G(\alpha'_{0})^{2}$), two-nucleon transfer can be viewed as the specific probe of pairing correlations in nuclei, in a similar way as Coulomb excitation, inelastic scattering and $\gamma$--decay are specific tools to probe (quadrupole) surface vibrations and rotations (see Fig. 2). In other words, specific information on $\alpha_0'$ and $\alpha_{dyn}$ can be obtained 
through two-nucleon transfer reactions. 

Because the correlation length associated with the nuclear Cooper pairs is
\begin{equation}
 \xi= \left\{ 
\begin{array}{l l}
  \cfrac{\hbar v_F}{2\Delta'},  & \quad (\alpha'_0 \neq 0),\\
 \\
  \cfrac{\hbar v_F}{2E_{corr}}, & \quad (\alpha'_0 = 0),\\ \end{array} \right.
\end{equation}
Cooper pair partners are correlated over distances considerably larger than nuclear dimensions ($\xi \approx 20-30$ fm, as compared to $R \approx 5$ fm). 
Consequently, from a nuclear structure point of view, Cooper pair transfer involves also regions in which the pairing interaction $G(x)$, $x$ representing e.g. the surface-surface distance between target and projectile (see e.g. \cite{Broglia:05c} Ch. III), vanishes, 
a situation already known in condensed matter in connection with the Josephson effect \cite{Josephson:62,Bardeen:62,Cohen:62}. This result, together with the fact that the depth of the single-particle potential $\vert V_0 \vert (\approx 50 \textrm{ MeV})$ is much larger than $G (\approx 25/A \textrm{ MeV})$, implies that, exception made for $Q$-value effects, successive transfer induced by the mean field single-particle (Saxon-Woods-like) potential, is expected to be, as a rule, the largest contribution to the two-nucleon transfer cross sections \cite{Udagawa:73,Chien:75,Segawa:75,Schneider:76,Takacsy:73,Takacsy:74,Hashimoto:78,Kubo:78,Bayman:82,Yagi:79,Igarashi:91,Becha:97}\footnote{Within this context it is of notice that the incoming proton (distorted) wave, in e.g. a $^{A+2}$Sn$(p,t)$ reaction,  is diffused by the scattering center, i.e. by the $^{A+2}$Sn target, into emergent distorted waves, in particular the one corresponding to the relative motion of a deuteron and of the $^{A+1}$Sn system after the interaction $v_{np}$ has acted for the first time. Even when these two systems are at relative distances $r=| \vec r_{^{A+1}\textrm{Sn}} - \vec r_{d} |$ much larger than the target radius, the Cooper pair wavefunction describing the pair correlation of the picked up neutron and of its partner in the $^{A+1}$Sn system, has a finite probability amplitude centered on the outgoing deuteron, and this is so not only  in the case of superfluid nuclei (like e.g. $^{120}$Sn) but also of normal nuclei (like e.g. $^{132}$Sn). This is in keeping with the fact that the stability, collectivity and associated correlation length associated with superfluid Cooper pairs and with pair addition and removal Cooper pairs are very similar, as discussed above. Making use of this finite amplitude, the interaction $v_{np}$ acting a second time (see (\ref{eq1_41}) below) can trigger the $^{A+1}$Sn Cooper pair partner to move into the deuteron leading to the triton, and completing in this way the successive transfer process. From this narrative, it is not surprising that the paper in which the probabilistic interpretation of Schr{\"{o}}dinger wavefunction was forcefully proposed, written by Born, describes a collisional process \cite{Born:26}.}.
The difficulties to absorb this simple result by nuclear structure practitioners partially stems from the fact that, neglecting reaction details, the two-particle transfer cross sections $gs \rightarrow gs$ can be schematically written as
\begin{equation}
 \sigma (gs \rightarrow gs) \sim \left\{ 
\begin{array}{l l}
  \vert \alpha'_{0} \vert^{2} & \quad (\alpha'_0 \neq 0),\\
  \vert \alpha_{dyn}\vert^{2} & \quad (\alpha'_0 = 0),\\ \end{array} \right. \\
\end{equation}
corresponding to the square modules of matrix elements of $P^{\dagger}$ and $P$, the associated spectroscopic amplitudes in the intrinsic coordinate system in gauge space $\mathcal{K}'$ being (see e.g. \cite{Broglia:73}, and Appendix \ref{Appendix:B})
\begin{subequations}
 \begin{equation}
  B(\nu \bar{\nu}; J=0)  \sim { }_{\mathcal{K}'}\langle BCS (A+2) \vert P'^{\dagger}_{\nu} \vert BCS (A) \rangle_{\mathcal{K}'},
 \end{equation}
where
 \begin{equation}
 { }_{\mathcal{K}'}\langle BCS (A+2) \vert P^{'\dagger}_{\nu} \vert BCS (A) \rangle_{\mathcal{K}'} =  U'_{\nu} (A) V'_{\nu} (A+2) 
\left( = { }_{\mathcal{K}'}\langle BCS (A) \vert P'_{\nu} \vert BCS (A+2) \rangle_{\mathcal{K'}} \right).
 \end{equation}
\end{subequations}

Now, thinking in terms of the transfer of a nucleon at a time, e.g. in the case of the reaction
\begin{equation}
 (A+2) + p \rightarrow F(\equiv A+1) +d \rightarrow A + t
\end{equation}
one can write, in connection with the first step ($ (A+2) + p \rightarrow F+d$),
\begin{equation}
 a^{'\dagger}_{\nu_{d}}a^{'}_{\nu} \vert BCS (A+2) \rangle \vert p \rangle \sim V'_{\nu} (A+2) \alpha^{\dagger}_{\bar{\nu}} \vert BCS (F) \rangle \vert d \rangle,
\end{equation}
and
\begin{equation}
 a^{'\dagger}_{\nu_{t}}a^{'}_{\bar{\nu}} \vert BCS (F) \rangle \vert d \rangle \sim U'_{\nu} (A) V'_{\nu}(A+2) \vert BCS (F) \rangle \vert d \rangle,
\end{equation}
in connection with the second step ($F+d\rightarrow A + t$), 
in keeping with the fact that
\begin{align}
&  a^{'\dagger}_{\nu} = U'_{\nu} \alpha^{\dagger}_{\nu} + V_{\nu}\alpha_{\bar{\nu}},          \nonumber \\
&  a^{'}_{\nu}        = U'_{\nu} \alpha_{\nu}        + V'_{\nu}\alpha^{\dagger}_{\bar{\nu}},    \nonumber \\
&  a^{'\dagger}_{\bar{\nu}}   = U'_{\nu} \alpha^{\dagger}_{\bar{\nu}}  - V'_{\nu}\alpha_{\nu},    \nonumber \\ 
&  a^{'}_{\bar{\nu}}   = U'_{\nu} \alpha_{\bar{\nu}}  - V'_{\nu}\alpha^{\dagger}_{\nu},    \nonumber
\end{align}
and that $\vert BCS \rangle$ is the quasiparticle vacuum. The primed quantities are referred to the intrinsic, body-fixed frame (see App. A and B).

Consequently, the associated (successive) two-nucleon transfer spectroscopic amplitude 
\begin{equation}
 B(\nu^2(0))=U_{\nu}(A) V_{\nu}(A+2) \left( = c_\nu \right),
 \label{eq.20}
\end{equation}
has the same dependence  on the BCS occupation numbers as that displayed by the amplitudes associated with  
simultaneous transfer (order parameter) \cite{Broglia:73}, and with the $c({\nu})$ amplitude entering in the Cooper pair wavefunction.
This last result is closely related to the smooth behavior of the BCS occupation parameters with mass number (see Appendix \ref{Appendix:B}).  
Summing up, pair coherence is maintained both in successive as well as in simultaneous transfer.

It is of notice that all the above results, which constitute the very essence of nuclear BCS, are not only inescapable, they are also almost tautological, at least for well bound nuclei. In fact, pairing in nuclei
does not affect neither the single-particle energies $\varepsilon_{\nu}$ (see in any case Eq. (\ref{vp}) and following discussion), nor the corresponding  wavefunctions $\phi_{\nu}(\vec r)$,
but only the  single-particle occupation probabilities. And this takes place in a small $(\Delta/\varepsilon_F \approx 5 \times 10^{-2}$) region around the Fermi energy. In this region,
and in keeping with the structure of the BCS wavefunction, which  takes into account the variety of excitations of pairs of particles $(\nu,\bar \nu'$), so as to produce  the most efficient
mixing of empty and occupied states leading to Cooper pairs, the only possible excitation mechanism of the nuclear superfluid, is that of breaking a Cooper pair,
individual  two-particle $(\nu \bar \nu)$ excitations being already taken into account in the BCS ground state. It is then neither surprising  that the amplitudes  of the Cooper pair wavefunction are $c(\nu) \sim U'_{\nu}V'_{\nu}$, nor that the absolute cross section for Cooper pair transfer are 
proportional to $\left (\sum_{\nu>0} U'_{\nu}V'_{\nu}\right)^2$. And least of all, it is not
surprising the fact these amplitudes and two-nucleon transfer processes involve not only $(nljm,nlj-m)$ configurations, but also $(nljm,n'lj-m)$ as well as 
$[(J(A)+n_1l_1j_1)_{J_1} \otimes (J'(a)+n_2l_2j_2)_{J_1}]_0$.
This is in keeping with the non-orthogonality of the single-particle wavefunctions describing the 
target and projectile  in the different channels ($a+A \to f+F \to b+B$). Within this context, 
not only $n$ and $n'$ are possible, due to the fact that partners of a Cooper pair feel different mean fields $(\phi^f_{n'ljm},\phi^F_{nljm})$, 
but also because   
a general nuclear structure treatment of pairing, will include 
Cooper-like correlations associated with multipole pairing (see e.g. \cite{Brink:05} Sect. 5.3 and refs. therein), correlations which, in the present case, have not a dynamical origin (one works with $H_p = - G P^{\dagger}P)$, but only a trivial
kinematical one.

\section{Reaction Mechanism}

In what follows we present the elements which enter the calculation of the absolute two-particle transfer differential cross section  in terms of 
the reaction 
\begin{equation}
A + t  \to B(\equiv A+2) + p ,  
\end{equation}
in which $A+2$ and $A$ denotes the mass number of even nuclei in their ground state. In other words, one concentrates on $L=0$ transfer.
The wavefunction of nucleus $A+2$ is written as,
\begin{equation}\label{eq1}
\Psi_{A+2}(\xi_A,\mathbf r_{A1},\sigma_1,\mathbf r_{A2},\sigma_2)=\psi_A(\xi_A)\sum_{l_i,j_i}[\phi^{A+2}_{l_i,j_i}(\mathbf r_{A1},\sigma_1,\mathbf r_{A2},\sigma_2)]^0_0 \quad,
\end{equation} 
product of the wavefunction describing the ground state of the nucleus $A$, 
the corresponding  relative (intrinsic) $3A - 3$ radial coordinates  being denoted $\xi_A$, and of the wavefunction of two-correlated nucleons  
\begin{equation}\label{eq2}
\phi^{A+2}_{l_i,j_i}(\mathbf r_{A1},\sigma_1,\mathbf r_{A2},\sigma_2)]^0_0 = 
\sum_{nm}a_{nm}\left[\varphi^{A+2}_{n,l_i,j_i}(\mathbf r_{A1},\sigma_1)\varphi^{A+2}_{m,l_i,j_i}(\mathbf r_{A2},\sigma_2)\right]^0_0
\end{equation} 
the wavefunctions $\varphi^{A+2}_{n,l_i,j_i}(\mathbf r, \mathbf \sigma)$ describing the single-particle motion of a nucleon in a mean field potential, e.g. a Sax\-on-Woods potential.
The spatial part of the two-neutron wavefunction in the triton can be written as, $\phi_t(\mathbf r_{p1},\mathbf r_{p2})=\rho(r_{p1})\rho(r_{p2})\rho(r_{12})$, 
$r_{p1}$ and $r_{p2}$ denoting the modulus of the relative coordinate of each of the two neutrons involved in the transfer process, measured with respect to the proton, 
while $r_{12}$ denotes the modulus of the relative coordinate of the two neutrons in the triton. 
The functions $\rho(r)$, as depicted in Fig. \ref{fig_triton}(a), are 
generated with the $p-n$ Tang--Herndon interaction \cite{Tang:65}
\begin{align}\label{eq.25}
v(r)&=-v_0\exp\left(-k(r-r_c)\right) \quad r>r_c\\
\label{eq.26}
v(r)&=\infty \quad r<r_c,
\end{align}
where $k=2.5$fm$^{-1}$ and $r_c=0.45$fm denotes the radius of the hard core. The depth $v_0$ is adjusted so as to reproduce the binding energy of the triton and of the deuteron respectively. This hard--core potential is also used in the above expressions as the $n-p$ interaction potential responsible for neutron transfer.

The two-particle transfer differential cross section is written as 
\begin{equation}
\frac{d\sigma}{d\Omega}=\frac{\mu_i\mu_f}{(4\pi\hbar^2)^2}\frac{k_f}{k_i}\left|T^{(1)}+T^{(2)}_{succ}-T^{(2)}_{NO}\right|^2.
\label{eq.28}
\end{equation}
The amplitudes appearing in it  describe the simultaneous,
\begin{subequations}
\begin{multline}\label{eq1_40}
T^{(1)}=2\sum_{l_i,j_i}\sum_{\sigma_1 \sigma_2}\int d\mathbf{r}_{tA}d\mathbf{r}_{p1}d\mathbf{r}_{A2}
  [\phi^{A+2}_{l_i,j_i}(\mathbf r_{A1},\sigma_1,\mathbf r_{A2},\sigma_2)]^{0*}_0\chi^{(-)*}_{pB}(\mathbf{r}_{pB})\\
 \times v(\mathbf{r}_{p1}) \phi_t(\mathbf r_{p1},\mathbf r_{p2})\chi^{(+)}_{tA}(\mathbf{r}_{tA}),
\end{multline}
successive
\begin{multline}\label{eq1_41}
T^{(2)}_{succ}=2\sum_{l_i,j_i}\sum_{l_f,j_f,m_f}\sum_{\substack{\sigma_1 \sigma_2\\\sigma'_1 \sigma'_2}}
\int d\mathbf{r}_{dF}d\mathbf{r}_{p1}d\mathbf{r}_{A2}
[\phi^{A+2}_{l_i,j_i}(\mathbf r_{A1},\sigma_{1},\mathbf r_{A2},\sigma_2)]^{0*}_0\chi^{(-)*}_{pB}(\mathbf{r}_{pB})
 v(\mathbf{r}_{p1})\\
 \times\phi_d(\mathbf r_{p1})\varphi^{A+1}_{l_f,j_f,m_f}(\mathbf r_{A2}) \int d\mathbf{r}'_{dF}d\mathbf{r}'_{p1}d\mathbf{r}'_{A2}G(\mathbf{r}_{dF},\mathbf{r}'_{dF})\\
 \times \phi_d(\mathbf r'_{p1})^*\varphi^{A+1*}_{l_f,j_f,m_f}(\mathbf r'_{A2}) \frac{2\mu_{dF}}{\hbar^2}v(\mathbf{r}'_{p2})   \phi_d(\mathbf r'_{p1})
 \phi_d(\mathbf r'_{p2}) \chi^{(+)}_{tA}(\mathbf{r}'_{tA}),
\end{multline}
and non-orthogonality
\begin{multline}\label{eq1_42}
T^{(2)}_{NO}=2\sum_{l_i,j_i}\sum_{l_f,j_f,m_f}\sum_{\substack{\sigma_1 \sigma_2\\\sigma'_1 \sigma'_2}}
\int d\mathbf{r}_{dF}d\mathbf{r}_{p1}d\mathbf{r}_{A2}
[\phi^{A+2}_{l_i,j_i}(\mathbf r_{A1},\sigma_1,\mathbf r_{A2},\sigma_2)]^{0*}_0\chi^{(-)*}_{pB}(\mathbf{r}_{pB})
 v(\mathbf{r}_{p1})\\
 \times \phi_d(\mathbf r_{p1})\varphi^{A+1}_{l_f,j_f,m_f}(\mathbf r_{A2})\int d\mathbf{r}'_{p1}d\mathbf{r}'_{A2}d\mathbf{r}'_{dF}\\
 \times\phi_d(\mathbf r'_{p1})^*\varphi^{A+1*}_{l_f,j_f,m_f}(\mathbf r'_{A2}) 
  \phi_d(\mathbf r'_{p1})\phi_d(\mathbf r'_{p2})\chi^{(+)}_{tA}(\mathbf{r}'_{tA}),
\end{multline}
\end{subequations}
contributions to the transfer process.
In these expressions, $\varphi^{A+1}_{l_f,j_f,m_f}(\mathbf r_{A1})$ are the wavefunctions describing the intermediate states
of the nucleus $F \equiv A+1$, generated as solutions of a Saxon-Woods potential, 
while 
$\phi_d(\mathbf r_{p2})$ is the wavefunction describing  the deuteron bound state (see Fig. \ref{fig_triton}(b)). 
We have chosen the so called post-post representation \cite{Broglia:05c}, in which the $p-n$ interaction appears twice in the successive amplitude.
The Green function $G(\mathbf{r}_{dF},\mathbf{r}'_{dF})$ propagates the intermediate channel $d,F$, and can be expanded in partial waves
\begin{equation}\label{eq7}
G(\mathbf{r}_{dF},\mathbf{r}_{dF}')=i\sum_{l}\sqrt{2l+1}
\frac{f_{l}(k_{dF},r_<)P_{l}(k_{dF},r_>)}{k_{dF}r_{dF}r_{dF}'}
\left[  Y^{l} (\hat r_{dF}) Y^{l} (\hat r_{dF}')\right]_0^0.
\end{equation}
The functions $f_{l}(k_{dF},r)$ and $P_l(k_{dF},r)$ are the regular and the irregular solutions of a Schr{\"{o}}dinger equation associated 
with  a suitable optical potential and an energy equal to the kinetic energy in the intermediate state. In most cases of interest, the result is hardly altered if 
one uses the same energy of relative motion for all the intermediate states. This representative energy is calculated when both nuclei  appearing in the
intermediate state are in their ground states. The validity of this approximation can break down in particular cases. For example, in the case in 
which some relevant intermediate states are strongly  off shell, in which case their  contribution is significantly quenched. An interesting situation can 
develop  when this situation becomes operative for all possible intermediate states, in which case  they can only be virtually populated, thus emphasizing the role of simultaneous transfer.  

\section{The isotopic chain $^{100}_{50}$S\lowercase{n}$_{50}$--$^{132}_{50}$S\lowercase{n}$_{82}$}

A collective mode is characterized by: 1) an enhanced cross section of transition probability; 2) a simple expression of its energy as a function of the quantum number characterizing the states connected by the transition. This quantum number is related to restoration of the symmetry violation (static or dynamical, e.g. particle number in the case of pairing 
rotations and vibrations, angular momentum in  the case of  e.g. quadrupole rotations and vibrations).

For example, in the case of a quadrupole rotational band of a 3D-deformed nucleus like, e.g. $^{152}$Dy, 1) corresponds to the $B(E2)$ transition probability, measured e.g. in the terms of single-particle Weisskopf units (of the order of $10^3$ in the example chosen), while 2) corresponds to $E_{I}=(\hbar^2/2\mathcal{J})I(I+1)$, $I$ being the angular momentum of the system ($I=0,2,4,...$). In the case of pairing rotational bands 1) corresponds to the absolute value of the two-nucleon differential cross section, measured in terms of the average pure two-particle units \cite{Broglia:73,Broglia:71,Broglia:72b} (typical value of the enhancement factor being, in the case of Sn-isotopes, of the order of $10^2$), while 2) corresponds to $E_N = (\hbar^2/2\mathcal{I})(N-N_0)^2$, $N$ being the number of particles associated with the condensate, namely with the neutrons in the case of ${}^{A}_{50}$Sn$_{N}$,
while $N_0$ is the mean number of neutrons representative of  the particle number  wavepacket describing the superfluid Sn-isotopes ($N_0 \approx $ 68, see discussion below).
Of these two quantities, 1) is arguably the most representative one. This is because accidental degeneracies or residual interactions may modify the energies without much altering the long-range correlation of the coherent state.\footnote{Within this context one can mention the fact that, was it not for $H_{sp}$, all pairspins would line up in the $(x,y)$-plane transverse to the gauge axis $z$ (see Appendix \ref{Appendix:Pair}), the pairspin alignment picture essentially becoming ``exact'' under such condition.} This is also the reason why, in what follows, use is made of the single j-shell model to discuss the basic features of the pairing rotational modes.

In this Section we present evidence of the accuracy with which 
the model of pairing rotations discussed in Section 2, together with the two-nucleon transfer reaction scheme summarized in the last section 
 allows for an overall quantitative description of the absolute value of the two-nucleon transfer cross sections,
when use is made   of global optical parameters to describe the (three) elastic channels involved in the process.
Consequently, the predictions given in Sect. \ref{Sect:PairVibration} concerning the pairing vibrational spectrum expected in connection 
with the two unstable closed shell systems $^{132}$Sn and the most exotic one $^{100}$Sn can be considered potentially important and likely quantitative.


\subsection{Pairing rotations}
\label{Sec.Rotations}
In Fig. \ref{fig4} we display the value of the absolute differential cross sections associated with the reactions ${}^{A+2}_{50}$Sn$^{ }(p,t){}^{A}_{50}$Sn$(gs)$ for which absolute measurements have been reported in the literature, in comparison with the experimental data \cite{Guazzoni:99,Guazzoni:04,Guazzoni:06,Guazzoni:08,Guazzoni:11,Guazzoni:12,Bassani:65} (see also \cite{Potel:11PRL,Potel:11PRL_erratum}). The corresponding integrated cross section are collected in Table \ref{Table_CS}. In all cases the contribution of the successive process is the dominant one. Examples of two-nucleon spectroscopic amplitudes obtained from BCS calculations  are displayed in Table 2 ($U,V$ for $^{120}$Sn$(p,t){}^{118}$Sn). 
They have been computed solving the gap and number equations with a monopole interaction acting on the 
bound orbitals, calculated  as the eigenfunctions of a standard parametrized Saxon-Woods potential,
and imposing that the gap reproduces the value obtained from the empirical odd-even mass differences for the various isotopes. 
The BCS spectroscopic amplitudes are in good agreement with those predicted by extended shell model calculation (see refs. \cite{Guazzoni:11,Guazzoni:12} and refs. therein).
The optical parameters in the entrance, intermediate, and final channel where taken from refs. in \cite{Guazzoni:99,Guazzoni:04,Guazzoni:06,Guazzoni:08,Guazzoni:11,Guazzoni:12} and from \cite{An:06} for the deuteron channel.

From the above results one can posit that theory provides an account of the experimental absolute differential cross section well within the experimental errors and, arguably, without free parameters. 

Let us now concentrate our attention on the value and on the structure of the probability amplitude for two nucleon, at $\vec r$ and $\vec{r}\hspace{0.07cm}'$ to belong to a Cooper pair, namely $\alpha'_{0}(\vec r, \vec{r}\hspace{0.07cm}') = \sum_{\nu>0} c_{\nu} \phi_{\nu}(\vec r) \phi_{\bar{\nu}}(\vec \vec{r}\hspace{0.07cm}')$ (see (\ref{eq.13}) and (\ref{eq.20})). That is, the nuclear structure component of the two-particle transfer cross section amplitude. To clarify the physics at the basis of the BCS description of pairing rotational bands, we discuss two scenarios for the case of the $^{120}$Sn$(p,t)^{118}$Sn reaction. In the first one, we consider all bound single--particle states, the cutoff energy being $E_{cutoff}=0$ MeV. In the second case one sets $E_{cutoff}=60$ MeV, discretizing the continuum inside a spherical box of 15 fm of radius. The BCS gap ($\Delta=1.47$ MeV; experimental value) and number ($N=70$) equations lead to $G=0.18$ MeV and $\lambda=-\,6.72$ MeV in the first case and $G=0.05$ MeV and $\lambda=-\,6.9$ MeV in the second one. 
The associated Cooper pair probability distributions in $r$-space are essentially identical (see Fig. \ref{fig.Cooper} and Tables \ref{Table:1} and \ref{Table:2}). It is then not surprising that they lead to essentially the same absolute value of the two--particle transfer cross section associated with the reaction $^{120}$Sn$(p,t)^{118}$Sn(gs).

Let us now repeat the argument, but this time in terms of $\sigma(gs \rightarrow gs) \sim (\Delta/G)^2$, as it customary done since the first publication which introduced it \cite{Yoshida:62}. Because the pairing gap has been fixed to reproduce the experimental value (1.47 MeV), one obtains in the case of $E_{cutoff}=0$MeV $(\Delta/G)^2\sim 70$ and $(\Delta/G)^2\sim 889$ in the case of  $E_{cutoff}=60$MeV. This result emphasizes the problem of working with an expression which contains explicitly the pairing coupling constant.

One could argue that such an objection could also be leveled off against the relation $\sigma\sim|\alpha_0|^2$. Note however, that a $(p,t)$ reaction would hardly feel the effect of contributions far removed from the Fermi surface $\lambda$. This is in keeping with the fact that transfer to levels lying far away from $\lambda$  will be unfavorable due to $Q$--value effects. If one argues in terms of the relative distance $r$ between target and projectile ($r\gg R_0$ for continuum--like contributions; $r<R_0$ for deeply bound--like contributions), the outcome is similar. In fact, for large distances the two--particle transfer form factor vanishes while at small distances the outgoing tritium will experience strong absorption (see Appendix \ref{Appendix:E}).

In fact, considering only the contribution to $\alpha'_0$ arising from the valence orbitals, that is, essentially those contributing to the ``naked'' vision of the Cooper pair wavefunctions, one obtains $\alpha'_0=2.12$ and $\alpha'_0=2.08$ respectively, and thus, a negligible squared relative difference between the two predicted cross sections, namely $(0.04/2.1)^2\approx 2\times 10^{-3}$.

Summing up, because the pair condensed state can be viewed as a coherent state which behaves essentially classically when viewed in terms of its building block (Cooper pair) the description of pairing rotational bands provided by the BCS model in terms of a coupling constant and an  energy cutoff can be considered essentially ``exact'' when probed with two-nucleon transfer processes, reactions which filter the inaccuracies of each individual $U_{\nu}V_{\nu}$ component, emphasizing the off-diagonal long range order provided by the phase coherence (cf. ref. \cite{Yang:62}). In fact, studying nuclear Cooper pair condensation in terms of e.g. single-nucleon transfer (see e.g. \cite{Gales:12,Idini:12}, the individual inaccuracies of the BCS occupation numbers cannot be averaged out.
As a consequence, the overall agreement between theory and experiment is much poorer than that reflected by e.g. the results collected in Table \ref{Table_CS}.
The above arguments provide further evidence of why two-nucleon transfer is the specific probe of pairing superfluidity.
%

%
In Fig. \ref{fig5} a quantity  closely related to the Sn--isotopes binding energy is reported as a function of the number of neutrons. Also displayed is the best parabolic fit to these energies, a quantity to be compared with
\begin{equation}
 E_N=\frac{\hbar^{2}}{2\mathcal{I}} (N-N_0)^{2},
\label{moment_inertia}
\end{equation}
namely the energy associated with the members of the pairing rotational band.

A simple estimate of the pairing rotational band moment of inertia is given by the single $j$--shell model (see e.g. \cite{Brink:05} App. H  $\hbar^{2}/2\mathcal{I} = G/4 \approx 25/(4 N_0)$ MeV). 


This estimate turns out to be rather accurate, even beyond expectation. Of notice that to the extent that one is discussing properties of a coherent state like that described by (\ref{eq.BCS}), for which $H_{sp}$ plays a secondary role (see discussion following Eq. (\ref{vp}) in Appendix \ref{Appendix:Pair}) this is not a surprising results. As can be seen from Fig. \ref{fig5}, the estimate (\ref{moment_inertia}) 
is rather accurate except close to  $N=$50 and $N=$ 82, in keeping with the fact that, as discussed before, the pairing deformed picture ($\alpha_0 \neq 0$) 
breaks down around closed shell ($\alpha'_0 = 0$), where a vibrational regime (associated with the dynamic distortion $\alpha_{dyn}$, 
see Eq. (\ref{sigma})) is expected to be valid.

Also reported in Fig. \ref{fig5}, are the integrated values of the measured absolute two--particle transfer cross sections. Naively, one would expect a marked constancy of these transitions, in keeping with the fact that the (pairing) rotational model implies a common intrinsic (deformed) state (Cooper pair condensate, see Eq. (\ref{eq.BCS})). On the other hand, due to 
the fact that the number of Cooper pairs contributing to the pairing distortion $\alpha'_0$ is rather small (less than 10), one expect strong fluctuations in this quantity ($\alpha'_0 \approx \sqrt{7}/7 \approx 0.4$) and consequently in the two--particle transfer cross section ($\sigma \sim \alpha_0^{'2}$, i.e. fluctuation in $\sigma$ of the order of 100 \%).

In keeping with the analogy presented in Fig. \ref{fig:analogy}, in the case of electromagnetic transition between members of a quadrupole rotational band one expects in heavy nuclei fluctuations of the order of ($\sqrt{250}/250$)$^{2}$, i.e. less than 1\%.
Within this context the average value of the absolute experimental cross section reported in Table \ref{Table_CS} is 1551 $\mu$b, while the average difference between experimental and predicted values is $81$ $\mu$b. Thus the discrepancies between theory and experiment are bound in the interval $0 \leq \vert \sigma_{exp}(i \rightarrow f)-\sigma_{th}(i \rightarrow f) \vert / \sigma_{exp}(i \rightarrow f) \leq 0.09$, the average discrepancy being 5$\%$.

In Fig. \ref{fig6} the excited, pairing rotational band associated with the average value of the $0^{+}$ pairing vibrational states with energy $\leq 3$ MeV, is displayed together with the best parabolic fit. Also given is the relative $(p,t)$ integrated cross section normalized with respect to the $gs \rightarrow gs$ transitions, a value which is in all cases $\leq 8 \%$, in overall agreement with the single j--shell estimate (see ref. \cite{Brink:05} App. H), given in the inset to the figure. The result testifies to the weak cross talk between pairing rotational bands and thus of the robust off-diagonal, long range order coherence of these modes.

\subsection{Pairing vibrational band in closed shell nuclei}
\label{Sect:PairVibration}

In Fig. \ref{fig7} we display the expected pairing vibrational spectrum (harmonic approximation, see refs. \cite{Brink:05,Bes:66,Broglia:73} and refs. therein) associated with the closed shell exotic nucleus $^{132}$Sn \cite{Cottle:10,Jones:10}, up to two-phonon states. Within this approximation, the one-phonon states are the pair addition $\vert a \rangle = | gs(^{134}$Sn$)\rangle$ and pair removal $\vert r \rangle = | gs(^{130}$Sn$)\rangle$  modes.
The two-phonon $0^+$ ($\vert pv(^{132}$Sn$)\rangle = | r \rangle \otimes |a\rangle= \vert 0^+ (^{132}$Sn$); 6.5\textrm{ MeV}\rangle$) pairing vibrational ($(2p-2h)$-like) state of $^{132}$Sn, is predicted at an excitation energy of 6.5 MeV (see Fig. \ref{fig:dispers}). 
The absolute two-particle transfer differential cross sections associated with $\vert a \rangle$ and $\vert r \rangle$, namely
\begin{eqnarray}
^{132}{\rm Sn} (t,p) ^{134} {\rm Sn} (gs),  \quad (E_{CM} = 20 {\rm MeV}), \label{eq.38} \\
^{132}{\rm Sn} (p,t) ^{130} {\rm Sn} (gs),  \quad (E_{CM} = 26 {\rm MeV}), \label{eq.39}
\end{eqnarray} 
are reported in the insets. Using detailed balance 
the reactions above 
\begin{eqnarray}
^{130}{\rm Sn} (t,p) ^{132} {\rm Sn} (0^+; 6.5 {\rm MeV}),  \quad (E_{CM} = 20 {\rm MeV}), \\
^{134}{\rm Sn} (p,t) ^{132} {\rm Sn} (0^+; 6.5 {\rm MeV}),  \quad (E_{CM} = 26 {\rm MeV}),
\end{eqnarray} 
are, within the harmonic approximation, equivalent to (\ref{eq.38}) and (\ref{eq.39}), exept for the relative flux which is determined by the ratio $k_f/k_i$.

Similar calculations to the ones discussed above have been carried out for the closed shell nucleus $^{100}$Sn, the results being collected in Fig. \ref{fig8}.
In this case the two-phonon $0^+$, pairing vibrational mode of $^{100}$Sn is expected, again within the harmonic approximation, at an excitation energy
of 7.1 MeV. As it emerges from  Figs. \ref{fig7} and \ref{fig8}, and 
at variance with  the 
pairing rotational scheme, the two-particle transfer cross section associated with the excited pairing vibrational state is of the same order
of magnitude than that connecting  the ground states. 

Within this context, it could be intriguing to check whether the reaction \linebreak $^{106}$Sn(p,t)$^{104}$Sn populates a $0^+$ state at an excitation energy of the order of 7 MeV. This state, can be written within the  (pairing vibration) harmonic approximation, as $\vert ^{104} {\rm Sn} (0^+; 7.1 {\rm MeV}) \rangle = \vert a \rangle \otimes \vert a \rangle \otimes \vert a \rangle \otimes \vert r \rangle$. Namely 
a four-phonon pairing vibrational state, where 
\begin{equation}
\vert a \rangle = \vert gs ( ^{102} {\rm Sn})  \rangle \quad  ; \quad \vert r \rangle = \vert gs (^{98} {\rm Sn})\rangle.
\end{equation}
In other words, the reaction $^{106}$Sn(p,t)$^{104}$Sn ($0^+$, 7.1 MeV) is, within the harmonic picture of pairing vibrations,
equivalent to the reaction $^{100}$Sn(p,t)$^{98}$Sn$(gs)$. It is of notice that a three-phonon pairing vibrational states 
has been observed \cite{Flynn:72} in the reaction $^{204}$Pb$(t,p)^{206}$Pb at an excitation energy of about 6 MeV, with $Q-$values and absolute differential cross sections
compatible with the excitation of the $^{208}$Pb pair addition mode, namely $^{208}$Pb(t,p))$^{210}$Pb$(gs)$. While in this case deviations from the harmonic prediction 
are modest (essentially , most of them arising from the presence of the valence orbital $p_{1/2}$ lying just below $\epsilon_F (^{208}$Pb$)$ \cite{Bortignon:78}), in the case of pairing vibrations based on $^{100}$Sn, anharmonicities are expected to be much stronger. This is in keeping with the fact  that $N=Z$ nuclei
display, as a rule, coexistence phenomena. That is, a strong competition between spherical and deformed $0^+$ states (cf. e.g. \cite{Donau:67,Barz:69,Wimmer:10} and refs. therein, see also \cite{Broglia:69}).

\section{Conclusions}


The microscopic nuclear structure (BCS) description of pairing rotational bands, together with the second order DWBA description of two-nucleon transfer reactions which include successive, simultaneous and non-orthogonality channels provide, arguably without free parameters, an overall account of Cooper pair transfer to superfluid nuclei. Inarguably, theory not only reproduces all reported $^{A+2}$Sn$(p,t){}^{A}$Sn$(gs)$ absolute cross section data within experimental errors, but it does so with uncertainties below the $10\%$ level.

The study of the pairing vibrational scheme around the starting and end points of the pairing rotational spectrum promises to provide new insight on pairing fluctuations and their anharmonicities, in situations of large neutron excess and of $N \sim Z$, i.e. around closed shell system $^{132}$Sn and of the, likely deformation coexistent $^{100}$Sn, respectively.

\clearpage
\fancyhead[LE,RO]{\bfseries\thepage}
\fancyhead[LO]{\bfseries Tables}
\fancyhead[RE]{\bfseries Tables}
\begingroup
\squeezetable

\begin{table}
\begin{center}
\begin{tabular}{ c|c|c|c|c|c|c|c|}

            &$\varepsilon_\nu$& $\epsilon_\nu$ & $E_{\nu}$ & $U'_{\nu}$ & $V'_{\nu}$ & $U'_{\nu} V'_{\nu}$ & $\Omega_\nu U'_{\nu} V'_{\nu}$ \\
\hline
$d_{5/2}$   & $-9.21$         &   -2.31        &  2.72     &   0.28     &   0.96     & 0.27                &  0.81  \\
\hline
$g_{7/2}$   & $-8.70$         &   -1.80        &  2.31     &   0.34     &   0.94     & 0.32                &  1.28  \\
\hline
$s_{1/2}$   & $-7.42$         &   -0.52        &  1.55     &   0.58     &   0.81     & 0.47                &  0.47  \\
\hline
$d_{3/2}$   & $-6.98$         &   -0.08        &  1.47     &   0.69     &   0.72     & 0.50                &  1.00  \\
\hline
$h_{11/2}$  & $-5.97$         &   1.07         &  1.75     &   0.88     &   0.48     & 0.42                &  2.52  \\
\hline
$f_{7/2}$   & $-1.87$         &   5.03         &  5.26     &   0.99     &   0.14     & 0.14                &  0.56  \\
\hline
$p_{3/2}$   & $-0.78$         &   6.12         &  6.32     &   0.99     &   0.12     & 0.12                &  0.24  \\
\hline

\end{tabular}
\caption{\protect Results for the valence shell lying closer to Fermi energy, of a BCS calculation, for $^{120}$Sn. The mean field used corresponds to a Saxon-Woods potential, in a 15 fm spherical box (continuum discretization).The pairing coupling constant used, $G=0.05$ MeV leads to the value of pairing gap obtained from the three point formula ($\Delta = 1.47$ MeV), summing the contributions of states from the $1s_{1/2}$ at -39 MeV to +60 MeV. The resulting Fermi energy computed by solving the BCS number equation is $\lambda= -6.9$ MeV. The quantity $\alpha'_0 = \sum_{\nu>0} U'_\nu V'_\nu$ ($=\sum_j \sum_{m>0}U'_j V'_j = \sum_j \Omega_j U'_j V'_j$) $\approx 6.1$ for the valence-shell space and $\alpha'_0 \approx 29.4$ for the whole single-particle space used in the calculation.}
\label{Table:1}
\end{center}
\end{table}
\begin{table}
\begin{center}
\begin{tabular}{ c|c|c|c|c|c|c|c|}

            &$\varepsilon_\nu$& $\epsilon_\nu$ & $E_{\nu}$ & $U'_{\nu}$& $V'_{\nu}$ & $U'_{\nu} V'_{\nu}$ & $\Omega_\nu U'_{\nu} V'_{\nu}$ \\
\hline
$d_{5/2}$   & $-9.21$       &   -2.45        &  2.90     &   0.26     &   0.96      & 0.25                &  0.75  \\
\hline
$g_{7/2}$   & $-8.70$       &   -1.98        &  2.48     &   0.32     &   0.95      & 0.30                &  1.20  \\
\hline
$s_{1/2}$   & $-7.42$       &   -0.70        &  1.63     &   0.54     &   0.84      & 0.45                &  0.45  \\
\hline
$d_{3/2}$   & $-6.98$       &   -0.26        &  1.49     &   0.64     &   0.77      & 0.49                &  0.98  \\
\hline
$h_{11/2}$  & $-5.97$       &    0.75        &  1.64     &   0.85     &   0.53      & 0.45                &  2.70  \\
\hline
$f_{7/2}$   & $-1.88$       &    4.84        &  5.04     &   0.99     &   0.15      & 0.15                &  0.60  \\
\hline
$p_{3/2}$   & $-0.78$       &    5.94        &  6.10     &   0.99     &   0.12      & 0.12                &  0.24  \\
\hline

\end{tabular}
\caption{\protect Results for the valence shell lying closer to Fermi energy, of a BCS calculation, for $^{120}$Sn. The mean field used corresponds to a Saxon-Woods potential, in a 15 fm spherical box (continuum discretization).The pairing coupling constant used, $G=0.18$ MeV leads to the value of pairing gap obtained from the three point formula ($\Delta = 1.47$ MeV), summing the contributions of states from the $1s_{1/2}$ at -39 MeV to +0 MeV. The resulting Fermi energy computed by solving the BCS number equation is $\lambda= -6.72$ MeV. The quantity $\alpha'_0 \approx 6.1$ (see caption to Table \ref{Table:1}) for the valence-shell space and $\alpha'_0 \approx 8.2$ for the whole single-particle space.}.
\label{Table:2}
\end{center}
\end{table}

\begin{table}
\begin{center}
\begin{tabular}{ c c|c|c|c|c|c|c|}
\multicolumn{8}{c}{$^{132}$Sn}                      \\
\multicolumn{1}{c}{ }&\multicolumn{1}{c}{$\Omega_j$}& \multicolumn{1}{c}{$\varepsilon_j$}&\multicolumn{1}{c}{$X_{rem}$}&\multicolumn{1}{c}{$Y_{add}$}& \multicolumn{1}{c}{$X_{rem}Y_{add}$} &  \multicolumn{1}{c}{$\sqrt{\Omega_j}\Lambda_{rem}$} & \multicolumn{1}{c}{$\sqrt{\Omega_j}\Lambda_{add}$} \\
\cline{3-8}
$g_{7/2}$   &  4  & $-9.78$    &  0.229  & 0.080    &  0.018 &  3.20 & 2.16 \\
\cline{3-8}
$d_{5/2}$   &  3  & $-9.01$    &  0.255  & 0.078    &  0.020 &  2.78 & 1.88 \\
\cline{3-8}
$s_{1/2}$   &  1  & $-7.68$    &  0.286  & 0.058    &  0.017 &  1.60 & 1.08 \\
\cline{3-8}
$h_{11/2}$  &  6  & $-7.52$    &  0.791  & 0.147    &  0.116 &  3.92 & 2.64 \\
\cline{3-8}
$d_{3/2}$   &  2  & $-7.35$    &  0.529  & 0.088    &  0.047 &  2.26 & 1.52 \\
\cline{3-8}
\multicolumn{1}{c}{ }  &  \multicolumn{1}{c}{ }   &  \multicolumn{1}{c}{ } & $Y_{rem}$&$X_{add}$ \\
\cline{3-8}
$f_{7/2}$   &  4  & $-2.44$    &  0.922  & 0.209    &  0.192 &  3.20 & 2.16 \\
\cline{3-8}
$p_{3/2}$   &  2  & $-1.59$    &  0.265  & 0.121    &  0.032 &  2.26 & 1.52 \\
\cline{3-8}
$h_{9/2}$   &  5  & $-0.88$    &  0.281  & 0.166    &  0.046 &  3.58 & 2.42 \\
\cline{3-8}
$p_{1/2}$   &  1  & $-0.78$    &  0.120  & 0.073    &  0.009 &  1.60 & 1.08 \\
\cline{3-8}
$f_{5/2}$   &  3  & $-0.44$    &  0.180  & 0.119    &  0.021 &  2.78 & 1.88 \\
\cline{3-8}
\\
\multicolumn{8}{c}{$^{100}$Sn}                      \\
\multicolumn{1}{c}{ }&\multicolumn{1}{c}{$\Omega_j$}& \multicolumn{1}{c}{$\varepsilon_j$}&\multicolumn{1}{c}{$X_{rem}$}&\multicolumn{1}{c}{$Y_{add}$}& \multicolumn{1}{c}{$X_{rem}Y_{add}$} &  \multicolumn{1}{c}{$\sqrt{\Omega_j}\Lambda_{rem}$} & \multicolumn{1}{c}{$\sqrt{\Omega_j}\Lambda_{add}$} \\
\cline{3-8}
$f_{5/2}$   &  3  & $-22.02$   &  0.353  & 0.116   &   0.041 & 9.22  & 4.72 \\
\cline{3-8}
$p_{3/2}$   &  2  & $-21.75$   &  0.300  & 0.098   &   0.029 & 7.52  & 3.84 \\
\cline{3-8}
$p_{1/2}$   &  1  & $-20.20$   &  0.282  & 0.082   &   0.023 & 5.32  & 2.72 \\
\cline{3-8}
$g_{9/2}$   &  4  &$-18.05$    &  1.156  & 0.248   &   0.286 & 11.90 & 6.08 \\
\cline{3-8}
\multicolumn{1}{c}{ }  &  \multicolumn{1}{c}{ }   &  \multicolumn{1}{c}{ } & $Y_{rem}$&$X_{add}$ \\
\cline{3-8}
$d_{5/2}$   &  3  & $-10.47$   &  0.461  & 0.803   &   0.370 & 9.22  & 4.72 \\
\cline{3-8}
$g_{7/2}$   &  4  & $-9.23$    &  0.427  & 0.502   &   0.214 & 10.64 & 5.44 \\
\cline{3-8}
$s_{1/2}$   &  1  & $-8.41$    &  0.188  & 0.192   &   0.036 & 5.32  & 2.72 \\
\cline{3-8}
$d_{3/2}$   &  2  & $-7.70$    &  0.242  & 0.226   &   0.055 & 7.52  & 3.84 \\
\cline{3-8}
$h_{11/2}$  &  6  & $-6.83$    &  0.377  & 0.325   &   0.123 & 13.04 & 6.66 \\
\cline{3-8}

\end{tabular}
\caption{\protect RPA wavefunctions of pair addition and removal mode of ${}^{132}$Sn (above) and ${}^{100}$Sn (below). 
Single particle energies have been taken from experimental values referenced in the National Nuclear Data Center.
The energy of the lowest pairing addition and removal  phonons in  ${}^{132}$Sn are respectively  
$W(A+2)=3.45$ MeV with $G(A+2)=0.131$ MeV and $W(A-2)= 3.06$ MeV with $G(A-2)=0.157$ MeV, the associated particle-pair vibration coupling strength (see Eq.(\ref{eq.XY})) being $\Lambda_{add}=\Lambda(A+2)=1.08$ MeV and $\Lambda_{rem}=\Lambda(A-2)=1.60$ MeV respectively.
The minimum of the dispersion relation, and thus the Fermi energy, are equal to $\lambda = -4.75$ MeV (see Fig. \ref{fig:dispers}(a)). The pair degeneracy of the single-particle space associated with ${}^{132}$Sn and ${}^{100}$Sn is $\Omega = 31$ and $\Omega = 27$ respectively.
The energy of the lowest pairing addition and removal  phonons in  ${}^{100}$Sn are respectively  
$W(A+2)=5.13$ MeV with $G(A+2)=0.290$ MeV and $W(A-2)= 1.96$ MeV with $G(A-2)=0.380$ MeV, the Fermi energy being $\lambda = -14.5$ MeV (see Fig. \ref{fig:dispers}(b)). The associated $\Lambda$ values being $\Lambda_{add}= 2.72$ MeV and $\Lambda_{rem}=5.32$ MeV.
The binding energy of $^{98}$Sn was assumed to be  $B({}^{98}$Sn$)= 794.24$ MeV, from the polinomial (4th grade) fit  of the binding energies of tin isotopic chain.}
\label{Table:Sn132_PV}
\end{center}
\end{table}

\begin{table}[h!]
\begin{center}
  \begin{tabular}{|c|c|c|}
 \cline{2-3} 
\multicolumn{1}{c|}{}                                                                & \multicolumn{2}{|c|}{$\sigma($gs$\rightarrow$ gs)}              \\
 \cline{2-3}
\multicolumn{1}{c|}{}                                                                & Theory         & Experiment${}^{c,d)}$                        \\

\hline 
$^{112}$Sn($p,t$)$^{110}$Sn, $E_{p}=26$ MeV                                         & 1301 ${}^{a)}$           & $1309 \pm 200 (\pm 14)$ ${}^{a)}$ \qquad  $[6^{\circ} \leq \theta \leq 62.7^{\circ}]$\\
\hline 
$^{114}$Sn($p,t$)$^{112}$Sn, $E_{p}=22$ MeV                                         & 1508 ${}^{a)}$           & $1519.3 \pm 228 (\pm 16.2)$ ${}^{a)}$ \qquad  $[7.64^{\circ} \leq \theta \leq 62.24^{\circ}]$\\
\hline 
$^{116}$Sn($p,t$)$^{114}$Sn, $E_{p}=26$ MeV                                         & 2078 ${}^{a)}$           & $2492 \pm 374 (\pm 32)$ ${}^{a)}$ \qquad $[4^{\circ} \leq \theta \leq 70^{\circ}]$  \\
\hline 
$^{118}$Sn($p,t$)$^{116}$Sn, $E_{p}=24.6$ MeV                                       & 1304 ${}^{a)}$           & $1345 \pm 202 (\pm 24)$ ${}^{a)}$ \qquad  $[7.63^{\circ} \leq \theta \leq 59.6^{\circ}]$\\
\hline 
$^{120}$Sn($p,t$)$^{118}$Sn, $E_{p}=21$ MeV                                         & 2190 ${}^{a)}$           & $2250 \pm 338 (\pm 14)$ ${}^{a)} $ \qquad $[7.6^{\circ} \leq \theta \leq 69.7^{\circ}]$\\
\hline
${}^{122}\textrm{Sn}(p,t){}^{120}$Sn, $E_{p}=26$ MeV                                & 2466 ${}^{a)}$           & $2505 \pm 376 (\pm 18)$ ${}^{a)}$ \qquad  $[2.5^{\circ} \leq \theta \leq 78.5^{\circ}]$\\
\hline 
$^{124}$Sn($p,t$)$^{122}$Sn, $E_{p}=25$ MeV                                         & 838  ${}^{a)}$           & $958 \pm 144 (\pm 15)$ ${}^{a)}$ \qquad  $[4^{\circ} \leq \theta \leq 57^{\circ}]$\\
\hline
\hline 
$^{112}$Sn($p,t$)$^{110}$Sn, $E_p=40$ MeV                                            & 3349 ${}^{b)}$ & $3715 \pm 1114$ ${}^{b)}$ \\
\hline
$^{114}$Sn($p,t$)$^{112}$Sn, $E_p=40$ MeV                                            & 3790 ${}^{b)}$ & $3776 \pm 1132$ ${}^{b)}$ \\
\hline
$^{116}$Sn($p,t$)$^{114}$Sn, $E_p=40$ MeV                                            & 3085 ${}^{b)}$ & $3135 \pm 940$ ${}^{b)}$ \\
\hline
$^{118}$Sn($p,t$)$^{116}$Sn, $E_p=40$ MeV                                            & 2563 ${}^{b)}$ & $2294 \pm 668$ ${}^{b)}$ \\
\hline
$^{120}$Sn($p,t$)$^{118}$Sn, $E_p=40$ MeV                                            & 3224 ${}^{b)}$ & $3024 \pm 907$ ${}^{b)}$ \\
\hline
$^{122}$Sn($p,t$)$^{120}$Sn, $E_p=40$ MeV                                            & 2339 ${}^{b)}$ & $2907 \pm 872$ ${}^{b)}$ \\
\hline
$^{124}$Sn($p,t$)$^{122}$Sn, $E_p=40$ MeV                                            & 1954 ${}^{b)}$ & $2558 \pm 767$ ${}^{b)}$ \\
\hline 
\end{tabular} 
\end{center}
%
%
%
%
%
%
%
%
%
%
\caption{Absolute cross section associated with the $^{A+2}$Sn$(p,t)^{A}$Sn$(gs)$ cross sections (i.e. between the members of the Sn-ground state pairing rotational band) calculated as described in the text, in comparison with the experimental findings. 
\protect\newline
   ${}^{a)}$ $\mu$b; the number in parenthesis corresponds to the statistical errors; the numbers in square brackets provide the angular range of integration of the absolute two-particle differential cross sections. \protect\newline
   ${}^{b)}$ $\mu$b/sr ($\sum_{i=1}^{N}(d\sigma/d\Omega)$; differential cross section summed over the few, $N=3-7$ experimental points) \protect\newline
   ${}^{c)}$ P. Guazzoni, L. Zetta, et al., Phys. Rev. \textbf{C 60}, 054603 (1999). \protect\newline
   P. Guazzoni, L. Zetta, et al., Phys. Rev. \textbf{C 85}, 054609 (2012), \protect\newline
   P. Guazzoni, L. Zetta, et al., Phys. Rev. \textbf{C 69}, 024619 (2004). \protect\newline
   P. Guazzoni, L. Zetta, et al., Phys. Rev. \textbf{C 74}, 054605 (2006). \protect\newline
   P. Guazzoni, L. Zetta, et al., Phys. Rev. \textbf{C 83}, 044614 (2011). \protect\newline
   P. Guazzoni, L. Zetta, et al., Phys. Rev. \textbf{C 78}, 064608 (2008). \protect\newline
   ${}^{d)}$ G. Bassani et al., Phys. Rev. \textbf{139}, (1965)B830.        }
\label{Table_CS}
\end{table}

%

\endgroup
\clearpage
\fancyhead[LE,RO]{\bfseries\thepage}
\fancyhead[LO]{\bfseries Figures}
\fancyhead[RE]{\bfseries Figures}

\begin{figure}[hbt!]
	\begin{center}
		\includegraphics[width=0.95\textwidth]{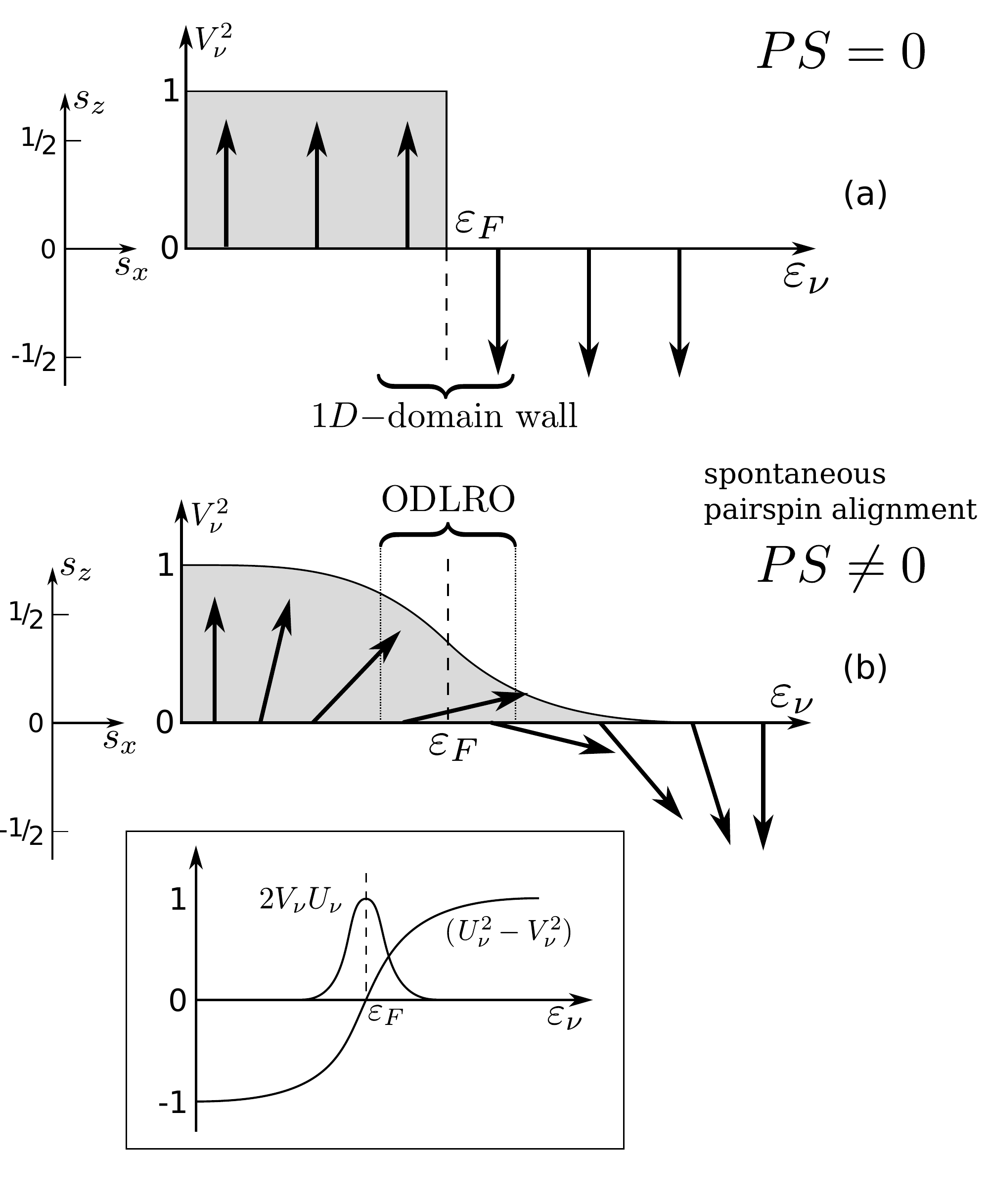}
	\end{center}
	\caption{(a) The occupancy $\langle N_{\nu} \rangle \sim V^{'2}_{\nu}$ of the 
single-particle states of the unperturbed system, in which the individual pairspins are aligned along the gauge ($z-$)axis (see Fig. A.1). 
This non--correlated system ($\alpha'_0=0$), displays zero pairspin alignment ($PS=0$), that is $\langle S_x \rangle =0$.
(b) The superconducting (nucleon superfluid) ground state displays Off-Diagonal 
Long Range Order (ODLRO, see Eq. (\ref{hnu})) and a finite value of the total pairspin ($PS\neq0;\alpha'_0 = \sum_{\nu>0} U'_{\nu} V'_{\nu}$),
i.e. $\langle s_x(\nu) \rangle \neq 0$, can be viewed as a one-dimensional domain wall (see also Fig. \ref{fig:A3} of Appendix \ref{Appendix:Pair}). The quantities in the inset, are the amplitudes with which $z-$ and $x-$components of the pairspin mix, leading
to a privileged  orientation in gauge space perpendicular to the gauge ($z-$)axis (see Fig. A.1 (b)).}
\label{fig:1}
\end{figure}

\begin{figure}
	\begin{center}
		\includegraphics[width=0.95\textwidth]{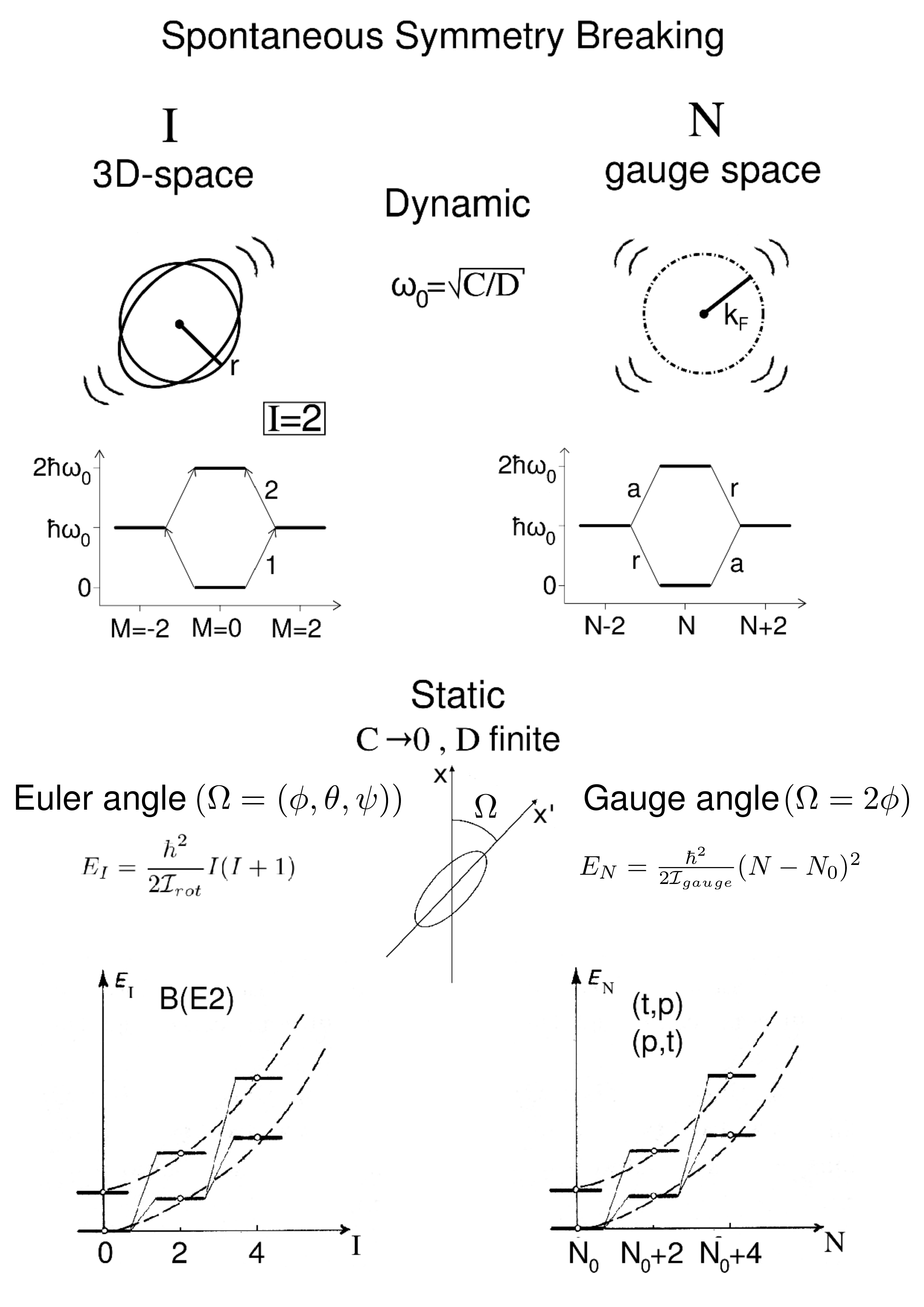}
	\end{center}
	\caption{Schematic representation of the nuclear structure consequences of spontaneous symmetry breaking of rotational and of gauge invariance (see also Table XI of ref. \cite{Broglia:73}).}
\label{fig:analogy}
\end{figure}

\begin{figure}
	\begin{center}
		\includegraphics[width=0.75\textwidth]{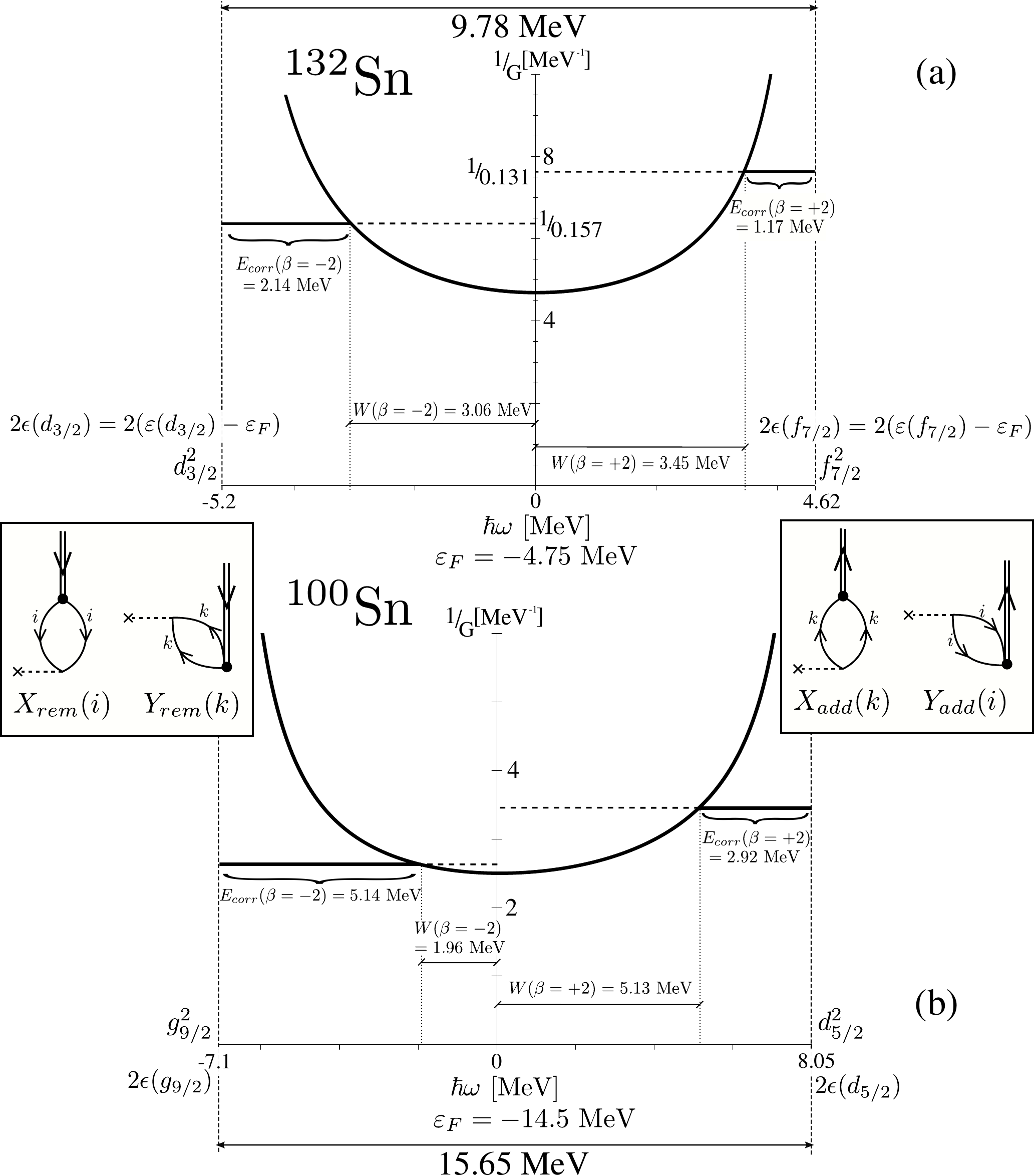}
	\end{center}
	\caption{Dispersion relation (see \cite{Bes:66}; see also \cite{Brink:05} Eq.(5.50)) associated with the pairing vibration  of $^{132}$Sn(a) and of $^{100}$Sn(b). 
(a) The part of the curve to the left of the minimum corresponds to the pair removal mode ($|gs(^{130}{\rm Sn})>$) while that to the right is associated with the pair addition mode ($|gs(^{134}{\rm Sn})\rangle$). The energy of the modes $W(A \pm 2)$ are measured from the minimum  of the dispersion relation, its values being explicitly indicated (see also caption 
to Table 2). The coupling constants $G(A\pm 2)$ used in the calculations are given in the caption of Table 2, where the $X$ and $Y$ amplitudes of the corresponding wavefunctions are displayed. As seen from the insets, the pairing vibrational modes blur the sharp distinction between occupied and empty states. In fact, the pair addition mode can be excited not only  by transferring two neutrons to levels lying above the Fermi energy, a process proportional to the $X_{add}(k)= \frac{(\sqrt{\Omega_{\nu}}/2) \Lambda_{add}}{2\epsilon_{k} - W_{add}}$ amplitude (inset to the right), but also to states lying below the Fermi energy, a process proportional to $Y_{add}(i)= \frac{(\sqrt{\Omega_{\nu}}/2) \Lambda_{add}}{2\epsilon_{i} + W_{add}}$. The associated values of the particle-pairing vibration coupling strength are $\Lambda_{add} = \Lambda(\beta= +2) = 1.08$ MeV and $\Lambda_{rem} = \Lambda(\beta= -2) = 1.6$ MeV. 
(b) The same as above, but for the closed shell system $^{100}$Sn. In this case $\Lambda_{add} = 2.72$ MeV and $\Lambda_{rem} = 5.32$ MeV.}
\label{fig:dispers}
\end{figure}

\begin{figure}
	\begin{center}
		\includegraphics[width=0.98\textwidth]{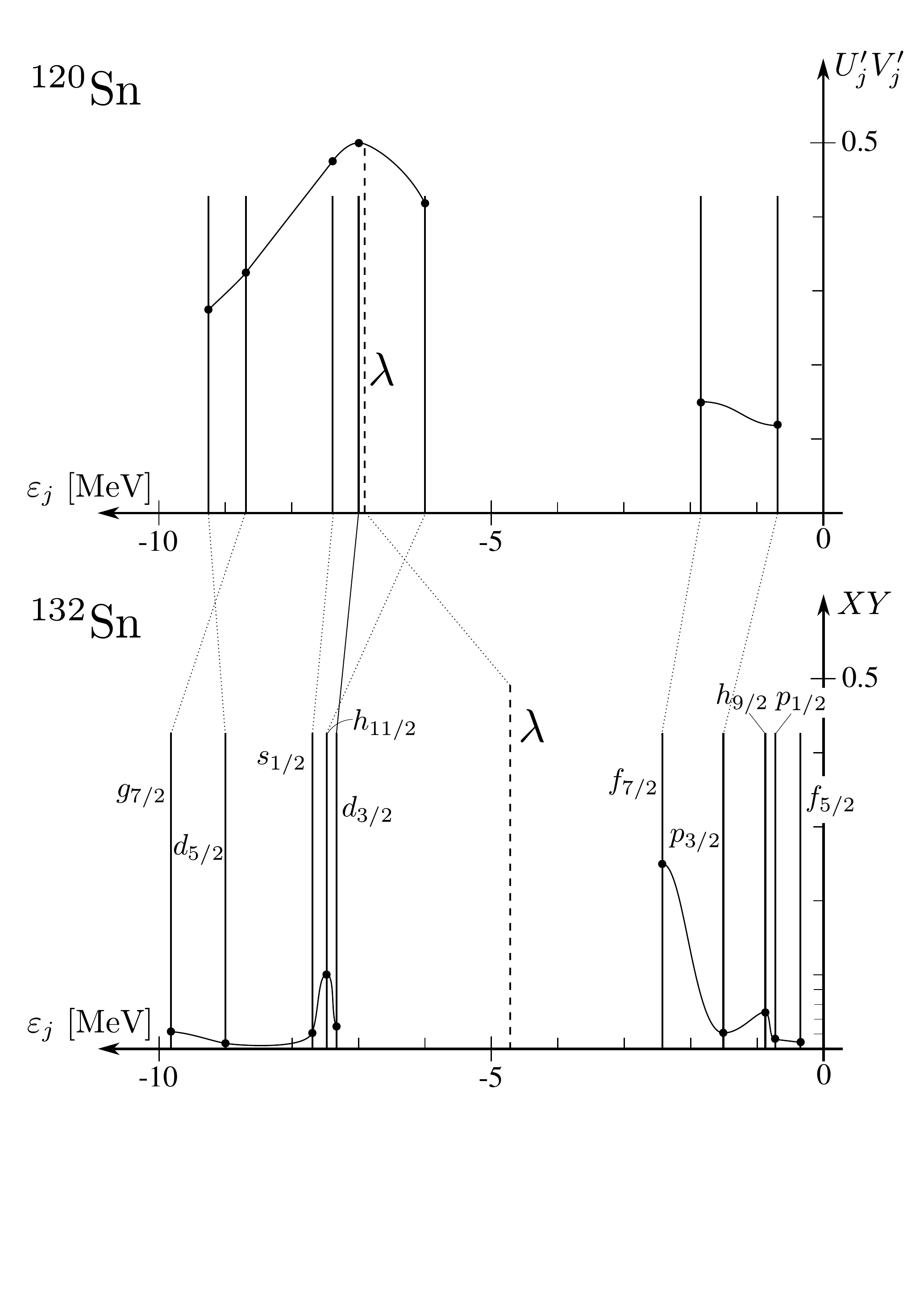}
	\end{center}
	\caption{(a) Value of the $U'V'$ products associated with $^{120}$Sn (see Table \ref{Table:2}).
(b) Value of the $XY$ products associated with the pair addition and removal modes of  $^{132}$Sn (see Table \ref{Table:Sn132_PV}).
}
\label{fig:xy}
\end{figure}

\begin{figure}
	\begin{center}
		\includegraphics[width=0.88\textwidth]{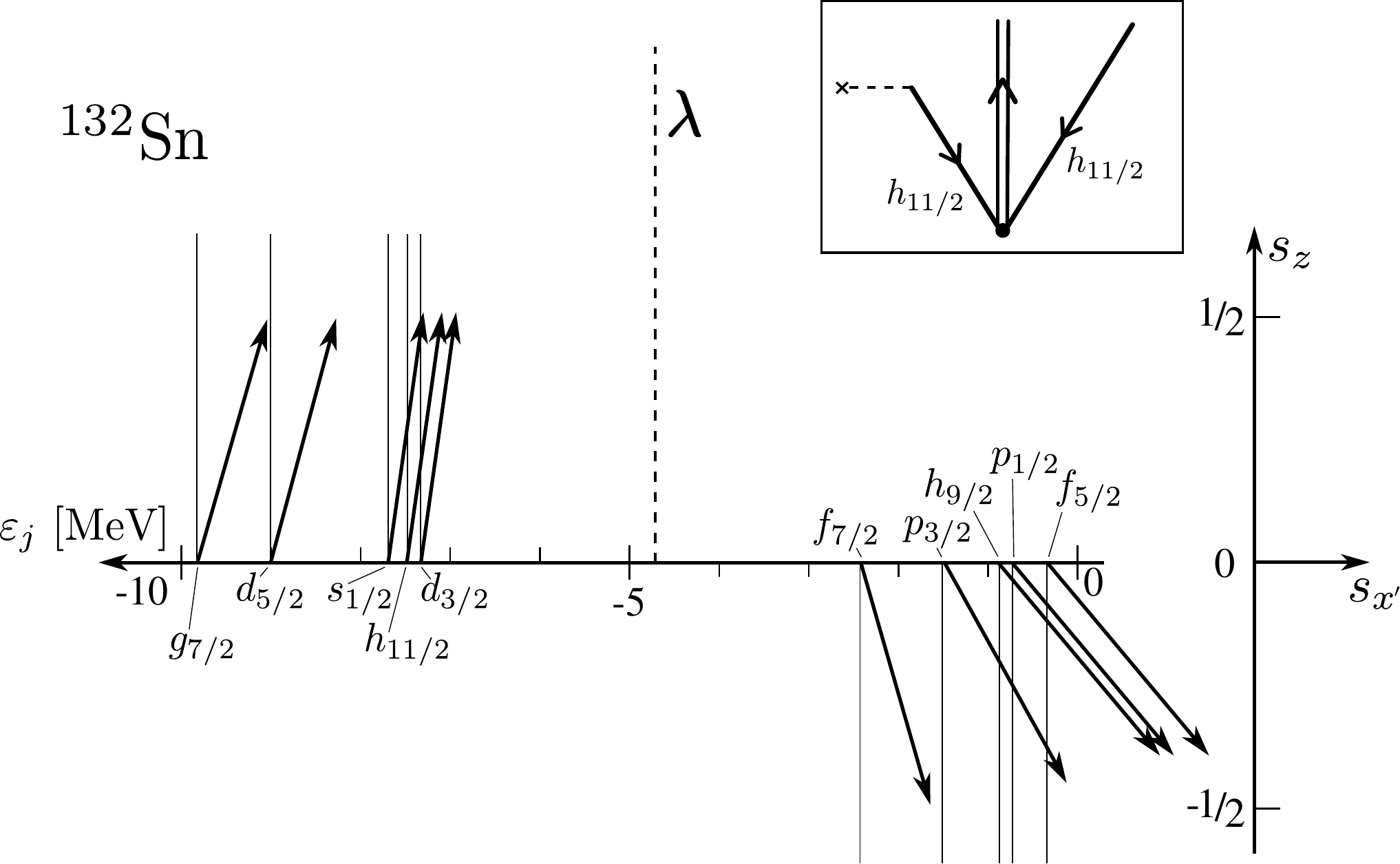}
	\end{center}
\caption{Schematic representation, in terms of pairspin alignment, of the dynamical ODLRO induced in the ground state of the doubly magic
nucleus $^{132}$Sn, by the zero-point fluctuations (ground state correlations) associated with pair addition and pair removal of the closed shell system (for comparison with the superfluid system $^{120}$Sn see Fig. \ref{fig:A3} of Appendix \ref{Appendix:Pair}). The pairspin states are defined in Eq. (\ref{pairspin_a}) ($\epsilon_{\nu} < \epsilon_F$) and in Eq. (\ref{pairspin_b}) ($\epsilon_{\nu} > \epsilon_F$), the $X$ and $Y$ amplitudes 
corresponding to Eq. (\ref{eq.XY}) (see Table 3).
The mean square root value of the angle of the pairspin measured from the gauge $(z-)$axis is fixed by the relation 
${\rm cos} \tilde \theta_{\nu} = \langle s_z(\nu) \rangle/|s|$, 
which for occupied states
is ${\rm cos} \tilde \theta_{\nu}= (1/2)/(3/4)^{1/2}$, i.e. $\tilde \theta_{\nu} =54.5^o$,  the average angle  of the precession cone 
of pairspins centered  around the $z-$axis. Pairspin states (\ref{pairspin_a}) and (\ref{pairspin_b}),  have been calculated
making use of the $X,Y$ values reported in Table \ref{Table:Sn132_PV} and of $\tilde{\theta_{\nu}}$ 
similar to the above one, but which for simplicity
was chosen equal to  be $45^{\circ}$. In the orientation of the pairspin reported in the figure, this fixed angle  has been subtracted. 
In other words, the $z-$axis has been rotated by $45^{\circ}$
into the $z'$-axis. Within the present scenario, it is expected that one-particle transfer reactions on $^{132}$Sn may e.g. excite the 
$h^{-1}_{11/2} \otimes gs(^{134}$Sn)
$2p-1h$ state of $^{133}$Sn (see inset) with a weak, but likely observable cross section. }
\label{fig:remadd}
\end{figure}

%
%
%

\begin{figure}
\centerline{
\includegraphics*[width=.55\textwidth,angle=0]{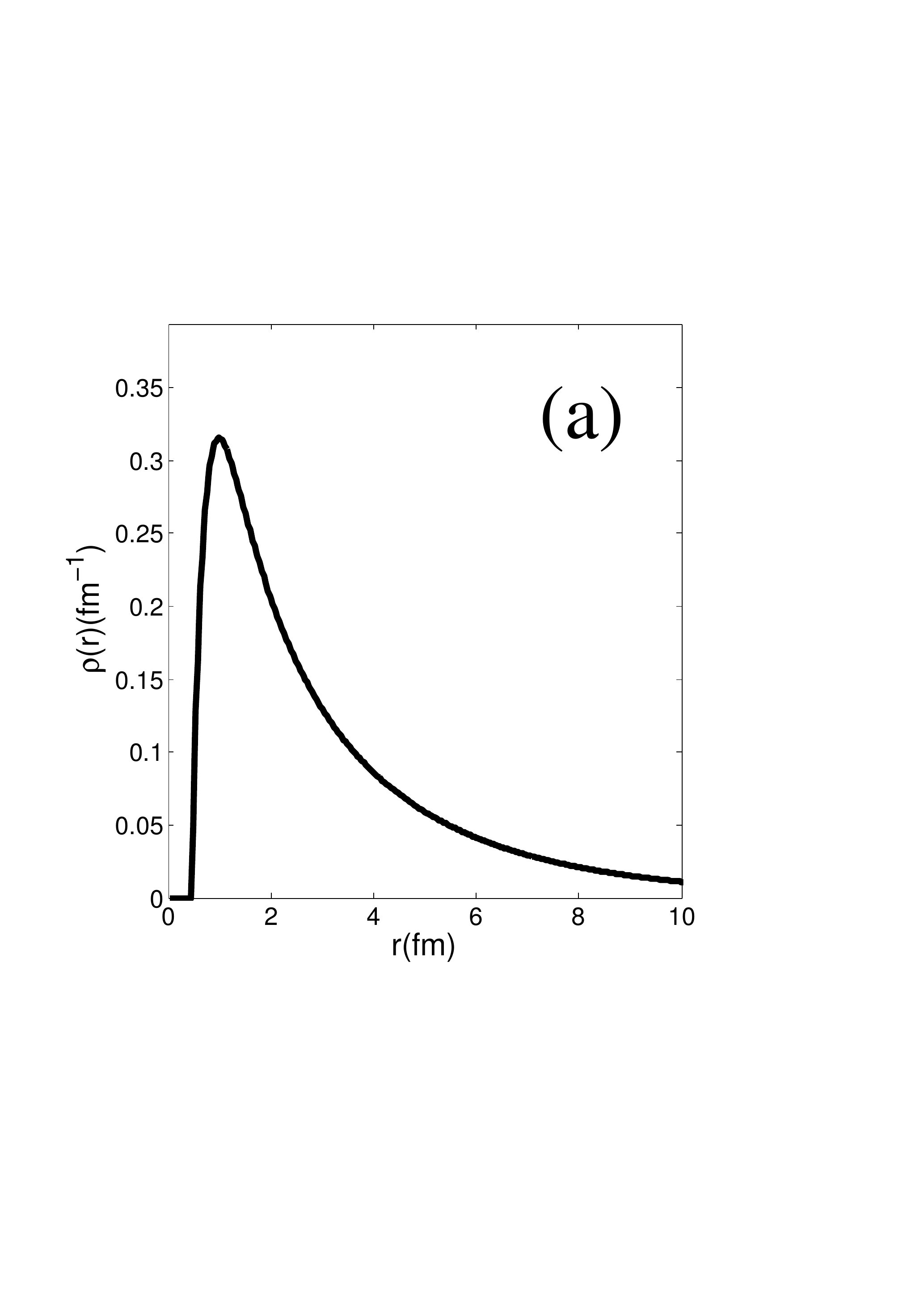}
\includegraphics*[width=.55\textwidth,angle=0]{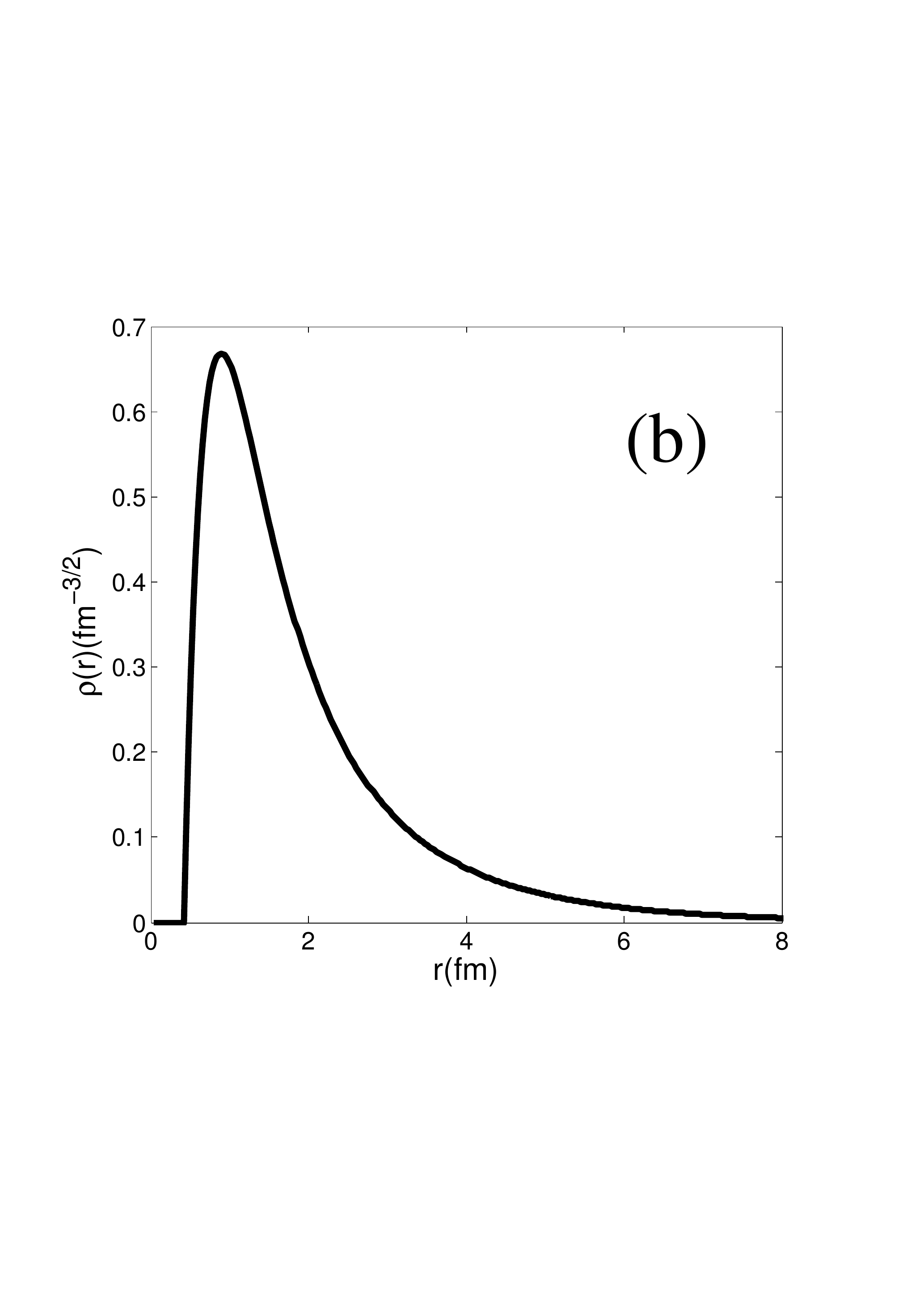} }
\caption{(a) Radial function $\rho(r)$ (hard core 0.45 fm) entering the triton wavefunction.
{(b) Radial} function $\rho(r)$ entering the deuteron wavefunction.}
\label{fig_triton}
\end{figure}


\begin{figure}
	\begin{center}
		\includegraphics[width=0.98\textwidth]{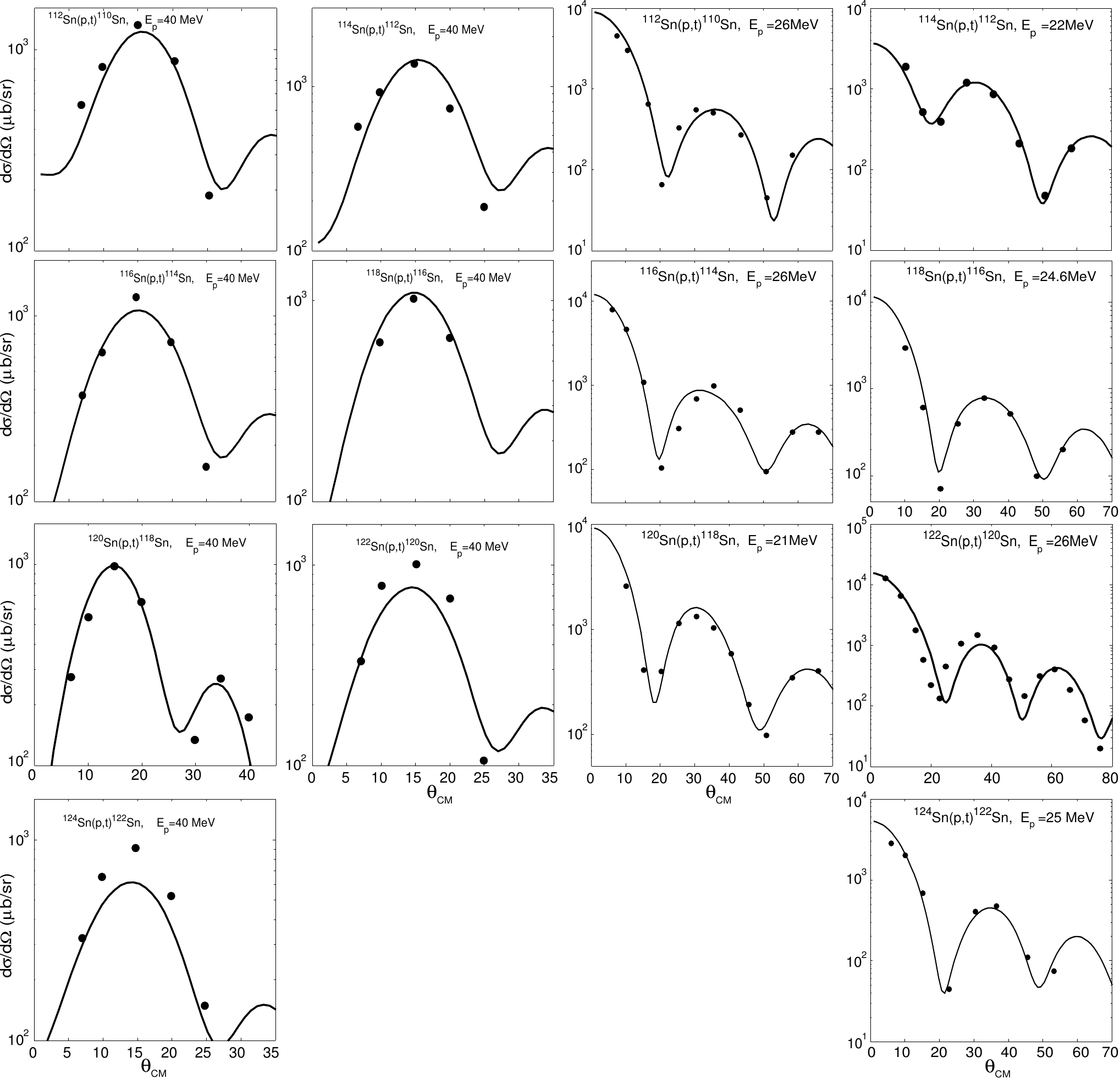}
	\end{center}
	\caption{Predicted absolute differential $^{A+2}$Sn$(p,t)^{A}$Sn$(gs)$ cross sections for bombarding energies $21$ MeV $\leq E_{p} \leq 26$ MeV, and $E_{p}=40$ MeV in comparison with the experimental data (see  \cite{Guazzoni:99,Guazzoni:04,Guazzoni:06,Guazzoni:08,Guazzoni:11,Guazzoni:12} and \cite{Bassani:65} respectively).}
\label{fig4} 
\end{figure}

\begin{figure}
	\begin{center}
		\includegraphics[width=0.98\textwidth]{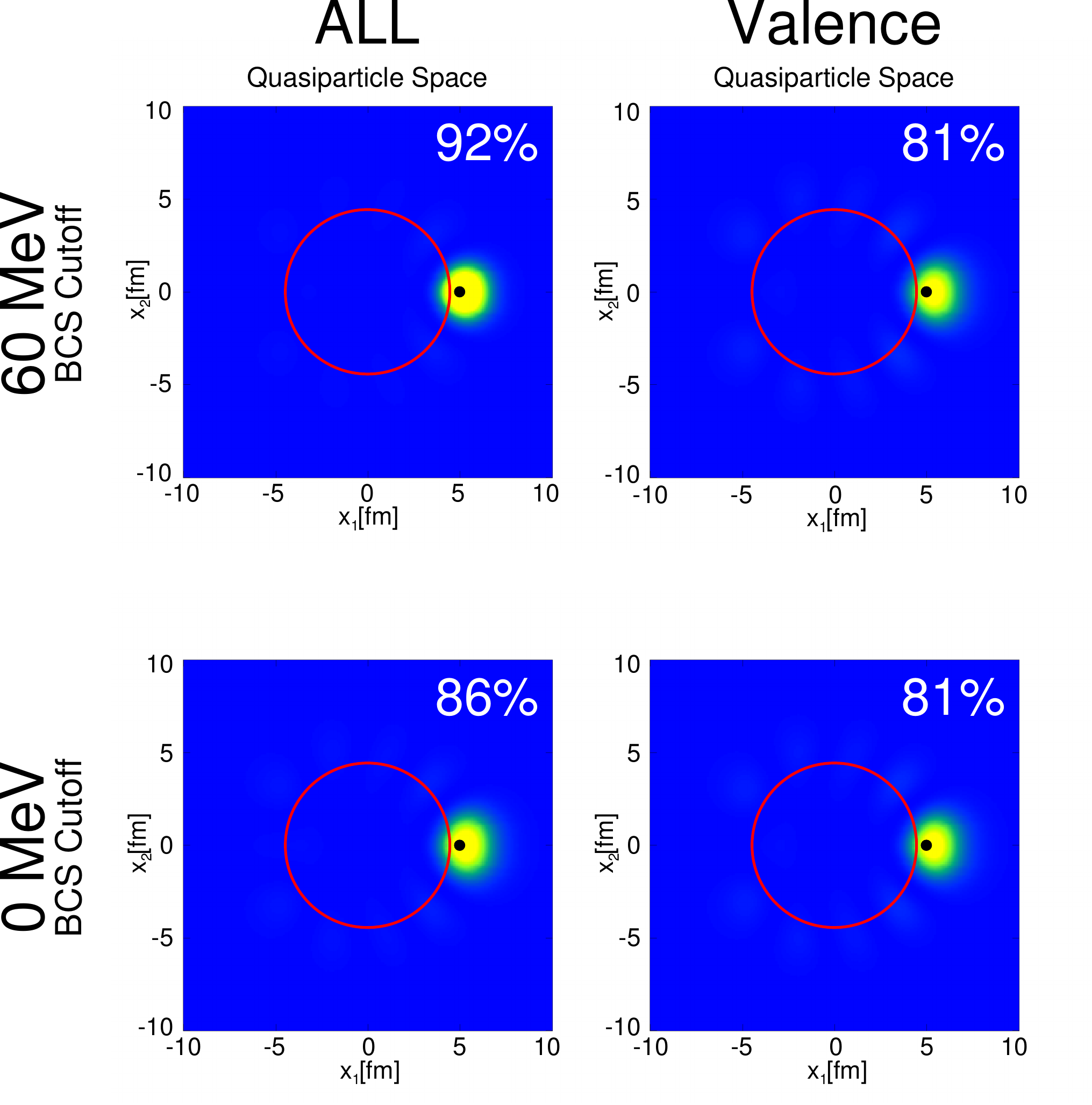}
	\end{center}
	\caption{(Color online) Spatial structure of a two-neutron Cooper pair of $^{120}$Sn (see (\ref{pairband}),(\ref{eq.B17}) and (\ref{eq.B23})). The modulus squared wavefunction $|\Psi_{0}(\vec{r}_1,\vec{r}_2)|^2 = |\langle \tilde{0} \vert \vec{r}_1,\vec{r}_2 \rangle|^2$ (see Tables \ref{Table:1} and \ref{Table:2}), multiplied by $16\pi^2 r^{2}_1 r^{2}_2$ and normalized to unity, is displayed as a function of the cartesian coordinates $x_{1}=r_2 {\rm cos} \theta_{12}$ and $x_{2}=r_2 {\rm sin} \theta_{12}$ of particle 2, for a fixed value of $r_1 = x_1 = 5$ fm (black dot) of particle 1, close to the surface of the nucleus (red circle).
  The numerical percentages correspond to the two-nucleon integrated density in a spherical box of radius 4 fm centered at the coordinates of the fixed particle. }
\label{fig.Cooper}
\end{figure}

\begin{figure}
	\begin{center}
		\includegraphics[width=0.98\textwidth]{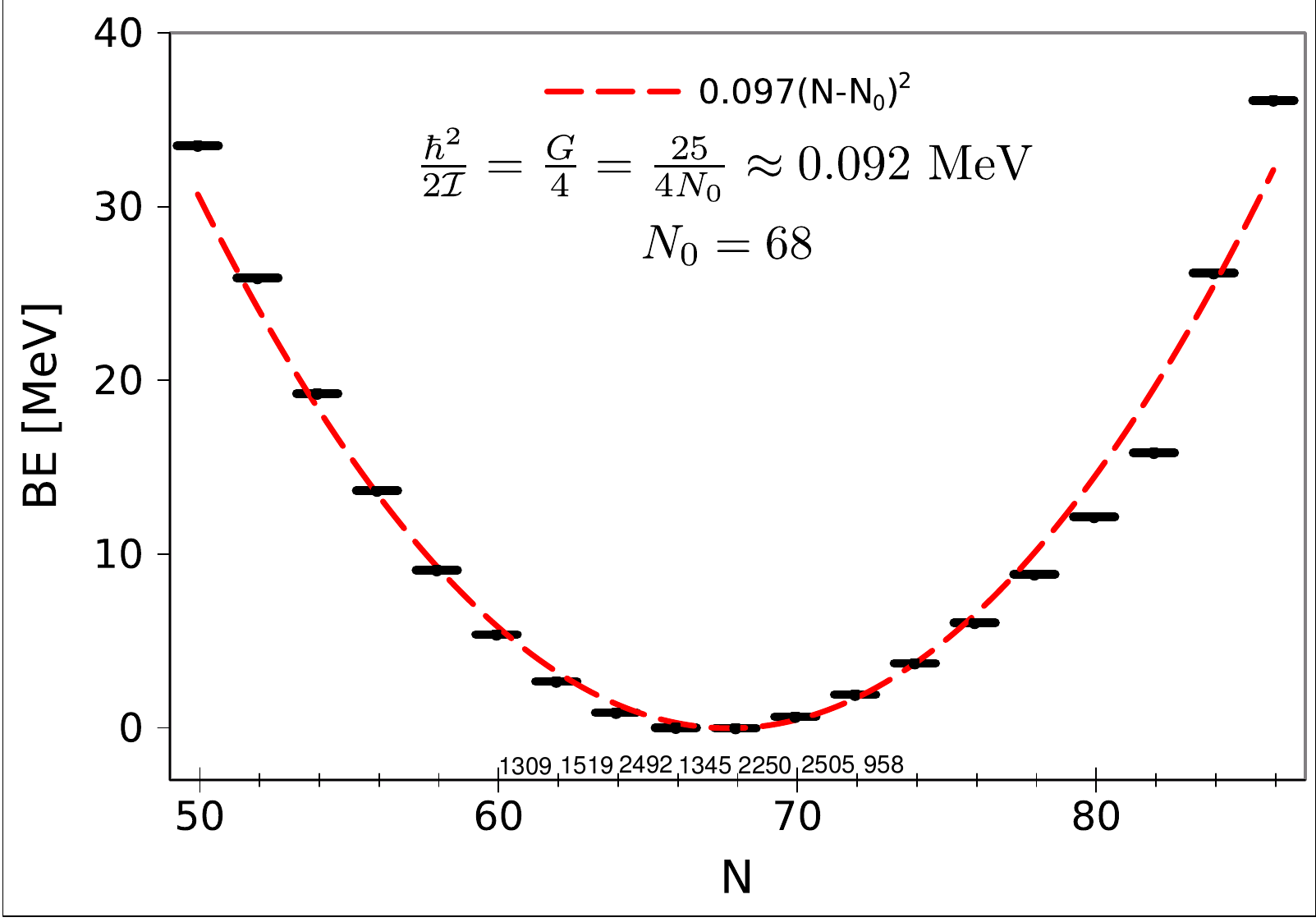}
	\end{center}
	\caption{(Color online) Pairing rotational band along the tin isotopes. The lines represent the energies 
calculated according to the expression  $\Delta B= B({}^{50+N}$Sn$_{N})-8.124N+46.33$ \cite{Brink:05},
subtracting the contribution of the single nucleon addition to the nuclear binding energy obtained by a linear fitting of the binding energies of the whole Sn--chain. The estimate of $\hbar^2/2\mathcal{I}$ was obtained using the single $j$-shell model (see e.g. \cite{Brink:05} App. H).The numbers given on the abscissa are the absolute values of the experimental $gs \to gs$ (in units of $\mu$b; see Table 4).}
\label{fig5}
\end{figure}


\begin{figure}
	\begin{center}
		\includegraphics[width=0.98\textwidth]{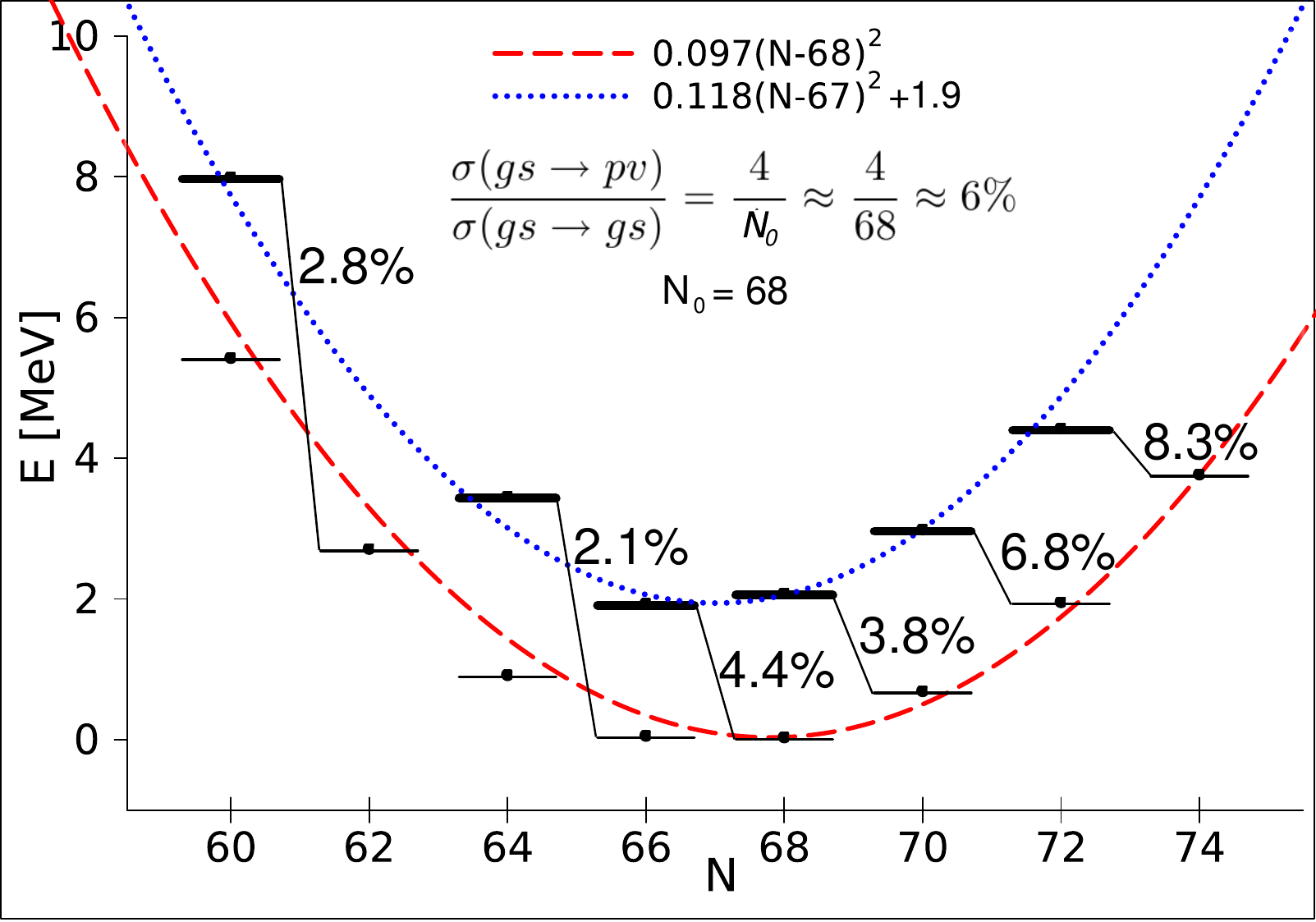}
	\end{center}
	\caption{(Color online) The weighted average energies ($E_{exc}=\sum_{i} E_i \sigma_i / \sum_{i} \sigma_i$) of the excited $0^+$ states  below 3 MeV in the Sn isotopic chain are shown on top of the pairing rotational band,
already displayed  in Fig. \ref{fig5}.  Also indicated is the percentage of cross section for two--neutron transfer to excited states, normalized to the 
cross sections populating the ground states. The estimate of the ratio of cross sections displayed on top of the figure was obtained making use of the single $j-$shell model (see e.g. \cite{Brink:05} App. H).} 
\label{fig6}
\end{figure}

\begin{figure}
	\begin{center}
		\includegraphics[width=0.98\textwidth]{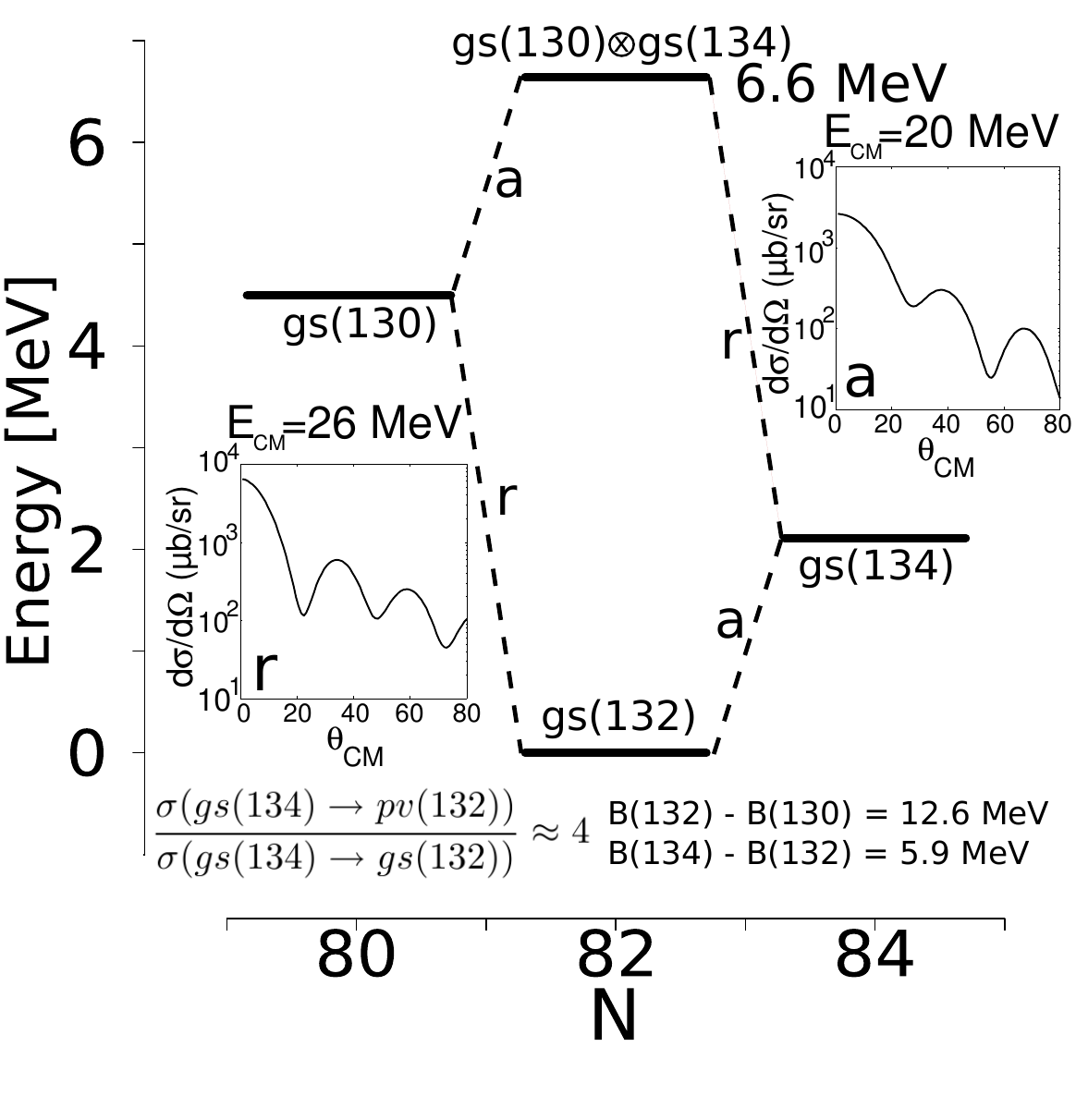}
	\end{center}
	\caption{Pairing vibrational scheme around ${}^{132}$Sn and calculated absolute reaction cross sections associated with the pairing  addition and removal modes. The two-nucleon transfer spectropic amplitudes, used in the calculation are collected in Table \ref{Table:Sn132_PV}. The optical model potential for the three channels, (namely $A+t \rightarrow (A+1) +d \rightarrow (A+2) + p$ or reversed) were taken from refs. \cite{Guazzoni:06} and \cite{An:06}.}
\label{fig7}
\end{figure}

\begin{figure}
	\begin{center}
		\includegraphics[width=0.98\textwidth]{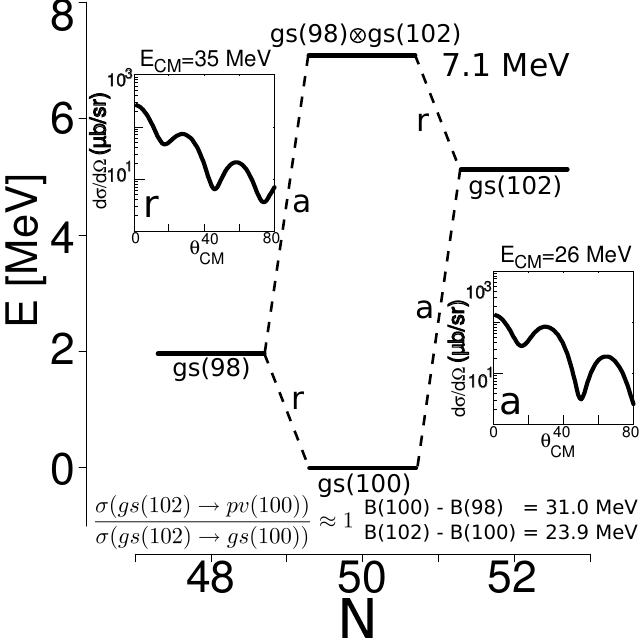}
	\end{center}
	\caption{Pairing vibrational scheme around ${}^{100}$Sn and calculated absolute reaction cross sections associated with the pairing  addition and removal modes. Concerning the spectroscopic amplitudes and optical parameters used in the calculations see caption to Fig.\ref{fig7}.}
\label{fig8}
\end{figure}

\clearpage
\fancyhead[LE,RO]{\bfseries\thepage}
\fancyhead[LO]{\bfseries\rightmark}
\fancyhead[RE]{\bfseries\leftmark}

\appendix
\section{Pair spin and domain wall}
\label{Appendix:Pair}
\setcounter{figure}{0}
\renewcommand{\thefigure}{A.\arabic{figure}}

\begin{figure}
\begin{center}
	\includegraphics[width=0.75\textwidth]{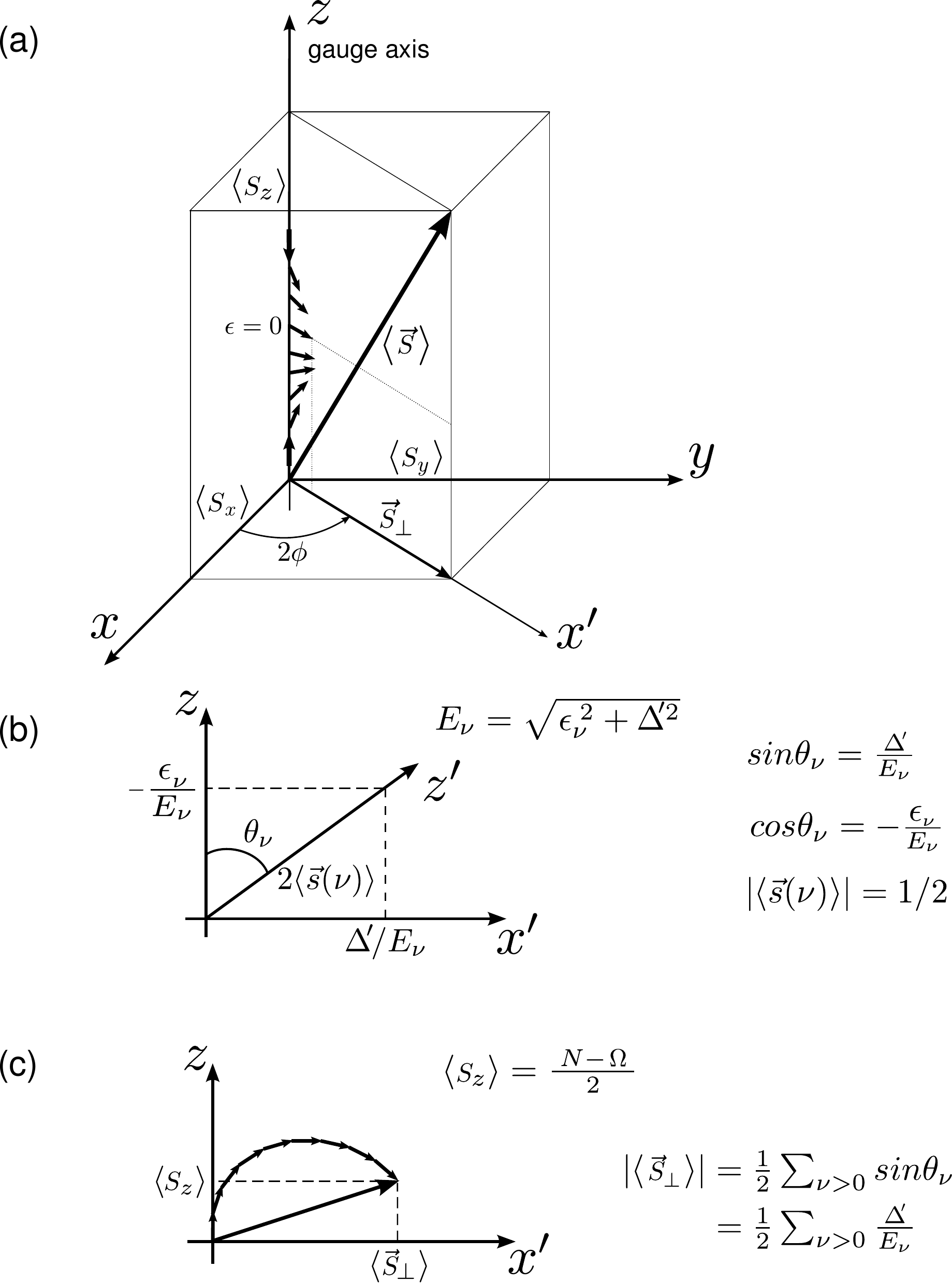}
\end{center}
        \caption{(a) In the presence of the pairing interaction, 
the pairspin vector $< \vec S>$ acquires a component in the $x,y$ plane.
It is then possible to define an intrinsic system ${\cal K'}$, which is obtained rotating the
laboratory system $\cal K$ by the gauge angle $2 \phi$ around the $z-$ (gauge) axis (positive angles correspond to counter clockwise rotations). The  contributions from
the individual pairspins (all lying in the ($z,x')$ plane) are schematically shown. The perpendicular component dominates
for states close to the Fermi energy ($\epsilon \approx 0$),
while the pairspins associated with states far from $\epsilon_F$ are aligned along the $z$-axis. 
(b) Contribution of a pairspin associated with single-particle states of energy $\epsilon_{\nu}$.
The pairspin  vector makes an angle  $\theta_{\nu}$  with the gauge $z-$axis
such that ${\rm sin} \theta_{\nu}= \Delta/E_{\nu}$ and $ cos \theta_{\nu} = - \epsilon_{\nu}/E_{\nu}$.
(c) The total pairspin vector is the sum of many individual contributions. The value of its projection on the $z-$axis
is equal to $(N-\Omega)/2$, where $N$ is the number of particles and $\Omega$ 
is the total pair degeneracy of the single-particle subspace considered to describe the system (see Eq. (\ref{vbcs})).}
        \label{fig:A1}
\end{figure}


In what follows we will discuss the mean field properties of the Hamiltonian 

\begin{equation}
	H_p -\lambda N = H'_{sp} + V_p,
\end{equation}
where
\begin{equation}
	H'_{sp} = \sum_{\nu>0} \epsilon_{\nu} {N}_{\nu} \;, 
\end{equation}
with
\begin{equation}
	\epsilon_{\nu} \equiv \varepsilon_{\nu} - \lambda 
\end{equation}
and
\begin{equation}
	{N}_{\nu} = a^+_{\nu} a_{\nu} + a^+_{\bar{\nu}} a_{\bar{\nu}} \;.
\end{equation}

The pairing interaction is defined as
\begin{equation}
V_p = - G P^+P,
\end{equation}
where 
\begin{equation}
	P^+ = \sum_{\nu>0} P^+_{\nu} \;,
\end{equation}
and
\begin{equation}
	P^+_{\nu} = a^+_{\nu} a^+_{\bar \nu}. 
\label{eq.P}
\end{equation}

The single-particle Hamiltonian $H'_{sp}$ is invariant under time reversal operations. As a consequence, orbitals
are twofold degenerate, the corresponding states being denoted $|\nu \rangle$ and $| \bar{\nu} \rangle$.

Because the operators ${N}_{\nu}$, $P^+_{\nu}$ and $P_{\nu}$ satisfy the commutation relations (cf. Appendix \ref{Appendix:C})
\begin{align}
	\left[ P^+_{\nu}, P_{\nu} \right] &= {N}_{\nu} - 1  \;, \label{comm1}\\
	\left[{N}_{\nu}-1, P^+_{\nu} \right] &= 2 P^+_{\nu} \;, 
\label{comm2}
\end{align}
and
\begin{equation}
	\left[ {N}_{\nu}-1, P_{\nu} \right] = - 2 P_{\nu} \;,
\label{comm3}
\end{equation}
one can define the $x$, $y$ and $z$ components of the pairspin operator $ \vec s(\nu)$ according to the relations,
\begin{equation}
	s_x(\nu) = \frac{1}{2} \left( P^+_{\nu} + P_{\nu} \right)  = \frac{1}{2} \left( \begin{array}{cc} 0  & 1 \\ 1 & 0 \end{array} \right) \;,
\quad s_y(\nu) = \frac{1}{2i} \left( P^+_{\nu} - P_{\nu} \right) = \frac{1}{2} \left( \begin{array}{cc} 0  &-i \\ i & 0 \end{array} \right) \;, 
\label{sxsy}
\end{equation}
and
\begin{equation}
	s_z(\nu) = \frac{1}{2} \left( { N_{\nu}} - 1 \right) = \frac{1}{2} \left( \begin{array}{cc} 1 & 0  \\ 0 &  -1 \end{array} \right) \;.
\label{sz}
\end{equation}

In fact, using the commutation relations (\ref{comm1})-(\ref{comm3}) stated above one obtains (cf. Appendix \ref{Appendix:C}),
\begin{align}
	\left[ s_x(\nu) , s_y(\nu) \right] &= i s_z(\nu)  \;, \\
	\left[ s_y(\nu) , s_z(\nu) \right] &= i s_x(\nu)  \;, \\
	\left[ s_z(\nu) , s_x(\nu) \right] &= i s_y(\nu)  \;.
\end{align}

Of notice that these $\vec{s} \equiv (s_x,s_y,s_z)$ operators although acting in an abstract, gauge space, are as real as the standard spin of electrons and nucleons. 
The $z$--component of pairspin pointing  up means ``occupied'' two-fold degenerate orbitals, pairspin pointing down means "empty", while a pairspin pointing  sidewise implies a certain phased linear combination of up and down (see Fig. \ref{fig:1} as well as Fig. \ref{fig:A1}). In keeping with this scenario,  the BCS ground state displays  a gradual rotation, 
like a domain wall, of the pairspin vectors across the Fermi surface.

\begin{figure}
\begin{center}
	\includegraphics[width=0.6\textwidth]{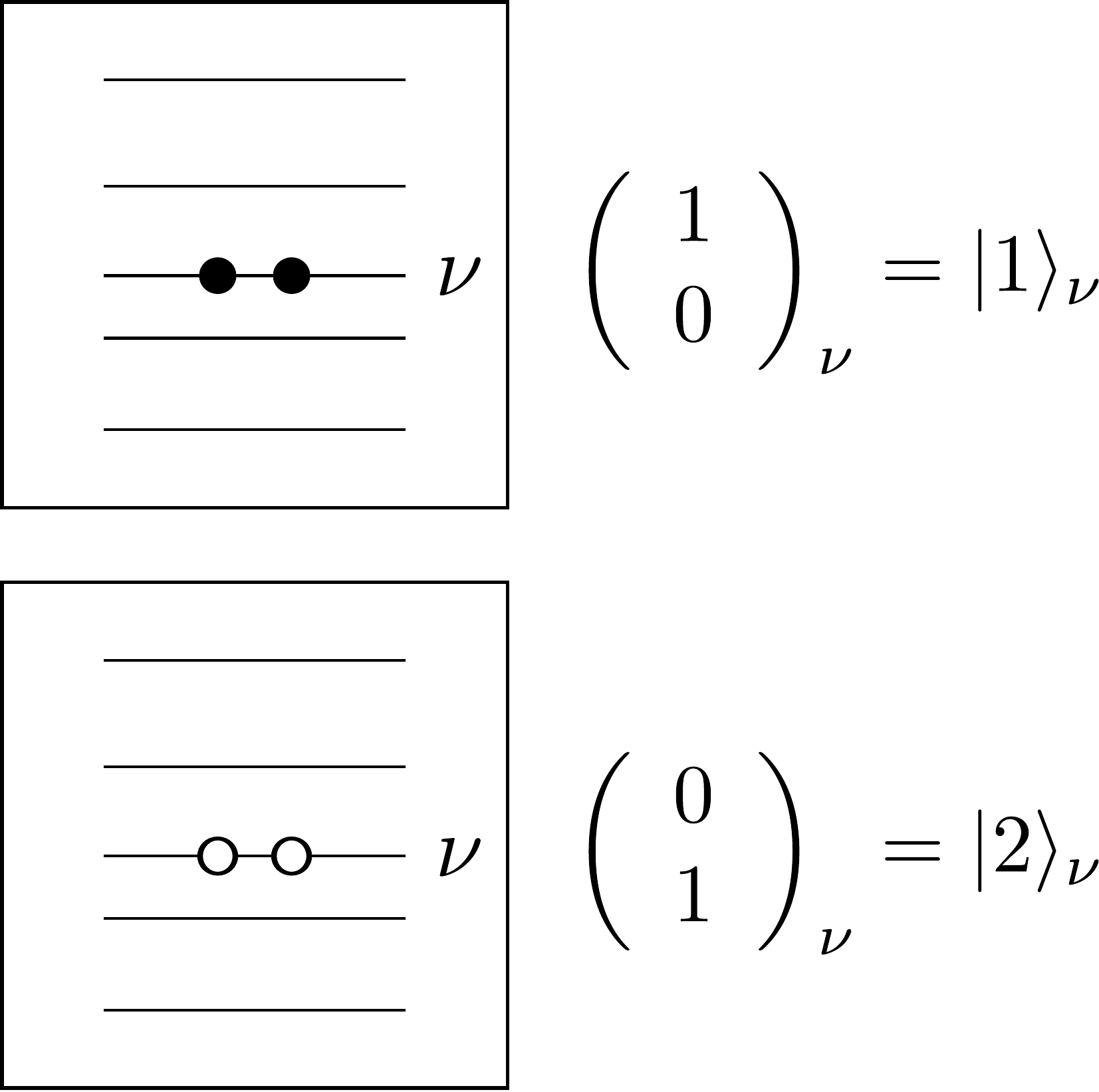}
\end{center}
        \caption{Schematic representation of the eigenvectors of $s_z(\nu)$ in pairspin space: states $\vert 1 \rangle_\nu$ and $\vert 2 \rangle_\nu$ can be viewed as a pair addition and a pair removal mode, at the level of individual pairs of time reversal states $(\nu,\bar \nu)$.}
        \label{fig:A2}
\end{figure}

The eigenvectors of $s_z(\nu)$  in pairspin space are 
\begin{subequations}
\begin{equation} |1\rangle_{\nu} = \left( 
\begin{array}{c} 1 \\ 0 \end{array} 
\right)_{\nu} \equiv  a^+_{\nu} a^+_{\bar \nu} |0\rangle_{\nu} \quad  \quad  \label{eig1} 
\end{equation}
{\rm and } 
\begin{equation}
 |2\rangle_{\nu} = \left( \begin{array}{c} 0 \\ 1 \end{array} \right)_{\nu} \equiv |0\rangle_{\nu}. \label{eig2}
\end{equation}
\end{subequations}
To better clarify the meaning of state $|1>_{\nu}$ and $|2>_{\nu}$, let us assume  to be working with a set of two-fold degenerate states $\nu_1,\nu_2 ...,$ each pair 
of levels connected by time reversal, e.g. $(\nu_1, \bar \nu_1)$. In the uncorrelated case ($G=0$), $|1 \rangle$ and $|2 \rangle$ 
can be viewed as fully occupied or fully empty states (see Fig. \ref{fig:A2}). That is, a  two-particle (filled) and a two-hole state (empty)
respectively (see. Eqs. (A.17) and (A.18) below; see also (A.22) and (A.23)). The same argumentation can be applied to each pair of $(m,-m)$ states connected by time reversal, 
of a general set of $(2j+1)$ degenerate single-particle states.
In other words, the system under consideration displays its pair addition and pair removal modes, at the level of individual pairs of time reversal states $(\nu, \bar \nu)$, building blocks of which Cooper pairs are built. Within this context, it is of notice that Cooper's model works equally well if one thinks of it in terms of a correlated two-hole state in the Fermi sea, BCS being an extension and, in a way, a natural melting of the two views, as required by quantum mechanics (zero point fluctuations (ZPF) which, within the present context can be interpreted in terms of ground state correlations (gsc)). The quantal nature of these correlations is further evidenced by the fact that the different components enter the correlated Cooper pair in terms of probability amplitudes (see in particular (\ref{eq.B17})).

In the basis (A.16), the operators $s_x,s_y$ and $s_z$ have the same matrix representation as the Pauli matrices except for a factor 1/2, that is, $\vec \sigma = 2 \vec s$. 
The action of $s_z$ and ${N}$ on the states (A.16) is given by 
\begin{align}
& s_z(\nu) \left( \begin{array}{c} 1 \\0 \end{array} \right)_{\nu} = 
\frac{1}{2} \left( \begin{array}{cc} 1 & 0  \\ 0 & -1 \end{array} \right) \left( \begin{array}{c} 1 \\ 0 \end{array} \right)_{\nu} = 
1/2  \left( \begin{array}{c} 1 \\0 \end{array} \right)_{\nu} \quad ; \quad \nonumber \\
& s_z(\nu) \left( \begin{array}{c} 0 \\1 \end{array} \right)_{\nu} = 
\frac{1}{2} \left( \begin{array}{cc} 1 & 0  \\ 0 &  -1 \end{array} \right) \left( \begin{array}{c} 0 \\ 1 \end{array} \right)_{\nu} = 
- 1/2  \left( \begin{array}{c} 0 \\1 \end{array} \right)_{\nu} \quad ; \quad \nonumber \\
& {N}_{\nu} \left( \begin{array}{c} 1 \\0 \end{array} \right)_{\nu} =
\left( \begin{array}{cc} 2  & 0  \\ 0 &  0 \end{array} \right)  \left( \begin{array}{c} 1 \\0 \end{array} \right)_{\nu} =
 2  \left( \begin{array}{c} 1 \\0 \end{array} \right)_{\nu} \label{szn} \nonumber \\
 & {N}_{\nu} \left( \begin{array}{c} 0 \\1 \end{array} \right)_{\nu} = 
\left( \begin{array}{cc} 2 &  0  \\ 0 &  0 \end{array} \right) \left( \begin{array}{c} 0 \\1 \end{array} \right)_{\nu}
= 0.
\end{align}

Inverting the relations (\ref{sxsy}), the pair operators $P^+$ and $P$ can be identified with  the raising and lowering operator in pairspace. In fact,
\begin{equation}
P^{+} = \sum_{\nu>0} (s_x(\nu)+ i s_y(\nu))=S_{x}+ i S_{y} \equiv  S_+,
\label{eq.A16}
\end{equation}
and
\begin{equation}
P  = \sum_{\nu>0} (s_x(\nu)- i s_y(\nu))=S_{x}- i S_{y} \equiv  S_-.
\label{eq.A17}
\end{equation}
In this space, i.e. the space subtended by the states $|1\rangle_{\nu} $ and $|2\rangle_{\nu}$, the operators $S_+ (\nu)$ and $S_-(\nu)$ are 
represented by the matrices 
\begin{equation}
\left( \langle i|S_+(\nu)|j \rangle \right)  = \left( \begin{array}{cc} <1|P^+_{\nu}| 1> &  <1|P^+_{\nu}|2> \\  <2|P^+_{\nu}| 1> & <2|P^+_{\nu}|2> \end{array} \right) =
\left( \begin{array}{cc} 0 & 1 \\ 0 & 0 \end{array} \right),
\end{equation}
 \begin{equation}
\left( \langle i|S_-(\nu)|j \rangle \right) = \left( \begin{array}{cc} <1|P_{\nu}| 1> & <1|P_{\nu}|2> \\  <2|P_{\nu}| 1> &  <2|P_{\nu}|2> \end{array} \right) =
\left( \begin{array}{cc} 0  & 0 \\ 1 & 0 \end{array} \right),
\end{equation}
while 
\begin{equation}
\left( \langle i|N(\nu) -1 |j \rangle \right) = \left( \begin{array}{cc} <\nu \bar \nu| (N_{\nu} -1)| \nu \bar \nu > & <\nu \bar \nu| (N_{\nu} -1)| 0>  \\  
<0| (N_{\nu} -1)| \nu \bar \nu > &  <0| (N_{\nu} -1)| 0> \end{array} \right) =  
\left( \begin{array}{cc} 1 & 0 \\ 0 & -1 \end{array} \right) . 
\end{equation}

Consequently
\begin{equation}
P^+_{\nu} \left(
\begin{array}{c}
0\\
1  \end{array} \right)_{\nu}  = 
\left( \begin{array}{cc}
0 & 1  \\
0 & 0  \end{array}  \right)
\left( \begin{array}{c}
0\\
1  \end{array} \right)_{\nu} =
\left( \begin{array}{c}
1 \\
0  \end{array} \right)_{\nu} 
\quad ; \quad
P^+_{\nu} \left(
\begin{array}{c}
1\\
0  \end{array} \right)_{\nu}  = 
\left( \begin{array}{cc}
0 & 1  \\
0 & 0  \end{array}  \right)
\left( \begin{array}{c}
1\\
0  \end{array} \right)_{\nu} = 0,
\label{actionpd}
\end{equation}

while

\begin{equation}
P_{\nu} \left(
\begin{array}{c}
0\\
1  \end{array} \right)_{\nu}  = 
\left( \begin{array}{cc}
0 & 0  \\
1 & 0  \end{array}  \right)
\left( \begin{array}{c}
0\\
1  \end{array} \right)_{\nu} = 0
\quad , \quad
P_{\nu} \left(
\begin{array}{c}
1\\
0  \end{array} \right)_{\nu}  = 
\left( \begin{array}{cc}
0 & 0  \\
1 & 0  \end{array}  \right)
\left( \begin{array}{c}
1\\
0  \end{array} \right)_{\nu} = 
\left( \begin{array}{c}
0 \\
1  \end{array} \right)_{\nu}. 
\label{actionp}
\end{equation}

The total pairspin in the $z$-direction is closely related  to the number operator 
\begin{equation}
S_{z} = \sum_{\nu>0} s_{z}(\nu) = \frac{1}{2}({{N}} - \Omega),
\end{equation}
where ${N} = \sum_{\nu>0} { N}_{\nu}$, and 
$\Omega$ is the total number of two-fold degenerated single-particle orbitals, associated with the single-particle space considered.
Within this context, see Eq. (\ref{vbcs}).

The two terms of the pairing Hamiltonian can now be rewritten, as 
\begin{equation}
H'_{sp} = \sum_{\nu >0} \epsilon_{\nu} \; {N}_{\nu} =  \sum_{\nu >0} \epsilon_{\nu} (1 + 2s_z(\nu)), 
\end{equation}
and 
\begin{align}
& V_p  =  - \sum_{\nu_1,\nu_2 >0} G P^+_{\nu_1} P_{\nu_2} \nonumber \\
& =  - G \left [ \sum_{\nu_1>0}  s_x(\nu_1) \sum_{\nu_2>0} s_x(\nu_2) 
+  \sum_{\nu_1>0}  s_y(\nu_1) \sum_{\nu_2>0} s_y(\nu_2) \right] -G \sum_{\nu_1 > 0}s_{z}(\nu_1) \nonumber \\
& = -  G  \sum_{\nu_1,\nu_2>0} (\vec s_{\perp}(\nu_1) \cdot \vec s_{\perp}(\nu_2))-G\sum_{\nu_1>0}s_z(\nu_1),
\label{vp}
\end{align}
where $s_{\perp}(\nu) = s_x(\nu) {\hat i} + s_y(\nu) {\hat j}$,
${\hat i}$ and ${\hat j}$ being unit vectors along the $x-$ and $y-$ directions. The last term is the contribution of the pairing interaction
to the single-particle mean field $H_{sp}$. Although it can easily be incorporated in this term, it is customary to neglect it, 
in keeping with the schematic nature of $V_p$, tailored to act on the pair space. On the other hand, this contribution is to be considered when comparing 
the solution of schematic models with exact solutions (see e.g. \cite{Bes:76a},\cite{Bes:76b}).

\begin{figure}
\begin{center}
	\includegraphics[width=0.95\textwidth]{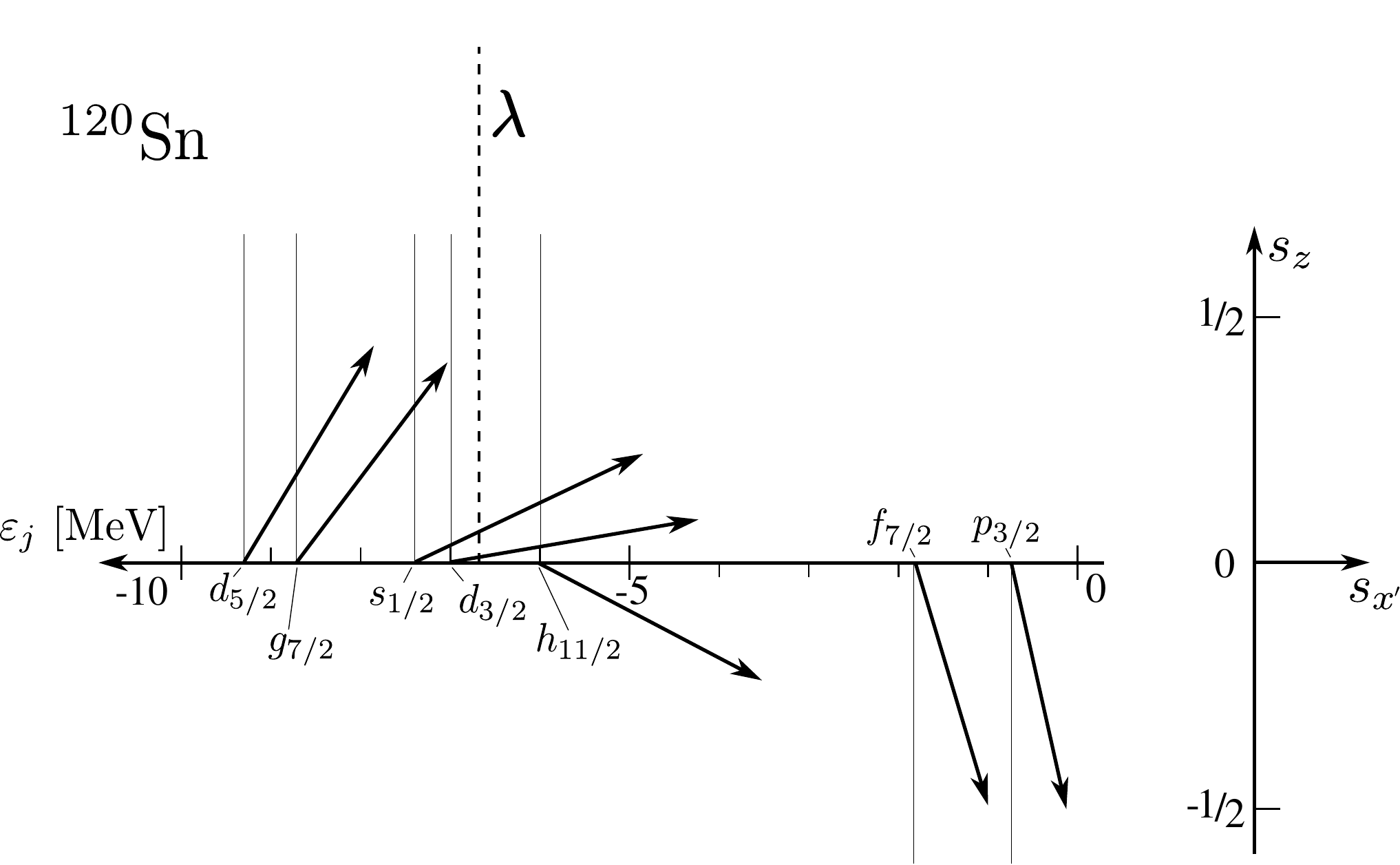}
\end{center}
        \caption{Pairspin distribution associated with the seven valence single-particle orbitals lying around the Fermi energy of $^{120}$Sn. 
        The calculations were carried out making use of the results displayed in Table \ref{Table:2}, in particular the values $U'_\nu V'_\nu$,  following the prescription discussed in the caption to Fig. \ref{fig:remadd}. }
        \label{fig:A3}
\end{figure}

The interaction $V_p$ is a spin-spin coupling, which seeks to align the transverse components of the pairspins. Was it not for $H_{sp}$, which depends on $S_z$, all pairspins would line up in the same direction (strong coupling limit), perpendicular to the $z$-axis, in keeping with the fact that $V_p$ is a function of only the $S_x$ and $S_y$ pairspin operator.
In the opposite limit, that is,  for $V_p = 0$, pairspin alignment is zero, that is, there is no component of the pairspin in the $(x,y)$ plane perpendicular to the $z$-axis (see Fig. 1(a)). From this figure it is clear that $H_{sp}$ counteracts pair spin alignment. Indeed $H_{sp}$ plays, in the nucleus, the role of a magnetic field in a solid, which tends to align the spins in the $z$-direction, or opposite to it, with a strength (as measured by $\varepsilon_{\nu}$ ) that increases in absolute value as a function of the energy of the twofold degenerate levels $(\nu,\bar{\nu})$ away from the Fermi energy $\varepsilon_F=\lambda$. Thus, for sufficiently large values of $\vert \varepsilon_{\nu} - \epsilon_F\vert $, of the order of 2 $\Delta$ (2-3 MeV), the single-particle energy as measured by $H_{sp}$ dominates (see Fig. 1 as well as Fig. \ref{fig:A1}(a) and \ref{fig:A3}).
This is also the reason why  calculations of pairing correlations in nuclei, which depend on small contributions
arizing from orbitals distant from the Fermi energy are to be handled with care, in particular  when comparing the calculated results with the experimental findings. 
Within this context see Sect.\ref{Sec.Rotations}, discussion connected with $\alpha'_0(\vec r_1, \vec r_2)$.
Close to the Fermi energy, $V_p$ is the overriding effect, and a large transverse pairspin is expected, as testified by the coherent sums 
in Eq. (\ref{vp}). While any single pairspin in these sums displays large quantal fluctuations, the total pairspin has a well defined magnitude and orientation, since the fluctuations of the constituent pairspins add quadratically.

The pairing Hamiltonian can be diagonalized in the mean field approximation, substituting one of the sums in Eq. (\ref{vp}) with its average value in the mean field ground state.
Let us assume that  the  average  value of the pairspin polarization vector   $|\langle \vec S_{\perp} \rangle| = {}_{\cal K'}\langle BCS | S_{\perp} | BCS \rangle_{\cal K'}$ 
(see. Eq. \ref{BCSphi} below) is non-zero, and choose a definite orientation for it in the $x-y$ plane (denoted $x'$), subtending 
an angle $2 \phi$ with the $x-$ axis  (definition of the body-fixed, intrinsic frame, see Fig. A.1).
The pair interaction is then replaced by a mean field, namely the pair field of  strength $\Delta' \equiv G |\langle\vec S_{\perp} \rangle|$. One can then write,
\begin{align}
 U_p & = - 2 G | \langle \vec S_{\perp} \rangle | \sum_{\nu > 0}  \left [ s_{x}(\nu) {\rm cos} 2\phi + s_y (\nu) {\rm sin} 2\phi \right]  \notag \\
& = - 2 \Delta' \sum_{\nu > 0}   \left [ s_{x}(\nu) {\rm cos} 2\phi + s_y (\nu) {\rm sin} 2\phi \right].
\label{phiansatz}
\end{align}
Making use of the relation given in (\ref{sxsy}) one finds, 
\begin{equation}
\Delta'(s_x(\nu) {\rm cos} 2\phi + s_y(\nu) {\rm sin} 2\phi) = \frac{1}{2} \left( P^{\dagger}_{\nu} \Delta' e^{-2 i \phi} + P_{\nu} \Delta' e ^{2 i \phi} \right) = 
\frac{1}{2} (P^{\dagger}_{\nu} \Delta + P_{\nu} \Delta^*),
\end{equation}
where we have introduced  $\Delta = e^{-2 i \phi} \Delta'$. Consequently,
\begin{equation}
U_p = - \Delta' \sum_{\nu > 0}  ( P^{\dagger'}_{\nu} + P'_{\nu}) = - \sum_{\nu>0} ( P^{\dagger}_{\nu} \Delta + P_{\nu} \Delta^* ),
\label{up}
\end{equation}
$P'^+, P'$ denoting the operators in the body-fixed frame of axis $(x',y')$ (cf. Fig. \ref{fig:A1}(a) and Appendix \ref{Appendix:B}) in which by definition 
$\phi = 0$, that is $\langle S'_{y} \rangle = 0$ and $\langle S'_{x} \rangle = \langle S'_{\perp} \rangle = \alpha'_0$. 


The total Hamiltonian then  becomes a sum  over individual pairspins, 
\begin{equation}
(H_p)_{MF} = \sum_{\nu > 0} h_{\nu},
\end{equation}
where 
\begin{equation}
h_{\nu} = \epsilon_{\nu} + \epsilon_{\nu}
\left( \begin{array}{cc} 1 & 0 \\ 0  & -1 \end{array} \right) -
\Delta' \left( \begin{array}{cc} \quad 0  & {\rm cos} 2 \phi - i {\rm sin} 2\phi \\  {\rm cos} 2 \phi + i {\rm sin} 2 \phi & \quad \quad 0 \end{array} \right). 
\label{hnu}
\end{equation}
It is of notice that thinking in terms of the independent (quasiparticle) densities the first term is connected with the normal and the second with the so called abnormal density respectively, leading to ODLRO.

Within this context,  in the mean field associated with the (diagonal) first term of the above equation (see also Eq. (A.2)), the particles which move independently of each other, are nucleons. In order that this can happen, all the nucleons must participate in a highly coherent ballet following a refined coreography, 
in such a way that each dancer moves as if he was alone in the scene, being fenced-in through a dancers-like wall, only when approaching the edges of the scene. As it has been stated
in the literature \cite{Mottelson:62}, it is a "rather unfortunate perversity"  which views independent  particle motion as antithetic  to nuclear collective motion.

In the case of the mean field described by the second term  of Eq. (\ref{hnu}), the entities which play the role of nucleons in the case above, 
are now pairspins (pair addition and pair removal ($\nu, \bar \nu$ modes)).
The only circumstance in which pairspins feel the pushings and pullings of the other pairspins, is when they try 
to adopt a different orientation 
but that defined by $ S_{\perp}$ (i.e.  $x'-$direction, see Fig. A.1), being forced to align back by a domain wall. 
In the present case, the ballet  is not performed  by single dancers in a scene, but  by couples  in a crowded dancing hall. 
In spite of such a less cultured setup, the coreography is even more refined than previously described. This is because the partners of each dancing couple can, 
not only when close to each
other,  but also when finding themselves at opposite  extremes of the dancing hall (coherence length), follow the other partners moves without missing a single step.
Such a coreography of strongly overlapping pairs translates, in the gauge (pairspin) space, into the definition of  a privileged orientation. 

This can be better seen by expressing the diagonalization condition (\ref{eq.4}), namely
\begin{subequations}
\begin{equation}
 \sum_{\nu>0} \epsilon_{\nu} N_{\nu} - \Delta' (P^{' \dagger}+P') + \frac{\Delta'^2}{G} = 
 \sum_{\nu>0} E_{\nu} \tilde N_{\nu} + \textit{const}
\end{equation}
in terms of the quasispin operators (\ref{sxsy}) (\ref{sz}) and
\begin{equation}
 s_{z'}(\nu) = -\frac{1}{2} (\widetilde N_{\nu} - 1).
\end{equation}
That is,
\begin{equation}
  \sum_{\nu > 0} \left[ \epsilon_{\nu} 2 s_{z}(\nu) - \Delta' 2s_x (\nu) \right] + \sum_{\nu>0}(\epsilon_\nu - E_\nu) + \frac{\Delta'^2}{G} 
= - \sum_{\nu > 0} E_{\nu} 2 s_{z'} (\nu) + \textit{const}.
\end{equation}
\end{subequations}
In a similar way in which $s_z(\nu)$ is diagonal in the independent (pair) particle situation, $s_{z'}(\nu)$ is diagonal in the quasiparticle representation, with eigeinvalues $1/2$ when acting on the corresponding occupied states ($\vert HF \rangle$ and $\vert BCS \rangle$ states respectively), and $-1/2$ when acting on the corresponding unoccupied states (levels above $\varepsilon_F$ and two quasiparticle states respectively).
Equating the (quasispin) operator terms and the c-number terms one obtains
\begin{subequations}
 \begin{equation}
  \textit{const} = \sum_{\nu > 0} (\epsilon_\nu - E_{\nu}) + \frac{\Delta'^2}{G},
 \end{equation}
and
 \begin{equation}
   - \frac{\epsilon_{\nu}}{E_{\nu}} s_{z}(\nu) + \frac{\Delta'}{E_{\nu}}s_{x}(\nu) = s_{z'}(\nu).
\label{eq.A34b}
 \end{equation}
\end{subequations}

%
The above equation determines the angle $\theta_{\nu}$ in the $z-x$ plane, which leads to independent quasispin motion. In the strong coupling limit ($|\Delta' | >> | \epsilon_{\nu} - \lambda|$), "magnetization" is total, all pairspins (pair addition and removal $(\nu, \bar \nu)$  pairs) pointing along the $x-$ direction.

As already stated in connection with Eq. (\ref{eq.4}) of the text, the c-number \textit{const} is equal to the ground state energy, also known as the $U$ term of $(H_{p})_{MF}$ (see e.g. \cite{Brink:05} Eq. (G.11) of App. G; see also below, subsection on ground state energy and pairing correlation energy).

In keeping with the fact that the quasispin states are normalized, and that one has chosen a representation in which $s_{y}(\nu) = 0$, the quasispin prefactors of Eq. (\ref{eq.A34b}) must fulfill the relation,
 \begin{equation}
   \left( -\frac{\epsilon_{\nu}}{E_{\nu}} \right)^2 + \frac{\Delta'^2}{E^2_{\nu}} = 1.
\label{eq.A35a}
 \end{equation}
It is of notice that the eigenvalue equation (\ref{hnu}),
\begin{equation}
\epsilon_{\nu} -\left( \begin{array}{cc} 
-\epsilon_{\nu}    &  \Delta' e^{-2i\phi} \\
\Delta' e^{2i\phi}  &  \epsilon_{\nu}        \end{array} \right) \left( \begin{array}{c} V_{\nu}\\ U_{\nu}\end{array} \right) =
 (\epsilon_{\nu} -  E)  \left ( \begin{array}{c} V_{\nu} \\ U_{\nu}\end{array} \right), 
\label{eq.A33}
\end{equation}
has the solutions $\pm E_{\nu}$, where (see also (\ref{eq.A35a}))
\begin{equation} 
E_{\nu}=\sqrt{ \epsilon_{\nu}^{2} + \Delta^{'2}}.
\label{eqpsol}
\end{equation} 
The positive sign corresponds to the lowest total pairspin energy ($\epsilon_{\nu} -  E_{\nu}$), while the negative sign corresponds to the excited states (2-quasiparticle states, cf. subsection below on excited states).
The relation between the $U_{\nu}$ and the $V_{\nu}$ components of the eigenvector can be deduced from equation (\ref{eq.A33}),
\begin{align}
 -\epsilon_{\nu}      V_{\nu} + \Delta' e^{-2i\phi}U_{\nu} = E_{\nu} V_{\nu}  \label{eq.A35}\\
  \Delta' e^{-2i\phi}V_{\nu} + \epsilon_{\nu}    U_{\nu} = E_{\nu} U_{\nu}.
\end{align}
Eq.( \ref{eq.A35}) leads to $(E_\nu+\epsilon_\nu) V_{\nu}= \Delta' e^{-2i\phi} U_{\nu}$, implying 
a phase difference  $-2 \phi$ between the $V_{\nu}$ and the $U_{\nu}$ occupation amplitudes,
This allows one to write the following relations
\begin{align}
U_{\nu}  & = U'_{\nu} e^{i \phi}, \label{unu}\\ 
V_{\nu}  & = V'_{\nu} e^{-i\phi}, \label{vnu}
\end{align}
with $U'_{\nu}$ and $V'_{\nu}$ being the moduli of $U_{\nu}$ and $V_{\nu}$ respectively.
These quantities can be calculated from  the square modulus of Eq. (\ref{eq.A35})
\begin{equation}
  (\epsilon_{\nu}+E_{\nu})^2 |V_{\nu}|^2 = \Delta^{'2} |U_{\nu}|^2.
\end{equation}
Making use of the  normalization relation $|V_{\nu}|^2=1-|U_{\nu}|^2$, one obtains
\begin{eqnarray}
U^{'2}_{\nu} =  & \frac{1}{2}  \left( 1 + \frac{\epsilon^{}_{\nu}}{E_{\nu}} \right), \nonumber \\ 
V^{'2}_{\nu} =  & \frac{1}{2}  \left( 1 - \frac{\epsilon^{}_{\nu}}{E_{\nu}} \right).
\end{eqnarray}  
Defining the angle $\theta_{\nu}$ 
according to 
\begin{equation}
{\rm cos} \theta_{\nu} = - \frac{\epsilon_{\nu}}{E_{\nu}} \quad , \quad
{\rm sin} \theta_{\nu}  = \frac{\Delta'}{E_{\nu}},
\end{equation}
one can rewrite the quasiparticle amplitudes as 
\begin{equation}
U'_{\nu} = {\rm sin} (\theta_{\nu}/2)  \quad  , \quad    V'_{\nu} = {\rm cos} (\theta_{\nu}/2),
\end{equation}
where the angle $\theta_{\nu}$ represents the angle between the direction of $\vec s(\nu)$  ($z'$-axis in Fig. \ref{fig:A1}(b)) and the $z-$axis. 
The average value of $s_x(\nu),s_y(\nu)$ calculated with these eigenfunctions, generalizations of the eigenvectors introduced before (see Eqs. (\ref{eig1}) and (\ref{eig2})), 
are given by 
\begin{align}
 \langle s_x(\nu) \rangle & =  ( V^*_{\nu} \; , \; U^*_{\nu}) s_x(\nu)  \left( \begin{array}{c} V_{\nu}\\ U_{\nu} \end{array} \right) 
 = \frac{1}{2} \left( U_{\nu} V^*_{\nu}+ V_{\nu} U^*_{\nu} \right) \notag \\
 & = {\rm sin} \frac{\theta_{\nu}}{2} \textrm{ cos} \frac{\theta_{\nu}} {2} \textrm{ cos} 2 \phi = U'_\nu V'_\nu \textrm{ cos}2\phi = \frac{\Delta'}{2E_{\nu}}\textrm{ cos}2\phi, \\
 \langle s_y(\nu) \rangle & =  ( V^*_{\nu} \; , \; U^*_{\nu}) s_y(\nu)  \left( \begin{array}{c} V_{\nu} \\ U_{\nu} \end{array} \right)  
= \frac{i}{2} \left ( U_{\nu} V^*_{\nu} - V^*_{\nu}U_{\nu} \right) \notag \\
 & = {\rm sin} \frac{\theta_{\nu}}{2} \textrm{ cos} \frac{\theta_{\nu}} {2} \textrm{ sin} 2 \phi = U'_\nu V'_\nu \textrm{ sin}2\phi = \frac{\Delta'}{2E_{\nu}}\textrm{ sin}2\phi,
\label{sxave}
\end{align}
and
\begin{align}
 \langle s_z(\nu) \rangle & =  ( V^*_{\nu} \; , \; U^*_{\nu}) s_z(\nu)  \left( \begin{array}{c} V_{\nu} \\ U_{\nu}\end{array} \right) 
 = \frac{1}{2} \left ( V_{\nu} V^*_{\nu}- U^*_{\nu} U_{\nu} \right) \notag \\
 & = \frac{1}{2} \left ({\rm cos}^2 \frac{\theta_{\nu}}{2} - {\rm sin}^2 \frac{\theta_{\nu}}{2} \right)=- \frac{1}{2}\left(U'^{2}_{\nu} - V'^{2}_{\nu} \right) = 
- \frac{\varepsilon_\nu}{2E_\nu}.
\end{align}
The dependence of $\langle s_x(\nu) \rangle$ and $\langle s_y(\nu) \rangle$ on the the angle $\phi$ is consistent with the ansatz (\ref{phiansatz}).
Furthermore, making use of Eq. (\ref{sxave}), one can determine the modulus of $S_\perp$ selfconsistently

\begin{equation}
\vert \sum_{\nu>0} \langle \vec s_{\perp} (\nu) \rangle \vert  = \vert \langle \vec S_{\perp} \rangle \vert = \sum_{\nu >0} \left[ {\rm cos} \frac{\theta_{\nu}}{2} {\rm sin}\frac{\theta_{\nu}}{2} \right] =
\sum_{\nu >0} U'_{\nu} V'_{\nu} = \alpha'_0,
\end{equation}
a relation which is closely connected with the BCS gap equation. In Fig. \ref{fig:A3} the pairspin distribution associated with the valence orbitals of $^{120}$Sn, calculated making use of the results collected in Table \ref{Table:2} and the scheme discussed in the caption to Fig. \ref{fig:remadd}, is displayed.

The wavefunction describing the ground state of the system is the product of all pairspins, each of which 
is a linear combination of the pair  addition and removal $(\nu, \bar \nu)$ modes with the corresponding weights $V_{\nu}$ and $U_{\nu}$
respectively, amplitudes which define their alignment in gauge space. That is, 
\begin{align}
& \prod_{\nu>0}\left( U_\nu |2\rangle_\nu + V_\nu |1\rangle_\nu \right) = \prod_{\nu>0} \left( U_{\nu}  \left( \begin{array}{c} 0 \\1 \end{array} \right) + 
             V_{\nu}  \left( \begin{array}{c} 1 \\0 \end{array} \right) \right)  = \prod_{\nu>0} \left(U_{\nu} + V_{\nu} a^{\dagger}_{\nu} a^{\dagger}_{\bar \nu} \right) |0\rangle  = \nonumber \\
& = \prod_{\nu>0} \left( e^{i \phi} U'_{\nu} + e^{-i \phi} V'_{\nu} a^{\dagger}_{\nu} a^{\dagger}_{\bar \nu} \right) |0\rangle = 
e^{i \Omega \phi} \prod_{\nu>0} \left(U'_{\nu} + V'_{\nu} e^{- 2 i \phi} a^+_{\nu} a^+_{\bar \nu} \right) |0 \rangle.
\end{align} 
 Leaving out the overall phase one can write 
\begin{eqnarray}
|BCS(\phi)\rangle_{\cal K} = \prod_{\nu>0}  \left( U'_{\nu} + V'_{\nu} e^{- 2 i \phi} a^+_{\nu} a^+_{\bar \nu} \right) |0\rangle  = \nonumber \\
\prod_{\nu>0} \left( U'_{\nu} + V'_{\nu} a^{'+}_{\nu} a^{'+}_{\bar \nu} \right)|0\rangle = |BCS(\phi=0) \rangle_{\cal K'},
\label{BCSphi}
\end{eqnarray}
where $\cal K$ and $\cal K'$ denote the laboratory and the body-fixed, intrinsic frame in which by definition has $\phi=0$ (see Fig. \ref{fig:B1}).

The creation operator in the intrinsic, body-fixed frame of reference is
\begin{equation}
a^{' \dagger}_{\nu} = {\cal G} a^{\dagger}_{\nu} {\cal G}^{-1} = e^{-i \phi} a^+_{\nu},
\end{equation}
where  ${\cal G} = e^{ - i N \phi}$.
is the gauge operator inducing rotations in the two-dimensional gauge space. 
The quantities $U'_{\nu}$ and $V'_{\nu}$ are real. 

In the independent particle limit, that is,
\begin{align}
 \lim_{\Delta \to 0} \prod_{\nu>0} \left ( U_{\nu} + V_{\nu} a^{\dagger}_{\nu} a^{\dagger}_{\bar \nu} | 0 \rangle \right) = 
\prod_{\nu>0,\nu < \nu_F} a^+_{\nu} a^+_{\bar \nu} |0 \rangle = a^+_{\nu_1}a^+_{\bar \nu_1} a^+_{\nu_2}a^+_{\bar \nu_2} ... a^+_{\nu_N}a^+_{\bar \nu_N} \vert 0 \rangle \notag \\
 = \frac{1}{\sqrt{2 N!}} \textrm{det} (1 \bar 1, 2 \bar 2 ... N \bar N) |0\rangle,
\end{align}
as expected.

Summing up, the instability of the Fermi surface associated with transverse pairspin 
polarization is associated with a many-body wavefunction, product of the individual pairspin states, 
\begin{equation}
\vert 0 \rangle_{\nu} = U_{\nu} | s_z (\nu)= -1/2 \rangle + V_{\nu} | s_z (\nu) = + 1/2 > = U_{\nu}|2>_{\nu}+V_{\nu}|1>_{\nu},
\end{equation}
superposition of pairspin up and down and therefore non axially symmetric with respect to the gauge axis ($z-$axis). 
In other words, a linear combination of pair addition and substraction modes mixed at the level of individual pairs of time-reversal 
states $(\nu,\bar \nu)$ (see Fig. \ref{fig:A2}). In the same way as $| ^{208}{\rm Pb} (gs)\rangle$ spends part of the time in the state
 $| ^{210}{\rm Pb} (gs)\rangle$ and part in $| ^{206}{\rm Pb} (gs)\rangle$, $|^{120}{\rm Sn(gs)}\rangle$ is a mixture of pair addition and removal states
$|1\rangle$ and $|2\rangle$, respectively.
The transverse polarization -
which inherently breaks gauge symmetry - arises from such a superposition, a phenomenon which is produced by the action 
of the ``external'' mean pair field $U_p$. 

The pairspin polarization 
may rotate collectively about the $z-$gauge axis. The azimuthal angle 
is therefore a dynamical variable
(pairing rotations). The static pair field constitutes a deformation that defines an orientation.
Through this deformation,  the system
spontaneously breaks away from axial symmetry, and the indeterminacy in the number of particles in a pair correlated state  is an inherent feature  of this symmetry breaking. 
The static deformation introduces a collective degree of freedom $\phi$ and gives the system the ability to rotate as a whole around the gauge axis.

When the mean field solution leads to $\langle \vec S_{\perp} \rangle = 0$, 
the intrinsic motion has axial symmetry and hence conserves particle number. 
In these systems, gauge invariance can also be broken dynamically, a phenomenon 
which gives rise to the pairing vibrational spectrum observed around closed shell 
nuclei, in terms of highly enhanced, single Cooper pair tunneling processes.

\subsection*{Excited states}

The BCS equation in the body-fixed frame,
\begin{equation}
 \left( \begin{array}{cc} \epsilon_\nu & -\Delta' \\ -\Delta' & -\epsilon_\nu \end{array} \right) \left( \begin{array}{c} V'_\nu \\ U'_\nu \end{array} \right) =
 -E\left( \begin{array}{c} V'_\nu \\ U'_\nu \end{array} \right),
\end{equation}
leads to the eigenvalue equation
\begin{equation}
 ( \epsilon_\nu + E ) ( \epsilon_\nu - E ) - \Delta'^2 = 0,
\end{equation}
with eigenvalues $E_{\nu} = \sqrt{\epsilon_{\nu}^{2} + \Delta'^2}$, 
corresponding to the ground state previously considered (cf. Eq. (\ref{eqpsol})) and $ - E_{\nu}$, corresponding to excited states.
The associated eigenvectors are 
\begin{equation} 
\vert {\rm gs} \rangle = \left( \begin{array}{c} V'_{\nu} \\   U'_{\nu} \end{array} \right)  \quad ,\quad 
\vert {\rm exc}\rangle = \left( \begin{array}{c} U'_{\nu} \\ - V'_{\nu} \end{array} \right). 
\end{equation}
The state $\vert \textrm{exc}\rangle$ is a two-quasiparticle state. It can be excited acting with $\alpha^+_{\nu} \alpha^+_{\bar \nu}$
on $|BCS\rangle_{\cal K}$. 
In fact, using  $\alpha^+_{\nu} = U'_{\nu} a'^+_{\nu} - V'_{\nu} a'_{\bar \nu} $ one finds (it is of notice that $s'_z(\nu)=s_z(\nu)$) 
\begin{equation}
\alpha^+_{\nu} \alpha^+_{ \bar \nu} = U'^2_{\nu} P'^+_{\nu} - V'^2_{\nu} P'_{\nu} + 2 U'_{\nu} V'_{\nu} s_z(\nu) = 
\left( \begin{array}{cc} U'_\nu V'_\nu & U'^2_\nu       \\ 
                        -V'^2_\nu      & - U'_\nu V'_\nu 
\end{array} \right), 
\end{equation}
and
\begin{equation}
\alpha_{\bar \nu} \alpha_{\nu} = U'^2_{\nu} P'_{\nu} - V'^2_{\nu} P'^+_{\nu} + 2 U'_{\nu} V'_{\nu} s'_z(\nu) = 
\left( \begin{array}{cc} U'_\nu V'_\nu & -V'^2_\nu        \\ 
                         U'^2_\nu      & -U'_\nu V_\nu
\end{array} \right).
\end{equation}
Making use of Eqs. (\ref{actionpd}-\ref{actionp}), one finds that 
\begin{equation}
\alpha^+_{\nu} \alpha^+_{\bar \nu} \vert \textrm{gs} \rangle = \alpha^+_{\nu} \alpha^+_{\bar \nu} \left( \begin{array}{c} V'_{\nu} \\ U'_{\nu} \end{array} \right) = 
\left( \begin{array}{cc} U'_\nu V'_\nu & U'^2_\nu       \\ 
                        -V'^2_\nu     & - U'_\nu V'_\nu 
\end{array} \right) \left( \begin{array}{c} V'_{\nu} \\ U'_{\nu} \end{array} \right) = \left( \begin{array}{c} U'_{\nu} \\  -V'_{\nu} \end{array} \right) = \vert \textrm{exc} \rangle,
\end{equation}
and 
\begin{equation}
\alpha^+_{\nu} \alpha^+_{\bar \nu} \vert \textrm{exc}\rangle = \alpha^+_{\nu} \alpha^+_{\bar \nu} \left( \begin{array}{c} U'_{\nu} \\  - V'_{\nu} \end{array} \right)= \left( \begin{array}{cc} U'_\nu V'_\nu & U'^2_\nu       \\ 
                        -V'^2_\nu     & - U'_\nu V'_\nu 
\end{array} \right) \left( \begin{array}{c} U'_{\nu} \\  - V'_{\nu} \end{array} \right) = 0.
\end{equation}
Analogously, one finds 
\begin{equation}
\alpha_{\bar \nu} \alpha_{\nu} \vert \textrm{gs} \rangle = 0,
\end{equation}
and 
\begin{equation}
 \alpha_{\bar \nu} \alpha_{\nu} \vert \textrm{exc} \rangle = \vert \textrm{gs} \rangle.
\end{equation}

We also remark that 
\begin{equation}
2i s'_y (\nu)= P'^+_{\nu} - P'_{\nu} = \alpha^+_{\nu} \alpha^+_{ \bar \nu} - \alpha_{\bar \nu} \alpha_{\nu}.
\end{equation}
One can then show that the operator 2$i s_y(\nu)$ acting once on the ground state 
produces the excited state and, acting twice gives back the ground state, but for a sign change, that is,
\begin{equation} 
2i s'_y(\nu) \vert {\rm gs} \rangle  = \vert {\rm exc} \rangle   \quad , \quad  2i s'_y(\nu) \vert {\rm exc} \rangle = - \vert {\rm gs}\rangle.
\end{equation}
Thus, two-quasiparticle excitation is equivalent to a rotation of the intrinsic system by an angle $\pi$ around the $y-$axis. 
This is in keeping with the fact that  $e^{\pi iS_{y}}=2iS_{y}$.

\subsection*{Ground state energy and correlations}
The energy of the ground state of the total system at the mean field level is obtained as the sum of the energy of each pair spin, $\epsilon_\nu - E_{\nu}$ (cf. eq. (\ref{eq.A33})), which is the eigenvalue diagonalizing the terms $\epsilon_{\nu}N_\nu - (P^{\dagger}\Delta + P \Delta^{*})$ of the mean field hamiltonian, plus the constant term $\frac{\Delta'^2}{G}$ (cf. eq. (\ref{eq.4})),
\begin{equation}
 E_{gs} = \sum_{\nu > 0} (\epsilon_\nu - E_{\nu}) +\frac{\Delta'^2}{G}.
\label{eq.E_gs}
\end{equation}
In the case of the non-interacting system ($\Delta \to 0$) $E_{\nu} \to \vert \epsilon_{\nu} \vert$, so that the system's ground state energy reads
\begin{equation}
 E^{0}_{gs} = \sum_{\nu > 0} (\epsilon_\nu - |\epsilon_\nu|).
\end{equation}
Consequently, the contributions corresponding to $\epsilon_\nu > 0$ (i.e. $\epsilon_\nu > \lambda$), cancel out, remaining only those of the states below the Fermi energy ($\epsilon_\nu < 0$). Thus,
\begin{equation}
 E^{0}_{gs} = 2 \sum_{\nu >0 ; \epsilon_{\nu} < 0} \epsilon_\nu,
\end{equation}
in keeping with the fact that in the present case  all the pairs below the Fermi energy fully occupied. An important quantity characterizing superfluid systems is the so called  correlation energy, $E_{corr}$. It is defined as the difference between the energy of the interacting system and that of the non-interacting one,
\begin{equation}
 E_{corr} = E_{gs} - E^{0}_{gs} = \sum_{\nu > 0} (\epsilon_{\nu} - E_{\nu}) + \frac{\Delta'^2}{G} - 2 \sum_{\nu >0 ; \epsilon_{\nu} < 0} \epsilon_{\nu}.
\end{equation}
Collecting the contribution arising from levels displaying $\epsilon_\nu > 0$ and $\epsilon_\nu < 0$, the above equation can be written as
\begin{equation}
 E_{corr} = \sum_{\nu >0 ; \epsilon_{\nu} > 0} (\epsilon_{\nu} - E_{\nu}) + \sum_{\nu >0 ; \epsilon_{\nu} < 0} (-\epsilon_{\nu} - E_{\nu}) +  \frac{\Delta'^2}{G}
          = \sum_{\nu >0 ; \epsilon_{\nu} > 0} (\epsilon_{\nu} - E_{\nu}) + \sum_{\nu >0 ; \epsilon_{\nu} < 0} (|\epsilon_{\nu}|- E_{\nu}) +  \frac{\Delta'^2}{G}.
\end{equation}
Assuming a symmetric distribution of levels around the Fermi energy, the above expression becomes
\begin{equation}
 E_{corr} = 2 \sum_{\nu >0 ; \epsilon_{\nu} > 0} (\epsilon_{\nu} - E_{\nu}) + \frac{\Delta'^2}{G}.
\end{equation}
It proofs useful to express the ground state energy in the term of occupation probabilities and potential interaction energy, namely as (see term $U$ eq. (G.11) of ref. \cite{Brink:05})
\begin{equation}
 E_{gs} = \sum_{\nu > 0} 2 \epsilon_{\nu} V'^2_{\nu} - \frac{\Delta'^2}{G}.
\end{equation}
Using the expression for $V'^2_{\nu}$
\begin{equation}
E_{gs} = \sum_{\nu > 0} \epsilon_{\nu} \left(1-\frac{\epsilon_{\nu}}{E_{\nu}} \right)  - \frac{\Delta'^2}{G},
\end{equation}
which may be rewritten as
\begin{align}
 E_{gs} & = \sum_{\nu > 0} \epsilon_{\nu} \left(1-\frac{\epsilon_{\nu}}{E_{\nu}} -\frac{E_{\nu}}{\epsilon_{\nu}}+ \frac{E_{\nu}}{\epsilon_{\nu}} \right) - \frac{\Delta'^2}{G}
          = \sum_{\nu > 0}(\epsilon_{\nu}-E_{\nu}) + \sum_{\nu > 0}\left(E_{\nu} - \frac{\epsilon^2_{\nu}}{E_{\nu}}\right)                   - \frac{\Delta'^2}{G} \notag \\
        & = \sum_{\nu > 0}(\epsilon_{\nu}-E_{\nu}) + \sum_{\nu > 0} \frac{\Delta'^2}{E_{\nu}}                                                - \frac{\Delta'^2}{G}
          = \sum_{\nu > 0}(\epsilon_{\nu}-E_{\nu}) + 2 \frac{\Delta'^2}{G}                                                                   - \frac{\Delta'^2}{G} \notag \\
        & = \sum_{\nu > 0}(\epsilon_{\nu}-E_{\nu}) + \frac{\Delta'^2}{G} ,
\end{align}
which coincides with (\ref{eq.E_gs}).

\section{Pairing rotational band wavefunction}
\label{Appendix:B}
\setcounter{figure}{0}
\renewcommand{\thefigure}{B.\arabic{figure}}

\begin{figure}[h]
 \centering
 \includegraphics[scale=0.1,keepaspectratio=true]{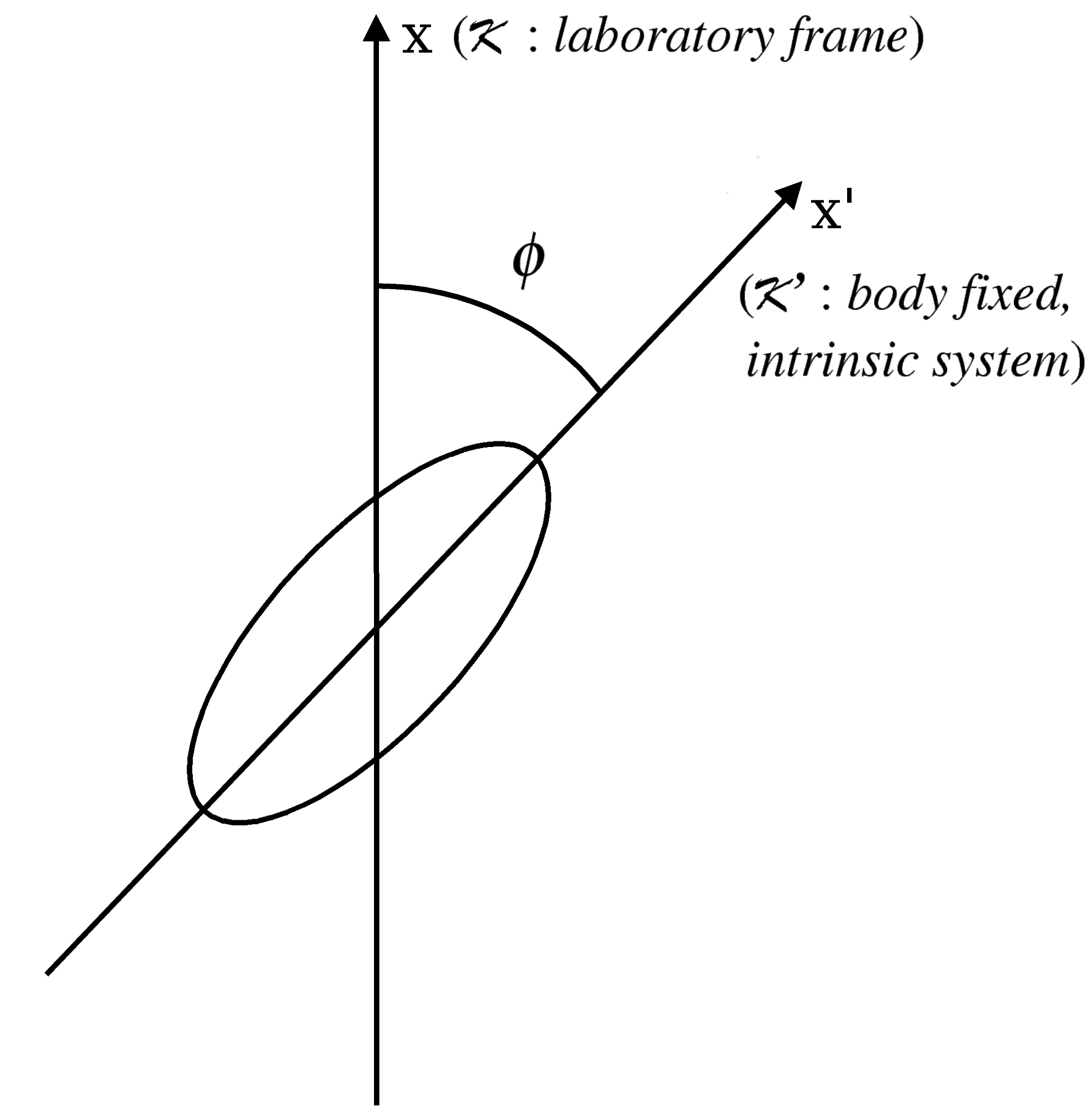}
 \caption{Schematic representation of the deformation in gauge space associated with a
superfluid nucleus leading to pairspin alignment 
(see also  Fig. \ref{fig:A1})}
 \label{fig:B1}
\end{figure}

Let us start by defining the operator inducing a gauge transformation. In keeping with the fact 
that the generator of such a transformation is the particle number operator, we will be dealing with rotations in a two-dimensional space. 

The operator inducing 
rotation in pairspin space about the gauge $z$-axis is given 
$ e^{- 2 i s_z({\nu}) \phi}$
The representation of this operator in pairspin space is 
\begin{equation}
\left( \begin{array}{cc} e^{-i \phi} & 0 \\ 0  & e^{i \phi} \end{array} \right).
\end{equation}
This operator  converts the state  $| \phi =0\rangle_{\nu} =\left( \begin{array}{c} V' _{\nu}\\ U_{\nu}' \end{array}\right)$, 
into the state rotated by an angle $\phi$ (cf. Eqs. (\ref{unu}),(\ref{vnu})):
\begin{equation}
\left( \begin{array}{cc}  e^{-i \phi} & 0 \\ 0 & e^{i \phi} \end{array} \right) \left( \begin{array}{c} V' _{\nu}\\ U_{\nu}' \end{array}\right) = 
 \left( \begin{array}{c} e^{-i \phi} V' _{\nu}\\e^{i \phi} U_{\nu}' \end{array}\right)
 \end{equation}
 In the following we shall use the operator ${\mathcal G}_{\nu}(\phi) = e^{- i \phi N_{\nu}}$ which differs from $e^{-2 i s_z(\nu)}$ only  by 
 an overall phase $e^{i \phi}$. Its representation in pairspin space is given by 
  \begin{equation}
\left( \begin{array}{cc} e^{-2 i \phi} & 0 \\ 0  & 1 \end{array} \right),
\end{equation} 
and its action on $|\phi=0>_{\nu} = \left( \begin{array}{c} V'_{\nu}\\U'_{\nu} \end{array}\right)$ is given by 
\begin{equation}
\left( \begin{array}{cc}  e^{- 2 i \phi} & 0 \\ 0 & 1 \end{array} \right) \left( \begin{array}{c} V' _{\nu}\\ U_{\nu}' \end{array}\right) = 
 \left( \begin{array}{c} e^{- 2 i \phi} V' _{\nu}\\ U'_{\nu} \end{array}\right),
 \end{equation}
 producing the  phase difference characterizing the rotated,  $|BCS(\phi)>$ state (cf. Eq. (A.51)).
 Calculating the average value of $P$ in the (rotated) state $\mathcal{G}(\phi) \vert \phi=0 \rangle_{\nu}$, that is 
 $\alpha_0 (\nu) = {\;}_{\nu}\langle \phi=0| {\cal G}^{-1}(\phi) P {\cal G}(\phi) |\phi=0\rangle_{\nu}$, one finds
 \begin{equation}
 (e^{2 i \phi}  V'_{\nu}  \quad U'_{\nu} ) \left( \begin{array}{cc} 0 & 0 \\ 1 & 0 \end{array}\right) \left( \begin{array}{c} e^{-2 i \phi} V'_{\nu}\\ U'_{\nu} \end{array} \right) = 
 e^{-2 i \phi} U'_{\nu} V'_{\nu}.
 \end{equation}

We can also calculate the same  average value 
by rotating the operator (Heisenberg representation), rather than acting on the state. 
The rotated operator is given by ${\cal G}_{\nu}^{\dagger}(\phi) P_{\nu}(\phi=0) {\cal G}_{\nu}(\phi)$.
The action of ${\cal G}_{\nu}^{\dagger}(\phi) P_{\nu}(\phi=0) {\cal G}_{\nu}(\phi)$ 
transforms the initial operator in the intrinsic frame $P_{\nu} (\phi=0)$, in which it has the average value 
$\alpha'_0 (\nu) = U_{\nu}'V_{\nu}'$ into the laboratory frame, in which its average value is $\alpha_0(\nu) = \alpha'_0(\nu) e^{-2 i \phi}$.  The inverse transformation,
from the laboratory into the intrinsic frame, is effected by the operator $({\cal G}^{\dagger}_{\nu}(\phi) P_{\nu}(\phi) {\cal G}_{\nu}(\phi))^{-1}$=  
${\cal G}_{\nu}(\phi) P_{\nu}(\phi) {\cal G}^{\dagger}(\phi)$.
In what follows, we shall use $P'_{\nu}  \equiv P_{\nu}(\phi=0)$,  and $P_{\nu} \equiv  P_{\nu}(\phi)$.

Similar considerations can be applied to the many-body wavefunctions. 
Let  us consider a wavefunction 
\begin{equation}
\Psi_{\mathcal{K}} = a^{\dagger}_1 a^{\dagger}_2 ... a^{\dagger}_{N_0} |0>,
\end{equation} 
with a fixed number of particles. Let us now apply $\mathcal{G}(\phi)$ to the creation operator 
\begin{equation}
a'^{\dagger}_{\nu} =  \mathcal{G}(\phi) a^{\dagger}_{\nu} \mathcal{G}^{-1} (\phi) = e^{-i \phi} a^{\dagger}_{\nu},
\label{eq.B7}
\end{equation}
$a'^{\dagger}$ being referred to the intrinsic, body-fixed reference system (see Fig. \ref{fig:B1}). Thus
\begin{equation}
\Psi_\mathcal{K} (\phi) = e^{i N_0 \phi} a'^{\dagger}_1 a'^{\dagger}_2 ... a'^{\dagger}_{N_0} | 0> =
e^{iN \phi} \Psi_{K'} (\phi=0),
\end{equation}
\begin{equation}
\Psi_{\mathcal{K}'} (\phi=0) = a'^{\dagger}_1 a'^{\dagger}_2 ... a'^{\dagger}_{N_0} | 0 >.
\end{equation}
In keeping with the fact that  
\begin{equation}
- i \frac{\partial \Psi_\mathcal{K} (\phi)} {\partial \phi} = -i \times i N_0 e^{iN\phi} \Psi_{\mathcal{K}'} (\phi=0) =
N_0 \Psi_\mathcal{K} (\phi),
\end{equation}
\begin{equation}
N = - i \frac{\partial}{\partial \phi}  \quad \quad  , \quad \quad  N  \; \Psi_\mathcal{K} (\phi) = N_0  \; \Psi_\mathcal{K} (\phi).
\end{equation}
This is equivalent to saying that 
\begin{equation}
[\phi,N] \psi = (\phi N  - N \phi) \psi =  \Phi \left( -i \frac{\partial}{\partial \phi} \psi \right ) + i  \frac{\partial}{\partial \phi} ( \phi \psi) =
+i \psi - i \phi \frac{\partial \psi}{\partial \phi} - i \phi \frac{\partial \psi}{\partial \phi} = i \psi, 
\end{equation}
i.e. $[\phi, N] = i$. The BCS wavefunction can then be written, in the intrinsic body-fixed frame, as  
\begin{equation}
|BCS (\phi=0)>_{\mathcal{K}'}\sim \prod_{\nu} \alpha_{\nu} |0 > \sim \prod_{\nu > 0} \alpha_{\nu}\alpha_{\bar \nu}|0>.
\end{equation}
The corresponding normalized wavefunction is then 
\begin{align}
\notag
& \vert BCS(\phi=0) \rangle_{\mathcal{K}'} = \prod_{\nu >0} (U'_{\nu} + V'_{\nu} a'^{\dagger}_{\nu} a'^{\dagger}_{\bar \nu} ) \vert 0 \rangle = \prod_{\nu} 
(U'_{\nu} + e^{- 2i \phi} V'_{\nu} a^{\dagger}_{\nu}a^{\dagger}_{\bar \nu}) |0> \\ 
&= \vert BCS(\phi)\rangle_{\mathcal{K}}= (\prod_{\nu >0} U'_{\nu}) \left( 1 + \frac{e^{-2i \phi}}{1!} \sum_{\nu >0} c_{\nu} a^{\dagger}_{\nu} a^{\dagger}_{\bar \nu} + \frac{e^{- 4 i \phi}}{2!} \left(\sum_{\nu >0} c_{\nu} a^{\dagger}_{\nu} a^{\dagger}_{\bar \nu} \right)^2 + .... \right),
\end{align}
where $c_{\nu} = V'_{\nu}/U'_{\nu}$. Thus,
\begin{align}
\notag
\vert N_0 \rangle & = \int d\phi e^{iN_0\phi} \vert BCS (\phi=0) \rangle_{\mathcal{K}'} = \int d \phi e^{iN_0\phi} \vert BCS(\phi) \rangle_{\mathcal{K}} \\ \notag
& = (\Pi_{\nu>0} U'_{\nu}) \int d \phi e^{iN_0\phi} (1+ ... + \frac{e^{-i N \phi}}{(N/2)!} \left( \sum_{\nu>0} c_{\nu} a^+_{\nu} a^+_{\bar \nu} \right)^{N/2} + ... ) \vert 0 \rangle \\
& \sim \left( \sum_{\nu>0} c_{\nu} a^{\dagger}_{\nu} a^{\dagger}_{\bar \nu} \right)^{N_0/2} \vert 0 \rangle.
\end{align}
It is of notice that  the factor $e^{iN_0\phi}$ above is equivalent to the transformation coefficient $e^{ipq}$ between $p$ and $q$ representations. 
Finally, the members  of a pairing rotational band  are described by the states,
\begin{equation}
\vert N_0 \rangle \sim \left ( \sum_{\nu>0} c_{\nu} a^{\dagger}_{\nu} a^{\dagger}_{\bar \nu} \right)^{N_0/2} \vert 0 \rangle.
\label{pairband}
\end{equation}
Now, from this relation it becomes clear that the Cooper pair wavefunction
\begin{equation}
\vert \tilde{0} \rangle = \sum_{\nu>0} c_{\nu} a^{\dagger}_{\nu} a^{\dagger}_{\bar \nu} \vert 0 \rangle,
\label{eq.B17}
\end{equation}
is to be interpreted to be valid for values of $\varepsilon_{\nu}$ close to $\varepsilon_F$, otherwise one risks not to be able 
to normalize it ($U_{\nu} \to 0$ for $\varepsilon_{\nu} << \varepsilon_F <0$ for deeply bound occupied states). This is in keeping with the fact that pair condensation in general, and nuclear superfluidity in particular, are associated with a modification of the Fermi surface within a narrow band around it $(\varepsilon_F \pm \Delta)$. In other words, pairspin alignment as described by BCS  implies that something unique takes place in the long wavelength limit of the spectrum, namely the appearance of a coherent state with almost (aside from the weak 
($\varepsilon_\nu \approx \varepsilon_F$) dealignment introduced by $H_{sp}$), perfect phase coherence. This state behaves essentially semiclassically, and 
its properties can hardly depend on the ultraviolet behavior of the system. In other words, $E_{cutoff}$ can be set to include only  the valence single-particle shells, adjusting $G$ to reproduce the value of the pairing gap. 
Within this context see also the discussion in Sect. 4 in connection with Fig. \ref{fig.Cooper} and Tables \ref{Table:1} and \ref{Table:2}.

In the case of a single $j-$shell (see e.g. \cite{Brink:05} App. I),
\begin{equation}
V' = \sqrt{\frac{N}{2 \Omega}} \quad \quad , \quad \quad U = \sqrt{1 -\frac{N}{2 \Omega}},
\label{vbcs}
\end{equation} 
thus
\begin{equation}
U'V' = \sqrt{\frac{N}{2\Omega} \left( 1 - \frac{N}{2\Omega} \right) },
\end{equation} 
while
\begin{equation}
\frac{V'}{U'} = \frac{ \sqrt{\frac{N}{2 \Omega}}}{\sqrt{1 - \frac{N}{2 \Omega}}} = \sqrt{\frac{N}{2\Omega -N}}.
\end{equation} 
For a number of particles considerably smaller than the full degeneracy of the single-particle subspace in which nucleons can correlate, that is for $N << 2 \Omega$,
one can write
\begin{equation}
U' V' = \sqrt{\frac{N}{2\Omega} \left (1 - \frac{N}{4 \Omega} \right)} \approx  \sqrt{\frac{N}{2 \Omega}},
\end{equation} 
and 
\begin{equation}
\frac{V'}{U'} = \sqrt{\frac{N}{2 \Omega} \frac{1}{\left( 1 - \frac{N}{2 \Omega} \right ) }} \approx \sqrt{\frac{N}{2\Omega}} \left (1+ \frac{N}{4 \Omega}\right) \approx
\sqrt{\frac{N}{2\Omega}} \approx U' V'.
\end{equation} 
Consequently
\begin{equation}
\vert \tilde 0 \rangle  \approx \sum_{\nu> 0} U_{\nu}V_{\nu} a^{\dagger}_{\nu}a^{\dagger}_{\bar \nu} \vert 0 \rangle,
\label{eq.B23}
\end{equation} 
in keeping with Eqs. (\ref{eq.12b}) and (\ref{eq.20}).

In what follows we work out some relations which are useful to calculate expectation values in the $\vert BCS \rangle$ state.

Making use of (\ref{eq.B7}) (i.e. $a'^{\dagger}_{\mu}=\mathcal{G} a^{\dagger}_{\mu} \mathcal{G}^{-1} = e^{-i \phi} a^{\dagger}_{\mu}$) one can write
\begin{equation}
 a^{\dagger}_{\mu} = e^{i\phi}a'^{\dagger}_{\mu}=e^{i\phi} \left\{
\begin{array}{l} 
a'^{\dagger}_{\nu} \quad (\mu = \nu), \\ 
a'^{\dagger}_{\bar \nu} \quad (\mu = \bar \nu),           
       \end{array} \right.
\end{equation}
thus
\begin{equation}
 a^{}_{\mu} = e^{-i\phi}a'^{}_{\mu}=e^{-i\phi} \left\{
\begin{array}{l} 
a'{}_{\nu} \quad (\mu = \nu), \\ 
a'{}_{\bar \nu} \quad (\mu = \bar \nu).
       \end{array} \right.
\end{equation}
Making use of (\ref{eq.alpha}) and (\ref{eq.UV}) one can write
\begin{equation}
 \alpha^{\dagger}_{\mu} = \left\{
\begin{array}{l}
U'_{\nu} a'^{\dagger}_{\nu} - V'_{\nu}a'_{\bar \nu} \quad (\mu=\nu), \\
U'_{\nu} a'^{\dagger}_{\bar \nu} + V'_{\nu}a'_{\nu} \quad (\mu=\bar \nu), \\
\end{array} \right.
\label{eq.B26}
\end{equation}
in keeping with the fact that $U'_{\mu}$ and $V'_{\mu}$ are real c-numbers, and thus $U'_{\mu}=U'_{\bar \mu}$ and $V'_{\mu}=V'_{\bar \mu}$, and consequently $U_{\bar \nu} = U_{\nu}$ and $V_{\bar \nu}=V_{\nu}$, as well as the fact that $a_{ \bar{\bar{\nu}} }= - a_{\nu}$, a consequence of the anti-unitary character of the time reversal operator. Within this context it is of notice that the intrinsic property of a nucleon  of being in a state with quantum numbers $(j,m)$ or $(j,-m)$ does not of course affect the gauge angle of rotation ($2\phi$, see also Fig. \ref{fig:A1}) defining the intrisinsic (body-fixed) frame of reference with respect to the laboratory system.
Taking the hermitian conjugate of the second case of eq. (\ref{eq.B26}) one obtains
\begin{equation}
 \alpha_{\bar \nu} = U'_{\nu} a'_{\bar \nu} + V'_{\nu} a'^{\dagger}_{\nu}.
\label{eq.B27}
\end{equation}
Multiplying the first entry of Eq. (\ref{eq.B26}) by $U'_{\nu}$ and (\ref{eq.B27}) by $V'_{\nu}$ one obtains,
\begin{align}
 U'_{\nu}\alpha^{\dagger}_{\nu} = U'^{2}_{\nu} a'^{\dagger}_{\nu} - V'_{\nu}U'_{\nu} a'_{\bar \nu}, \notag \\
 V'_{\nu}\alpha_{\bar \nu}      = V'_{\nu}U'_{\nu} a'_{\bar \nu}  + V'^{2}_{\nu} a'^{\dagger}_{\nu}. \notag
\end{align}
Summing these two expressions and making use of (\ref{eq.Norm}) one obtains
\begin{subequations}
 \begin{equation}
  a'^{\dagger}_{\nu} = U'_{\nu} \alpha^{\dagger}_{\nu}+ V'_{\nu}\alpha_{\bar \nu},
\label{eq.B28a}
 \end{equation}
which is equivalent to
\begin{equation}
  a^{\dagger}_{\nu} = U_{\nu} \alpha^{\dagger}_{\nu}+ V^{*}_{\nu}\alpha_{\bar \nu}.
 \end{equation}
\end{subequations}
Similarly, multiplying the hermitian conjugate of the first entry of Eq. (\ref{eq.B26}) by $V'_{\nu}$ and the second entry by $U'_{\nu}$  and subtracting the resulting expressions leads to
\begin{subequations}
 \begin{equation}
 a'^{\dagger}_{\bar \nu} = U'_{\nu} \alpha^{\dagger}_{\bar \nu} - V'_{\nu}\alpha_{ \nu},
\label{eq.B29a}
 \end{equation}
which is equivalent to
\begin{equation}
  a^{\dagger}_{\bar \nu} = U_{\nu} \alpha^{\dagger}_{\bar \nu} -  V^{*}_{\nu}\alpha_{\nu} \;.
 \end{equation}
\end{subequations}
One then obtains
\begin{subequations}
\begin{equation}
 P^{\dagger} = \sum_{\nu > 0} a^{\dagger}_{\nu} a^{\dagger}_{\bar \nu} 
= \sum_{\nu > 0} \left\{ U^{2}_{\nu} \alpha^{\dagger}_{\nu} \alpha^{\dagger}_{\bar \nu} - (V^{*}_{\nu})^{2} \alpha_{\bar \nu}\alpha_{\nu} - U_{\nu}V^{*}_{\nu}(\alpha^{\dagger}_{\nu}\alpha_{\nu} + \alpha^{\dagger}_{\bar \nu} \alpha_{\bar \nu}) + U_{\nu}V^{*}_{\nu}  \right\},
\end{equation}
\begin{equation}
 P = \sum_{\nu > 0} a_{\bar \nu} a_{\nu} 
= \sum_{\nu > 0} \left\{ (U^{*}_{\nu})^{2} \alpha_{\bar \nu} \alpha_{\nu} - V^{2}_{\nu} \alpha^{\dagger}_{\nu}\alpha^{\dagger}_{\bar \nu} - U^{*}_{\nu}V_{\nu}(\alpha^{\dagger}_{\nu}\alpha_{\nu} + \alpha^{\dagger}_{\bar \nu} \alpha_{\bar \nu}) + U^{*}_{\nu}V_{\nu}  \right\},
\end{equation}
\end{subequations}
thus
\begin{align}
 \alpha_0 & = \langle BCS \vert P \vert BCS \rangle = \sum_{\nu > 0} U^{*}_{\nu}V_{\nu} = e^{-2i\phi} \sum_{\nu > 0} U'_{\nu}V'_{\nu} = e^{-2i\phi} \alpha'_0 \notag \\
 & = \left(  \sum_{\nu>0}U_{\nu}V^{*}_{\nu} \right)^{*} = \langle BCS \vert P^{\dagger} \vert BCS \rangle^{*}.
\end{align}
Summing up
\begin{equation}
 \alpha_0 = \sum_{\nu > 0}U^{*}_{\nu}V_{\nu},
\end{equation}
and
\begin{equation}
 \alpha'_{0} = \sum_{\nu > 0}U'_{\nu} V'_{\nu},
\end{equation}
leading to
\begin{equation}
 \Delta = G \alpha_0 = e^{-2i\phi} G \alpha'_{0} = e^{-2i\phi} \Delta'.
\end{equation}
An alternative derivation of the above relations can be obtained by inserting the expressions of $a'^{\dagger}_{\nu}$ and $a'^{\dagger}_{\bar \nu}$ obtained from (\ref{eq.B28a}) and (\ref{eq.B29a}) into
\begin{equation}
  P = \sum_{\nu > 0}a_{\bar \nu} a_{\nu} =  e^{-2i\phi} \sum_{\nu > 0} a'_{\bar \nu}a'_{\nu} = e^{-2i\phi} P',
\end{equation}
which leads to 
\begin{align}
 P & = e^{-2i\phi} \sum_{\nu > 0} (U'_{\nu} \alpha_{\bar \nu} - V'_{\nu} \alpha^{\dagger}_{\nu})(U'_{\nu} \alpha_{\nu} + V'_{\nu} \alpha^{\dagger}_{\bar \nu}) \notag \\
   & = e^{-2i\phi} \sum_{\nu > 0} \left\{ U'^{2}_{\nu} \alpha_{\bar \nu} \alpha_{\nu} - V'^{2}_{\nu} \alpha^{\dagger}_{\nu}\alpha^{\dagger}_{\bar \nu} - U'_{\nu}V'_{\nu}(\alpha^{\dagger}_{\nu}\alpha_{\nu} + \alpha^{\dagger}_{\bar \nu} \alpha_{\bar \nu}) + U'_{\nu}V'_{\nu}  \right\}.
\end{align}

\section{Commutation relations}
\setcounter{figure}{0}
\renewcommand{\thefigure}{C.\arabic{figure}}
\label{Appendix:C}

Making use of the relations 
\begin{equation}
[AB,C] = A[B,C] + [A,C]B,
\end{equation}
\begin{equation}
[C,AB]= A[C,B] + [C,A]B,
\end{equation}
\begin{equation}
[AB,C] = A \{B,C\} - \{A,C\}B,
\end{equation}
and of the definitions
\begin{equation}
P^+_{\nu} = a^+_{\nu} a^+_{\bar \nu} \quad \quad ; \quad \quad  P_{\nu} = a_{\bar \nu} a_{\nu},
\end{equation}
one can calculate the commutator
\begin{equation}
[P_{\nu},P^+_{\nu}] = [a_{\bar \nu}a_{\nu}, a^+_{\nu}a^+_{\bar \nu}] = a^+_{\nu} [a_{\bar \nu} a_{\nu}, a^+_{\bar \nu}] +
[a_{\bar \nu} a_{\nu}, a^+_{\nu}] a^+_{\bar \nu} 
\end{equation}
\begin{equation}
= a^+_{\nu} \left ( - \{ a_{\bar \nu}, a^+_{\bar \nu}  \} a_{\nu} \right ) +
\left( a_{\bar \nu} \{ a_{\nu},a^+_{\nu} \} \right) a^+_{\bar \nu}
\end{equation}
\begin{equation}
= -a^+_{\nu}a_{\nu}  + a_{\bar \nu} a^+_{\bar \nu} = 1 - (a^+_{\nu}a_{\nu} + a^+_{\bar \nu} a_{\bar \nu}) = 1 - N_{\nu},
\end{equation}
where
\begin{equation}
N_{\nu} = a^+_{\nu}a_{\nu} + a^+_{\bar \nu} a_{\bar \nu}.
\end{equation}
Similarly,
\begin{eqnarray}
[a^+_{\nu}a_{\nu},P^+_{\nu}]= [a^+_{\nu} a_{\nu}, a^+_{\nu} a^+_{\bar \nu} ]  = &  a^+_{\nu} [a^+_{\nu}a_{\nu},a^+_{\bar \nu}] + [a^+_{\nu}a_{\nu},a^+_{\nu}]a^+_{\bar \nu} = \\
a^+_{\nu} \left\{a_{\nu},a^+_{\nu} \right\} a^+_{\bar \nu} = a^+_{\nu}a^+_{\bar \nu} &
\end{eqnarray}
and 
\begin{eqnarray}
[a^+_{\bar \nu}a_{\bar \nu}, P^+] = [a^+_{\bar \nu} a_{\bar \nu} , a^+_{\nu} a^+_{\bar \nu} ]= & a^+_{\nu} [a^+_{\bar \nu} a_{\bar \nu}, a^+_{\bar \nu}] 
 + [a^+_{\nu} a_{\bar \nu}, a^+_{\nu} ] a^+_{\bar \nu} \\
= a^+_{\nu} (a^+_{\bar \nu} \{ a_{\bar \nu}, a^+_{\bar \nu} \} ) = a^+_{\nu} a^+_{\bar \nu} &  .
\end{eqnarray}
Thus
\begin{equation}
[ N_{\nu}, P^+_{\nu}] = 2 P^+_{\nu}.
\end{equation}
Consequently,
\begin{equation}
[N_{\nu},P_{\nu}] = - 2P_{\nu}.
\end{equation}
Summing up,
\begin{equation}
[ P^+_{\nu},P_{\nu}] = N_{\nu} - 1
\end{equation}
\begin{equation}
[ N_{\nu} -1, P^+_{\nu}] =  2 P^+_{\nu}
\end{equation}
and
\begin{equation}
[N_{\nu}-1, P_{\nu}] = - 2 P_{\nu}.
\end{equation}
Making use of these relations and of the definitions
\begin{equation}
s_x(\nu) = \frac{1}{2} (P^+_{\nu} + P_{\nu}) \quad \quad , \quad \quad s_y(\nu) = \frac{1}{2i} (P^+_{\nu} - P_{\nu} ),
\end{equation}
and
\begin{equation}
s_z(\nu) = \frac{1}{2} (N_{\nu} - 1),
\end{equation}
one obtains
\begin{align}
[s_x(\nu),s_y(\nu)] & =  [\frac{1}{2} (P^+_{\nu} + P_{\nu}) , \frac{1}{2i} (P^+_{\nu} - P_{\nu})] =
\frac{1}{4i} ( - [P^+_{\nu},P_{\nu}] + [P_{\nu},P^+_{\nu}]) \notag \\
& = \frac{1}{2i} [P_{\nu},P^+_{\nu}] = \frac{1}{2i} (1 - N_{\nu}) = i s_z(\nu),
\end{align}
\begin{align}
[s_y(\nu),s_z(\nu)] & = \frac{1}{4i} ([P^+_{\nu}, (N_{\nu}-1)] - [P_{\nu},(N_{\nu}-1)] = - \frac{1}{4i} (2P^+_{\nu} + 2P_{\nu}) \notag \\
& =  - \frac{1}{2i} (P^+_{\nu} +P_{\nu}) = i s_x(\nu),
\end{align}

\begin{align}
[s_z(\nu),s_x(\nu)] & =  \frac{1}{4} [(N_{\nu}-1), P^+_{\nu}+P_{\nu}] = \frac{1}{4} ( (N_{\nu}-1), P^+_{\nu}] +
[(N_{\nu}-1),P_{\nu}] ) \notag \\
& = - \frac{1}{2} (P_{\nu} - P^+_{\nu}) = + i s_y(\nu).
\end{align}

\section{Generalized Rigidity in Gauge Space}
\setcounter{figure}{0}
\renewcommand{\thefigure}{D.\arabic{figure}}
\label{Appendix:D}

Generalized rigidity in gauge space implies that if one pushes, with the help of a field that changes the number of particles in two, one of the  poles 
of a deformed system in gauge space (see e.g. Fig. \ref{fig:B1}), system which can be viewed as a  wavepacket in particle number,
the other pole reacts rigidly to the push, and the system starts rotating as a whole (pairing rotational band). Similarly, if it was already in rotation it changes its rotational frequency from $\omega = \lambda(N_0)/\hbar$ to $\omega' = \lambda (N_0 \pm 2)/\hbar$, $\lambda$ being the Lagrange multiplier which, in BCS theory, is closely connected with the particle number equation.

If the gauge space image is not sufficiently concrete to create a physical picture of the process, let us think of a quadrupole deformed nucleus whose intrinsic state is described in terms of the Nilsson intrinsic state. Making use of a proton beam which acts upon one of the poles, the system reacts as a whole and starts rotating with a frequency associated with one of the allowed values of the angular momentum, the transfer quantum corresponding to an energy inversely proportional to the moment of inertia.

One may argue that the reaction of the pole not acted upon by the external field is not instantaneous but takes place only after an interval of time,  compatible with the propagation of information  in the nuclear medium, has elapsed. This parlance is not even wrong, as a wavefunction, in particular that describing the intrinsic ground state of a superfluid nucleus (see Eq. (\ref{eq.BCS})),
is not a matter function but a probability amplitude function \footnote{It is of notice that similar arguments are at the basis of the discussion of Bohr with Schr{\"{o}}dinger and De Broglie (matter waves), let alone with Einstein (instantaneous) "information" propagation in connection with a single photon hitting  a screen after having gone through  a
single slit of a box full of photons 
(see e.g. D. Lindley , {\it Uncertainty, Einstein, Heisenberg, Bohr and the struggle for the soul of science}, Doubleday Publishing Group, New York (2007)). }
with perfect phase coherence throughout. Within the quadrupole deformed nucleus analogy, generalized rigidity implies that the nucleus reacts rigidly as a whole to the 
action of the proton field acting on a pole\footnote{To avoid arguments such that the wavelength of the external hadronic field (beam) allows it to act on both poles, one  can work in terms of a Gedanken experiment,  
in which the proton bombarding energy is about 1.6 GeV ($\lambda \approx 1$ fm$\ll R$).}, even if the moment of inertia of the associated rotational band is that of superfluid nuclear matter, and thus considerably smaller than the rigid moment  of inertia. 

The specific experiment to study the consequences  (emergent properties) resulting from a spontaneous breaking of symmetry (e.g. of rotational invariance) is a  probe which itself violates  the symmetry in question.
Now, while most of the devices we find in a well equipped nuclear laboratory violate rotational invariance -- think for example of a proton beam line defining a privileged orientation in 3D-space --  one does not find many which violate gauge invariance. In other words, while rulers and goniometers defy empty space isotropy and homogeneity, one does not usually walk around with instruments which do not have a fixed number of particles.

\begin{figure}[hbt!]
	\begin{center}
		\includegraphics[width=0.82\textwidth]{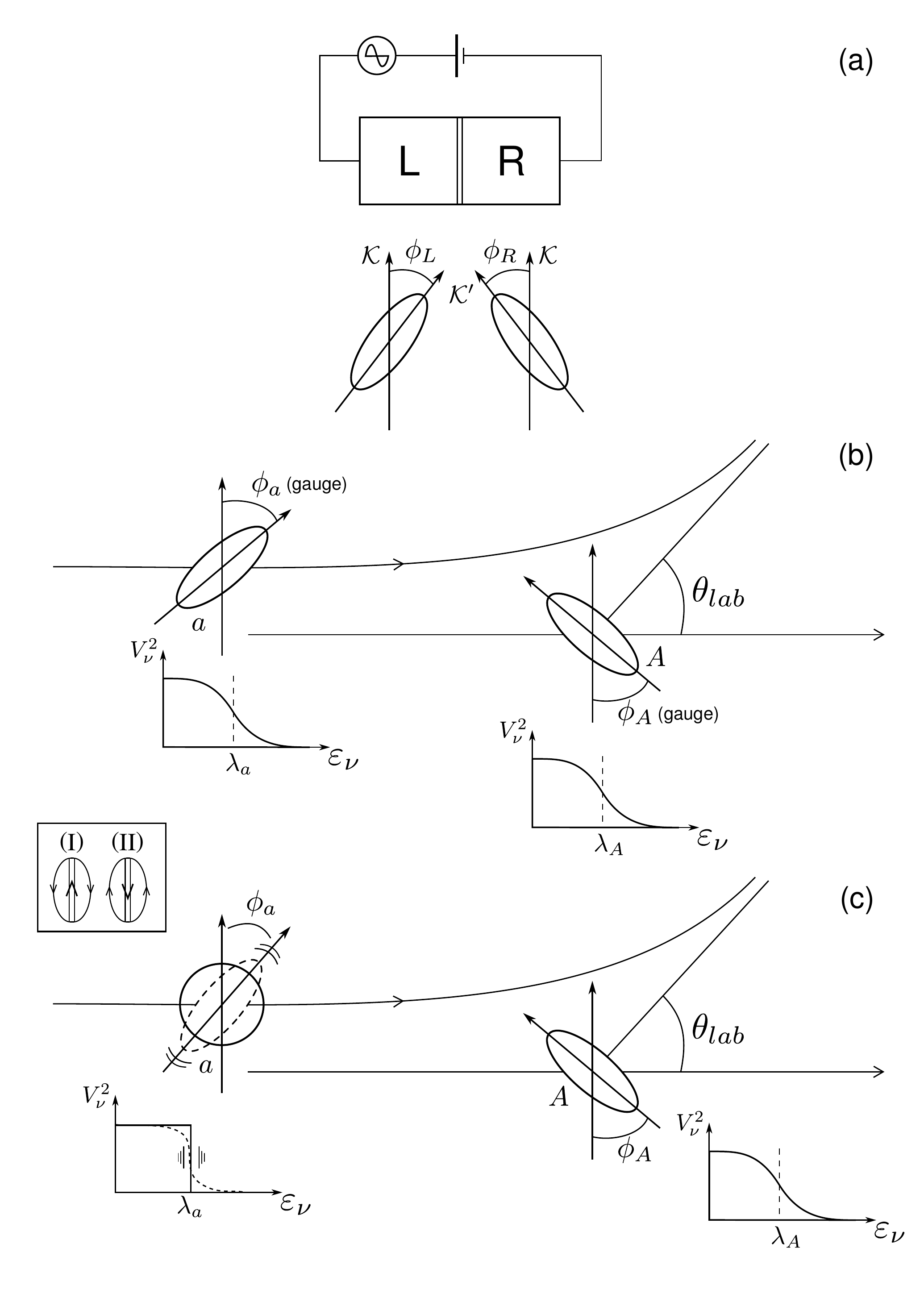}
	\end{center}
\caption{ Josephson junction between left ($L$) and right ($R$) superconductors weakly coupled through a thin (DC biased) oxide layer;
(b) time dependent nuclear Josephson junction established at about the distance of closest approach by the superfluid nuclei $a$ and $A$ in the reaction 
$a + A \to (a \pm n) + A (-+ n)$, where $n = 0,1,2 ...$ is the number of Cooper pairs transferred in the process;
(c) dynamical time dependent Josephson junction between a closed shell system $a$ displaying strong pairing vibrations (pair addition and pair removal mode) 
and a superfluid  nucleus $A$. In the inset (left) the ground  zero point fluctuations of the system $a$ 
associated with the pair addition  (arrowed double line pointing up) and pair removal mode (arrowed double line pointing down) are shown.}
\label{Fig.D1}
\end{figure}

This was the real importance of the Josephson effect (see Fig. \ref{Fig.D1}), which provided a simple, and quantitative accurate answer to the question: how does one measures the gauge phase of a superconductor? The answer is, with the help of another superconductor displaying also an unknown but nonetheless well defined gauge phase. Establishing a weak coupling (oxide layer) so that electrons can tunnel, one at a time, between the junction. If the first system can be viewed  as a rotor in gauge space, the second  one can equally well be represented  in this way.  Biasing the junction with a constant potential difference, will lead to a two-rotor-coupled system (through pair transfer across the junction), rotating with frequencies which differ by $e \Delta V /\hbar = (\lambda_1 - \lambda_2) /\hbar$. Such a system will display a resonant behavior (alternating current  with frequency $ e \Delta V t/\hbar$) provided the rotor, deformed  system picture in gauge space, is applicable. The fact that this effect provides  the most accurate measure of $(e /\hbar)$ available, testifies to the validity of the deformed rotor picture in gauge space associated with the $BCS$ wavefunction (\ref{eq.BCS}).

It is sobering  that this is so, in keeping with the fact that it was Bardeen the most strenuous  opponent to Josephson's ideas, even more so if one is reminded that the choice argument of such opposition was based on the fact  that the pairing gap $\Delta = G \alpha_0$ vanishes at the junction. Now, the order parameter of BCS theory (as well as the justification of Cooper pair model) is $\alpha_0 = \langle P^+ \rangle = \langle P \rangle^*$, namely the  condensed pair field. Electrons may tunnel  one at a time,  without obliterating  the validity of pair transfer, as the  coherence length ($\xi = \hbar v_F /2 \Delta)$ is much larger than  typical junction  dimensions,  diverging at the junction (kind of extreme $^{11}$Li-halo-like phenomenon (see e.g. \cite{Potel:10}, \cite{Potel:12}) within the condensed matter framework). 

Within this scenario one can posit that, in the nuclear case, the equivalent of the Josephson junction device or better , the setup of an experiment which can measure differences in gauge phases, allows for embodiments which are not possible within the field of condensed matter physics. This is related to the fact that fluctuations in finite many-body systems, in general and pairing vibrations in normal nuclei in particular,  are not only quantitatively but also qualitatively stronger than in condensed matter. 

Consequently, not only a collision between two superfluid nuclei (see Fig. \ref{Fig.D1}(b)) can be viewed as a time dependent generalization of a Josephson junction (see. Fig. \ref{Fig.D1}(a)).
Also a normal-superfluid nuclear reaction can provide similar information, as a closed shell system displays a very collective pairing vibrational spectrum which can be viewed  as  large amplitude dynamical gauge symmetry violating mode (see Fig. \ref{Fig.D1}(c)).

Let us consider the ground state of a (light) closed shell system. Within the present discussion, it can be written as 
\begin{equation}
\widetilde{\vert a(gs) \rangle} = \alpha \vert a(gs) \rangle + \beta \vert (a-2)(gs) \otimes (a+2) (gs) \rangle
\label{eq.D1}
\end{equation}
where $\vert (a-2)(gs) \rangle$ and $\vert (a+2)(gs) \rangle$, are the pair  removal and pair addition modes, while $\alpha^2 + \beta^2 = 1$ is the normalization condition. In other words, the closed-shell system is part of the time in states with two more (see (I) inset Fig. \ref{Fig.D1}), or two less correlated nucleons
(see (II) inset Fig. \ref{Fig.D1}). The two-hole uncorrelated states which, arguably, ensure particle number correlation, in fact describes 
the ground state correlations  (backwardsgoing amplitudes, see e.g. Table \ref{Table:Sn132_PV} Y-amplitudes as well as insets in Fig. \ref{fig:dispers}) associated with the pair addition mode which, dynamically, deforms
the nucleus  in gauge space defining a transient, privileged orientation. The same can be said concerning the two-particle uncorrelated system shown in the inset (II) of Fig. \ref{Fig.D1}(c).

Let us now return to  the analogy with quadrupole rotational and vibrational bands. Going away from closed shell nuclei in medium heavy systems the energy of the first $2^+$ state lowers in energy, the corresponding period  becoming longer, the associated amplitudes larger (see e.g. \cite{Brink:05} Ch.7). Eventually, after the quantal phase transition has taken place, the rotation of the system as a whole  can be viewed as a very low frequency quadrupole vibrational mode (for which the restoring force vanishes while inertia remains finite), which dynamically defines a privileged quadrupole deformation symmetry axis (and thus an associated set of Euler angles), orientation which after each period changes orientation, with a frequency  inversely proportional to the inertia of the mode. 

In a similar way the state (\ref{eq.D1}) defines dynamically, a privileged orientation in gauge space which specifically can probe the corresponding static  
quantity of  a superfluid target nucleus $A$ (see Fig. \ref{Fig.D1}(c)), that is,
\begin{equation}
 a+A \rightarrow \left\{ a((a+2) \otimes (a-2)(gs))+   A \left( \sum_N c_N \vert N \right)  \rangle \right\} \rightarrow \left\{ 
\begin{array}{l}
  (a-2)(gs) + (A+2)(gs),\\
  (a+2)(gs) + (A-2)(gs),\\ \end{array} \right.
\end{equation}
where $A \left( \sum_N c_N \vert N \right)$ labels the ground state of  a superfluid nucleus, e.g. of $^{120}$Sn which can be viewed 
as a wavepacket in neutron number, the scattering state within curly brackets being a virtual  set of states each displaying a dynamical or a static privledged orientation 
in gauge space and thus a gauge phase.
A particular embodiment of such a reaction can be 
\begin{equation}
{}^{9}_{3}\textrm{Li}_{6}+{}^{120}_{50}\textrm{Sn}_{70}\rightarrow \left\{ 
\begin{array}{l}
  {}^{7}_{3}\textrm{Li}_{4}+{}^{122}_{50}\textrm{Sn}_{72},\\
 {}^{11}_{3}\textrm{Li}_{8}+{}^{118}_{50}\textrm{Sn}_{68},\\ \end{array} \right.
\end{equation}
in keeping with the fact that $N=6$ corresponds to a (parity inversion) magic number, $\vert^9{\rm Li} (gs)\rangle$ and  $\vert^{11}{\rm Li} (gs)\rangle$ being the pair removal and pair addition modes of $^9$Li (see \cite{Broglia:12ArXiv,Potel:11}). 

\section{Kramers-Kronig dispersion relation}
\setcounter{figure}{0}
\renewcommand{\thefigure}{E.\arabic{figure}}
\label{Appendix:E}

The nuclear superfluid phase, and associated pairing rotational band behaves, because of its ODLRO essentially as a classical (coherent) state. This is the basic reason which is at the basis of the results displayed in Fig. \ref{fig4} and Table \ref{Table_CS}, that is the remarkable quantitative accuracy with which theory provide an overall account of the absolute value of the experimental findings.

There is, however, a second reason for the higher accuracy  with which one  can, in principle, predict absolute two-nucleon transfer cross sections, as compared with one-nucleon transfer cross sections, or, simply, (although, arguably, less well defined see e.g. \cite{Mahaux:85,Dickhoff:05,Jennings:11}) absolute values  of the single-particle spectroscopic factors. 
This is connected with the ambiguity and eventual lack of consistency of empirically determined  optical parameters. Typical of the first  type of limitations, is connected with the fact that the depth of the real part  of the optical potential can have  different, commensurable values, all leading to  the same phase shift (e.g. socalled 
Igo's ambiguity \cite{Igo:59}). Concerning  the second point (consistency) , one is reminded of the fact that  the real and the  imaginary part of  the optical potentials $(U+iW)$ controlling off- and on -the energy shell processes, are the  real and imaginary part of the nuclear mass operator (sum of polarization  and correlation contributions), referred to, also,  as the  nuclear dielectric function \cite{Mahaux:85}. Consequently $U$ and $W$ must fulfill the Kramers-Kronig dispersion relation (cf. refs. \cite{Kronig:26,Kramers:27}, see also \cite{Mahaux:85}).   
This is a bidirectional mathematical relation, connecting the real and imaginary parts 
of any complex function that is analytic in the upper half of the  complex plane. The Kramers-Kronig  is a rather fundamental relation, in that it is strictly related to causality. Summing up, the real and imaginary parts of the optical potentials empirically determined from a global  elastic scattering fitting, should  respect the above  mentioned 
dispersion relation. This condition  is only marginally fulfilled in a number of cases, thus introducing uncertainties difficult to control. In keeping with  the fact that, theoretically, $U(r)$ results from  the convolution of the nucleon-nucleon interaction with  the nuclear density $\rho(r)$, and that this quantity  is best  known around the nuclear surface,  in a similar  way in which $W(r)$ receives important contributions from  the interweaving  of single-particle motion  and collective surface vibrations,  one can posit that  
the lack of consistency between $U$ and $W$ are likely to be more serious for 
values of $r$ larger and smaller than the nuclear radius  (i.e. $ r > R_0$ and $r < R_0$). The phase coherence of the nuclear condensate implies  that Cooper pair wavefunctions, 
closely related to the effective  two-particle transfer nuclear formfactors,  are rather compact objects, essentially concentrated on the nuclear surface (see e.g. 
Fig. \ref{fig.Cooper}). Consequently,  the main contribution to the absolute two-particle differential cross section  arises from values of $r \approx R_0$ (nuclear surface), region in which  the optical potential is best known. On the other hand,  the formfactor associated with single-particle transfer is a standard mean field wavefunction (in any case as far as the main peak  of the single-particle  strength function -- one-pole approximation --  is concerned). Consequently, the absolute one-nucleon transfer differential cross section can depend, in  an important way, on the knowledge of the optical potential inside the nuclear volume. 

The above discussion can also be related to the fluctuation-dissipation theorem (cf. eg. refs. \cite{Nyquist:28,Callen:51}), closely related to the Kramers-Kronig dispersion relation.  
While it is true that single-particle motion emerges from the same features leading to collective nuclear vibrations \cite{Mottelson:62} -  
the independent particle shell structure being a result of a 
collective and concerted motion of all the nucleons allowing free motion to each single of them, and letting themselves felt through the pushing and pulling only when a 
particle tries to leave the system, forcing it to  bounce elastically  off the nuclear surface -  the picture becomes less stable as soon as the particle-vibration coupling strength is  switched on, leading to real, dressed particles (mechanism also contributing to the imaginary part of the optical potential). Within this context one can posit that the associated 
single-particle  strength function in general and thus its centroid and width in particular, depend on a delicate interplay of spin- and non spin- flip matrix elements, 
large and small energy denominators, and the like. On the other hand the strength function associated with a Cooper pair, being  
a coherent object behaving almost semiclassically, resents much less of the above mentioned inhomogeneous damping phenomena. 

Summing up, even without totally consistent optical potentials in the Kramers-Kronig sense, it is likely  that one can calculate the absolute value 
of the two-particle transfer cross section between members of pairing rotational
(vibrational) bands with high accuracy. On the other hand, in the case of one-nucleon transfer processes, this possibility is likely to be restricted to the centroid and width of the (main peak) single-particle strength function, 
the integrated area being affected, as a rule, by little controllable, non-specific background effects, difficult to estimate and/or remove. 
Such limitations will likely be more important in the case of strongly fragmented single-particle states.

\clearpage
\section*{References}
\fancyhead[LE,RO]{\bfseries\thepage}
\fancyhead[LO]{\bfseries References}
\fancyhead[RE]{\bfseries References}

\bibliographystyle{ieeetr}
\bibliography{./nuclear}{}

\begin{thebibliography}{10}

\bibitem{Mayer:55}
M.~Mayer and J.~Jensen, {\em Elementary Theory of Nuclear Shell Structure}.
\newblock Wiley, New York, NY, 1955.

\bibitem{Cottle:10}
P.~Cottle, ``{Doubly Magic Tin},'' {\em Nature}, vol.~465, p.~430, 2010.

\bibitem{Jones:10}
K.~L. Jones, A.~S. Adekola, D.~W. Bardayan, J.~C. Blackmon, K.~Y. Chae, K.~A.
  Chipps, J.~A. Cizewski, L.~Erikson, C.~Harlin, R.~Hatarik, R.~Kapler, R.~L.
  Kozub, J.~F. Liang, R.~Livesay, Z.~Ma, B.~H. Moazen, C.~D. Nesaraja, F.~M.
  Nunes, S.~D. Pain, N.~P. Patterson, D.~Shapira, J.~F. Shriner, M.~S. Smith,
  T.~P. Swan, and J.~S. Thomas, ``{The magic structure of $^{132}$Sn explored
  through the single--particle states of $^{133}$Sn},'' {\em Nature}, vol.~465,
  p.~454, 2010.

\bibitem{Haufler:91}
R.~Haufler, L.-S. Wang, L.~Chibante, C.~Jin, J.~Conceicao, Y.~Chai, and
  R.~Smalley, ``Fullerene triplet state production and decay: {R2PI} probes of
  {$C_{60}$} and {$C_{70}$} in a supersonic beam,'' {\em Chemical Physics
  Letters}, vol.~179, no.~5–6, p.~449, 1991.

\bibitem{Mahaux:85}
C.~Mahaux, P.~F. Bortignon, R.~A. Broglia, and C.~H. Dasso, ``Dynamics of the
  shell model,'' {\em Phys. Rep.}, vol.~120, p.~1, 1985.

\bibitem{Bogoliubov:58}
N.~Bogoljubov, ``On a new method in the theory of superconductivity,'' {\em Il
  Nuovo Cimento (1955-1965)}, vol.~7, 1958.

\bibitem{Valatin:58}
J.~Valatin, ``Comments on the theory of superconductivity,'' {\em Il Nuovo
  Cimento (1955-1965)}, vol.~7, 1958.

\bibitem{Anderson:58}
P.~W. Anderson, ``Random--{P}hase {A}pproximation in the theory of
  superconductivity,'' {\em Phys. Rev.}, vol.~112, p.~1900, 1958.

\bibitem{Bohr:89}
A.~Bohr and O.~Ulfbeck, ``Quantal structure of superconductivity. {G}auge
  angle,'' {\em 1st Tops\o{}e Summer School, AEK Ris\o{}, Denmark
  (unpublished)}, 1988.
\newblock \url{http://www.risoe.dtu.dk/rispubl/reports_INIS/RISOM2756.pdf}.

\bibitem{Brink:05}
D.~Brink and R.~A. Broglia, {\em Nuclear Superfluidity}.
\newblock Cambridge: Cambridge University Press, 2005.

\bibitem{Bes:66}
D.~R. B{\`{e}}s and R.~A. Broglia, ``Pairing vibrations,'' {\em Nucl. Phys.},
  vol.~80, p.~289, 1966.

\bibitem{Broglia:73}
R.~A. Broglia, O.~Hansen, and C.~Riedel, ``Two--neutron transfer reactions and
  the pairing model,'' {\em Adv. Nucl. Phys.}, vol.~6, p.~287, 1973.
\newblock http://merlino.mi.infn.it/repository/BrogliaHansenRiedel.pdf.

\bibitem{Nikam:87a}
R.~S. Nikam and P.~Ring, ``Manifestation of the {B}erry phase in diabolic pair
  transfer in rotating nuclei,'' {\em Phys. Rev. Lett.}, vol.~58, p.~980, 1987.

\bibitem{Nikam:87b}
R.~Nikam, P.~Ring, and L.~Canto, ``A nuclear squid: Diabolic pair transfer in
  rotating nuclei,'' {\em Phys. Lett. B}, vol.~185, no.~3, p.~269, 1987.

\bibitem{Shimizu:89}
Y.~R. Shimizu, J.~D. Garrett, R.~A. Broglia, M.~Gallardo, and E.~Vigezzi,
  ``Pairing fluctuations in rapidly rotating nuclei,'' {\em Rev. Mod. Phys.},
  vol.~61, p.~131, 1989.

\bibitem{Note1}
Within this context one can also mention a different (although not directly
  pertinent to the Sn-isotopes studied in the present paper) analogy between
  nuclear and metallic normal state properties which has important consequences
  on Cooper pair stability. Bad conductors, that is bad single-particle
  electronic metals (like e.g. Pb, Sn and Hg), display stable Cooper pair
  condensation at low temperatures, becoming superconductors, while good
  conductors , independent electron motion metals, (e.g. Au, Ar, Cu) do not. In
  the nuclear case, arguably, one of the best nuclear embodiments of Cooper's
  model is the {$ |gs({}^{11} $Li$ )> $} state, pair addition mode of $^9$Li.
  The associated.

\bibitem{Broglia:71}
R.~A. Broglia, C.~Riedel, and T.~Udagawa, ``Coherence properties of two-neutron
  transfer reactions and their relation to inelastic scattering,'' {\em Nucl.
  Phys. A}, vol.~169, p.~225, 1971.

\bibitem{Broglia:72b}
R.~A. Broglia, C.~Riedel, and T.~Udagawa, ``Sum rules and two-particle units in
  the analysis of two-neutron transfer reactions,'' {\em Nucl. Phys. A},
  vol.~184, p.~23, 1972.

\bibitem{Note2}
Within this context one can mention the fact that if instead of pairspins with
  two projections, one studies the properties of finite systems which depend on
  the alignment of a pairspin with twenty components , like protein evolution
  and structure in which each projection corresponds to one of the twenty
  natually occurring aminoacids, the fluctuations of pair spin and thus of
  emergent properties, become even richer and subtler than in the case shown in
  Fig. 5 (see e.g. ref. \cite {Broglia:12ArXiv} and refs. therein). From a
  technical point of view, but not only, the situation may be analogous to that
  resulting moving from tinkering with the Ising model, to confront oneself
  with Potts model.

\bibitem{Broglia:05c}
R.~A. Broglia and A.~Winther, {\em Heavy Ion Reactions, 2nd ed.}
\newblock Boulder: Westview Press, Perseus Books, 2005.

\bibitem{Josephson:62}
B.~D. Josephson, ``Possible new effects in superconductive tunnelling,'' {\em
  Phys. Lett.}, vol.~1, p.~251, 1962.

\bibitem{Bardeen:62}
J.~Bardeen, ``Tunneling into superconductors,'' {\em Phys. Rev.}, vol.~9,
  p.~147, 1962.

\bibitem{Cohen:62}
M.~H. Cohen, L.~M. Falicov, and J.~C. Phillips, ``Superconductive tunneling,''
  {\em Phys. Rev. Lett.}, vol.~8, p.~316, 1962.

\bibitem{Udagawa:73}
T.~Udagawa and D.~Olsen, ``Note on the “forbidden” $(p, t)$ excitation of
  unnatural parity final states from 0+ targets via multistep processes,'' {\em
  Physics Letters B}, vol.~46, p.~285, 1973.

\bibitem{Chien:75}
W.~S. Chien, C.~H. King, J.~A. Nolen, and M.~A.~M. Shahabuddin, ``Unnatural
  parity transitions in $^{22}\mathrm{Ne}(p,~t)^{20}\mathrm{Ne}$,'' {\em Phys.
  Rev. C}, vol.~12, p.~332, 1975.

\bibitem{Segawa:75}
H.~Segawa, K.~I. Kubo, and A.~Arima, ``Two-step analysis of the
  $^{18}\mathrm{O}(p,~t)^{16}\mathrm{O}$ ${2}^{-}$ excitation and phase
  relations in the nonorthogonal term,'' {\em Phys. Rev. Lett.}, vol.~35,
  p.~357, Aug 1975.

\bibitem{Schneider:76}
M.~Schneider, J.~Burch, and P.~Kunz, ``Competition of two-step processes in the
  reactions {$^{60,62}$Ni$(p, t)$} leading to unnatural parity states,'' {\em
  Physics Letters B}, vol.~63, p.~129, 1976.

\bibitem{Takacsy:73}
N.~B. de~Takacsy, ``Two-step mechanism in the reaction
  $^{208}\mathrm{Pb}(p,~t)$,'' {\em Phys. Rev. Lett.}, vol.~31, p.~1007, 1973.

\bibitem{Takacsy:74}
N.~D. Takacsy, ``On the contribution from a two-step mechanism, involving the
  sequential transfer of two neutrons, to the calculation of $(p, t)$ reaction
  cross sections,'' {\em Nuclear Physics A}, vol.~231, p.~243, 1974.

\bibitem{Hashimoto:78}
N.~Hashimoto and M.~Kawai, ``The {$(p,d)$} {$(d,t)$} process in strong
  {$(p,t)$} reactions,'' {\em Prog. Theor. Phys.}, vol.~59, p.~1245, 1978.

\bibitem{Kubo:78}
K.~Kubo and H.~Amakawa, ``Energy dependence of two-step ($p,t$) cross
  sections,'' {\em Phys. Rev. C}, vol.~17, p.~1271, 1978.

\bibitem{Bayman:82}
B.~F. Bayman and J.~Chen, ``One-step and two-step contributions to two-nucleon
  transfer reactions,'' {\em Phys. Rev. C}, vol.~26, p.~1509, 1982.

\bibitem{Yagi:79}
K.~Yagi, S.~Kunori, Y.~Aoki, K.~Nagano, Y.~Toba, and K.~I. Kubo, ``Anomalous
  analyzing powers for strong (${p}_{\mathrm{pol}}$,$t$) ground-state
  transitions and interference between direct and ($p$,$d$) ($d$,$t$)
  sequential process,'' {\em Phys. Rev. Lett.}, vol.~43, p.~1087, 1979.

\bibitem{Igarashi:91}
M.~Igarashi, K.~ichi Kubo, and K.~Yagi, ``Two-nucleon transfer reaction
  mechanisms,'' {\em Physics Reports}, vol.~199, no.~1, 1991.

\bibitem{Becha:97}
M.~B. Becha, C.~O. Blyth, C.~N. Pinder, N.~M. Clarke, R.~P. Ward, P.~R. Hayes,
  K.~I. Pearce, D.~L. Watson, A.~Ghazarian, M.~D. Cohler, I.~J. Thompson, and
  M.~A. Nagarajan, ``The $^{40}${C}a($t,p)^{42}${C}a reaction at triton
  energies near 10 {MeV} per nucleon,'' {\em Phys. Rev. C}, vol.~56, p.~1960,
  Oct 1997.

\bibitem{Note3}
Within this context it is of notice that the incoming proton (distorted) wave,
  in e.g. a $^{A+2}$Sn$(p,t)$ reaction, is diffused by the scattering center,
  i.e. by the $^{A+2}$Sn target, into emergent distorted waves, in particular
  the one corresponding to the relative motion of a deuteron and of the
  $^{A+1}$Sn system after the interaction $v_{np}$ has acted for the first
  time. Even when these two systems are at relative distances $r=| \protect
  \mathaccentV {vec}17Er_{^{A+1}\protect \textrm {Sn}} - \protect \mathaccentV
  {vec}17Er_{d} |$ much larger than the target radius, the Cooper pair
  wavefunction describing the pair correlation of the picked up neutron and of
  its partner in the $^{A+1}$Sn system, has a finite probability amplitude
  centered on the outgoing deuteron, and this is so not only in the case of
  superfluid nuclei (like e.g. $^{120}$Sn) but also of normal nuclei (like e.g.
  $^{132}$Sn). This is in keeping with the fact that the stability,
  collectivity and associated correlation length associated with superfluid
  Cooper pairs and with pair addition and removal Cooper pairs are very
  similar, as discussed above. Making use of this finite amplitude, the
  interaction $v_{np}$ acting a second time (see (\ref {eq1_41}) below) can
  trigger the $^{A+1}$Sn Cooper pair partner to move into the deuteron leading
  to the triton, and completing in this way the successive transfer process.
  From this narrative, it is not surprising that the paper in which the
  probabilistic interpretation of Schr{\"{o}}dinger wavefunction was forcefully
  proposed, written by Born, describes a collisional process \cite {Born:26}.

\bibitem{Tang:65}
Y.~C. Tang and R.~C. Herndon, ``Form factors of {$^{3}$}{H} and {$^{4}$}{H}e
  with repulsive--core potentials,'' {\em Phys. Lett.}, vol.~18, p.~42, 1965.

\bibitem{Note4}
Within this context one can mention the fact that, was it not for $H_{sp}$, all
  pairspins would line up in the $(x,y)$-plane transverse to the gauge axis $z$
  (see Appendix \ref {Appendix:Pair}), the pairspin alignment picture
  essentially becoming ``exact'' under such condition.

\bibitem{Guazzoni:99}
P.~Guazzoni, M.~Jaskola, L.~Zetta, A.~Covello, A.~Gargano, Y.~Eisermann,
  G.~Graw, R.~Hertenberger, A.~Metz, F.~Nuoffer, and G.~Staudt, ``{Level
  structure of $^{120}${S}n: High resolution ($p,t$) reaction and shell model
  description},'' {\em Phys. Rev. C}, vol.~60, p.~054603, 1999.

\bibitem{Guazzoni:04}
P.~Guazzoni, L.~Zetta, A.~Covello, A.~Gargano, G.~Graw, R.~Hertenberger, H.-F.
  Wirth, and M.~Jaskola, ``High-resolution study of the $^{116}${S}n$(p,t)$
  reaction and shell model structure of $^{114}${S}n,'' {\em Phys. Rev. C},
  vol.~69, no.~2, p.~024619, 2004.

\bibitem{Guazzoni:06}
P.~Guazzoni, L.~Zetta, A.~Covello, A.~Gargano, B.~F. Bayman, G.~Graw,
  R.~Hertenberger, H.-F. Wirth, and M.~Jaskola, ``Spectroscopy of
  {$^{110}$}{S}n via the high-resolution {$^{112}$}{S}n{$(p,t)$} {$^{110}$}{S}n
  reaction,'' {\em Phys. Rev. C}, vol.~74, p.~054605, 2006.

\bibitem{Guazzoni:08}
P.~Guazzoni, L.~Zetta, A.~Covello, A.~Gargano, B.~F. Bayman, T.~Faestermann,
  G.~Graw, R.~Hertenberger, H.-F. Wirth, and M.~Jaskola, ``$^{118}\mathrm{Sn}$
  levels studied by the $^{120}\mathrm{Sn}$($p$, $t$) reaction: High-resolution
  measurements, shell model, and distorted-wave born approximation
  calculations,'' {\em Phys. Rev. C}, vol.~78, p.~064608, 2008.

\bibitem{Guazzoni:11}
P.~Guazzoni, L.~Zetta, A.~Covello, A.~Gargano, B.~F. Bayman, T.~Faestermann,
  G.~Graw, R.~Hertenberger, H.-F. Wirth, and M.~Jaskola, ``High-resolution
  measurement of the $^{118,124}${S}n($p$,$t$)$^{116,122}${S}n reactions:
  Shell-model and microscopic distorted-wave {B}orn approximation
  calculations,'' {\em Phys. Rev. C}, vol.~83, no.~4, p.~044614, 2011.

\bibitem{Guazzoni:12}
P.~Guazzoni, L.~Zetta, A.~Covello, A.~Gargano, B.~F. Bayman, G.~Graw,
  R.~Hertenberger, H.-F. Wirth, T.~Faestermann, and M.~Jask\'ola, ``High
  resolution spectroscopy of ${}^{112}${S}n through the
  ${}^{114}${S}n($p$,$t$)${}^{112}${S}n reaction,'' {\em Phys. Rev. C},
  vol.~85, p.~054609, 2012.

\bibitem{Bassani:65}
G.~Bassani, N.~M. Hintz, C.~D. Kavaloski, J.~R. Maxwell, and G.~M. Reynolds,
  ``($p,~t$) ground-state {$L=0$} transitions in the even isotopes of {Sn} and
  {Cd} at 40 {MeV}, {$N=62~\mathrm{to}~74$},'' {\em Phys. Rev.}, vol.~139,
  p.~B830, 1965.

\bibitem{Potel:11PRL}
G.~Potel, F.~Barranco, F.~Marini, A.~Idini, E.~Vigezzi, and R.~A. Broglia,
  ``Calculation of the transition from pairing vibrational to pairing
  rotational regimes between magic nuclei $^{100}\mathrm{Sn}$ and
  $^{132}\mathrm{Sn}$ via two-nucleon transfer reactions,'' {\em Phys. Rev.
  Lett.}, vol.~107, p.~092501, 2011.

\bibitem{Potel:11PRL_erratum}
G.~Potel, F.~Barranco, F.~Marini, A.~Idini, E.~Vigezzi, and R.~A. Broglia,
  ``Erratum: Calculation of the transition from pairing vibrational to pairing
  rotational regimes between magic nuclei $^{100}\mathrm{Sn}$ and
  $^{132}\mathrm{Sn}$ via two-nucleon transfer reactions [phys. rev. lett. 107,
  092501 (2011)],'' {\em Phys. Rev. Lett.}, vol.~108, p.~069904, 2012.

\bibitem{An:06}
H.~An and C.~Cai, ``Global deuteron optical model potential for the energy
  range up to 183 {M}e{V},'' {\em Phys. Rev. C}, vol.~73, p.~054605, 2006.

\bibitem{Yoshida:62}
S.~Yoshida, ``Note on the two-nucleon stripping reaction,'' {\em Nucl. Phys.},
  vol.~33, p.~685, 1962.

\bibitem{Yang:62}
C.~N. Yang, ``Concept of off-diagonal long-range order and the quantum phases
  of liquid {H}e and of superconductors,'' {\em Rev. Mod. Phys.}, vol.~34,
  p.~694, 1962.

\bibitem{Gales:12}
S.~Gal{\`{e}}s, ``Study of {BCS} occupation numbers and spectroscopic factors
  from one nucleon transfer reactions,'' in {\em Fifty Years of Nuclear {BCS}}
  (R.~A. Broglia and V.~Zelevinsky, eds.), World Scientific Publishing Company,
  2012.
\newblock to be published.

\bibitem{Idini:12}
A.~Idini, F.~Barranco, and E.~Vigezzi, ``Quasiparticle renormalization and
  pairing correlations in spherical superfluid nuclei,'' {\em Phys. Rev. C},
  vol.~85, p.~014331, 2012.

\bibitem{Flynn:72}
E.~R. Flynn, G.~J. Igo, and R.~A. Broglia, ``Three-phonon monopole and
  quadrupole pairing vibrational states in $^{206}${P}b,'' {\em Phys. Lett. B},
  vol.~41, p.~397, 1972.

\bibitem{Bortignon:78}
P.~Bortignon, R.~Broglia, and D.~Bes, ``On the convergence of the nuclear field
  theory perturbation expansion for strongly anharmonic systems,'' {\em Phys.
  Lett. B}, vol.~76, p.~153, 1978.

\bibitem{Donau:67}
F.~D{\"{o}}nau, K.~Hehl, C.~Riedel, R.~Broglia, and P.~Federman, ``Two-nucleon
  transfer reaction on oxygen and the nuclear coexistence model,'' {\em Nuclear
  Physics A}, vol.~101, no.~3, pp.~495 -- 512, 1967.

\bibitem{Barz:69}
H.~Barz, K.~Hehl, C.~Riedel, and R.~Broglia, ``The structure of $^{42}${C}a and
  $^{42}${S}c investigated by two-nucleon transfer reactions,'' {\em Nuclear
  Physics A}, vol.~126, p.~577, 1969.

\bibitem{Wimmer:10}
K.~Wimmer, T.~Kr\"oll, R.~Kr\"ucken, V.~Bildstein, R.~Gernh\"auser, B.~Bastin,
  N.~Bree, J.~Diriken, P.~Van~Duppen, M.~Huyse, N.~Patronis, P.~Vermaelen,
  D.~Voulot, J.~Van~de Walle, F.~Wenander, L.~M. Fraile, R.~Chapman,
  B.~Hadinia, R.~Orlandi, J.~F. Smith, R.~Lutter, P.~G. Thirolf, M.~Labiche,
  A.~Blazhev, M.~Kalk\"uhler, P.~Reiter, M.~Seidlitz, N.~Warr, A.~O.
  Macchiavelli, H.~B. Jeppesen, E.~Fiori, G.~Georgiev, G.~Schrieder,
  S.~Das~Gupta, G.~Lo~Bianco, S.~Nardelli, J.~Butterworth, J.~Johansen, and
  K.~Riisager, ``Discovery of the shape coexisting ${0}^{+}$ state in
  $^{32}\mathrm{Mg}$ by a two neutron transfer reaction,'' {\em Phys. Rev.
  Lett.}, vol.~105, p.~252501, Dec 2010.

\bibitem{Broglia:69}
R.~A. Broglia, C.~Riedel, and T.~Udagawa, ``Nuclear spectroscopy on deformed
  nuclei with two-neutron transfer reactions,'' {\em Nucl. Phys. A}, vol.~135,
  p.~561, 1969.

\bibitem{Bes:76a}
D.~R. B{\`{e}}s, R.~A. Broglia, G.~G. Dussel, R.~J. Liotta, and H.~M.
  Sof{\'{\i}}a, ``The nuclear field treatment of some exactly soluble models,''
  {\em Nucl. Phys. A}, vol.~260, p.~1, 1976.

\bibitem{Bes:76b}
D.~R. B{\`{e}}s, R.~A. Broglia, G.~G. Dussel, R.~J. Liotta, and H.~M.
  Sof{\'{\i}}a, ``Application of the nuclear field theory to monopole
  interactions which include all the vertices of a general force,'' {\em Nucl.
  Phys. A}, vol.~260, p.~27, 1976.

\bibitem{Mottelson:62}
B.~R. Mottelson, ``Selected topics in the theory of collective phenomena in
  nuclei,'' in {\em International School of Physics ``Enrico Fermi'' Course
  {XV}, Nuclear Spectroscopy} (G.~Racah, ed.), (New York), p.~44, Academic
  Press, 1962.

\bibitem{Note5}
It is of notice that similar arguments are at the basis of the discussion of
  Bohr with Schr{\"{o}}dinger and De Broglie (matter waves), let alone with
  Einstein (instantaneous).

\bibitem{Note6}
To avoid arguments such that the wavelength of the external hadronic field
  (beam) allows it to act on both poles, one can work in terms of a Gedanken
  experiment, in which the proton bombarding energy is about 1.6 GeV ($\lambda
  \approx 1$ fm$\ll R$).

\bibitem{Potel:10}
G.~Potel, F.~Barranco, E.~Vigezzi, and R.~A. Broglia, ``{Evidence for phonon
  mediated pairing interaction in the halo of the nucleus $^{11}$Li},'' {\em
  Phy. Rev. Lett.}, vol.~105, p.~172502, 2010.

\bibitem{Potel:12}
G.~Potel and R.~A. Broglia, ``Pairing correlations with single {Cooper} pair
  transfer to individual quantal states,'' in {\em Fifty Years of Nuclear
  {BCS}} (R.~A. Broglia and V.~Zelevinsky, eds.), World Scientific Publishing
  Company, 2012.
\newblock in print, arXiv:1206.1640v1.

\bibitem{Broglia:12ArXiv}
R.~A. Broglia, ``More is different: 50 years of nuclear {BCS},'' in {\em Fifty
  Years of Nuclear {BCS}} (R.~A. Broglia and V.~Zelevinsky, eds.), World
  Scientific Publishing Company, 2012.
\newblock in print, arXiv:1206.1523v1.

\bibitem{Potel:11}
G.~{Potel}, A.~{Idini}, F.~{Barranco}, E.~{Vigezzi}, and R.~{Broglia},
  ``{Single Cooper pair transfer in stable and in exotic nuclei},'' {\em ArXiv
  e-prints}, June 2011.
\newblock [nucl-th] 0906.4298.

\bibitem{Dickhoff:05}
W.~Dickhoff and D.~Van~Neck, {\em Many-Body Theory Exposed!: Propagator
  Description of Quantum Mechanics in Many-Body Systems}.
\newblock World Scientific, 2005.

\bibitem{Jennings:11}
B.~Jennings, ``Non-observability of spectroscopic factors,'' {\em
  arXiv:1102.3721v1}.

\bibitem{Igo:59}
G.~Igo, ``Optical-model analysis of excitation function data and theoretical
  reaction cross sections for alpha particles,'' {\em Phys. Rev.}, vol.~115,
  p.~1665, 1959.

\bibitem{Kronig:26}
R.~Kronig, ``On the theory of dispersion of {X}-rays,'' {\em J. Opt .Soc. Am.},
  vol.~12, pp.~547--557, 1926.

\bibitem{Kramers:27}
H.~Kramers, ``La diffusion de la {lumi\`{e}re} par les atomes,'' {\em Atti
  Cong. Intern. Fisica (Transactions of Volta Centennial Congress) Com.},
  vol.~2, pp.~545--559, 1927.

\bibitem{Nyquist:28}
H.~Nyquist, ``Thermal agitation of electric charge in conductors,'' {\em Phys.
  rev.}, vol.~32, pp.~110--113, 1928.

\bibitem{Callen:51}
H.~Callen and T.~Welton, ``Irreversibility and generalized noise,'' {\em Phys.
  Rev.}, vol.~83, p.~34, 1951.

\end{thebibliography}

\end{document}